\documentclass[aps,prd,preprint,groupedaddress,longbibliography,nofootinbib]{revtex4-1}

\usepackage{amsmath}
\usepackage{graphicx}
\usepackage{hyperref}
\newcommand{\Sb}{S_\mathrm{bare}}

\newcommand{\lb}{\left\lbrace}
\newcommand{\rb}{\right\rbrace}
\newcommand{\fs}[1]{\hbox{$#1$\kern-0.5em\raise0.3ex\hbox{/}}}

\newcommand{\vvev}[1]{\left\langle\kern-0.3em\left\langle #1
    \right\rangle\kern-0.3em\right\rangle} 
\newcommand{\Ld}[1]{\frac{\overrightarrow{\delta}}{\delta #1}}
\newcommand{\Rd}[1]{\frac{\overleftarrow{\delta}}{\delta #1}}
\newcommand{\Tr}{\mathrm{Tr}\,}
\newcommand{\G}{\mathcal{G}}
\newcommand{\F}{\mathcal{F}_\Lambda}
\newcommand{\N}{\mathcal{N}}
\newcommand{\Op}{\mathcal{O}}
\newcommand{\E}{\mathcal{E}}
\newcommand{\SL}{S_\Lambda}
\newcommand{\RL}{R_\Lambda}

\newcommand{\nt}{\notag}

\allowdisplaybreaks

\begin{document}


\title{Explicit construction of the energy-momentum tensor\\ in the
  large $N$ limit of a four-Fermi theory}


\author{Carlo~Pagani}
\email[]{carlo.pagani@guest.ung.si}
\affiliation{University of Nova Gorica, Vipavska 13, Nova Gorica, SI-5000 EU, Slovenia}

\author{Hidenori Sonoda}\email[]{h-sonoda@pobox.com}
\affiliation{Independent Researcher, Kobe 658-0063,
  Japan}


\date{\today}

\begin{abstract}
  Using the exact renormalization group (functional RG) formalism, we
  construct the energy-momentum tensor of a four-Fermi theory in the
  Euclidean space of dimensions $2 < D < 4$ in the large $N$ limit,
  where $N$ is the number of Dirac fields.  The energy-momentum tensor
  is a functional of the Dirac fields, and its cutoff dependence is
  controlled by the exact RG equation.  In the zero cutoff limit the
  functional reduces to the one-particle-irreducible effective action
  with a single insertion of the energy-momentum tensor.  At a finite
  cutoff the energy-momentum tensor satisfies the Ward-Takahashi
  identity for translation invariance, and its antisymmetric part is
  determined by rotation invariance.  We also show that the critical
  theory is conformally invariant by deriving the corresponding
  Ward-Takahashi identity.  This implies the conformal invariance of
  any critical theory belonging to the same universality class at
  large distances.
\end{abstract}


\maketitle

\section{Introduction}{\label{introduction}}

This paper is a natural extension of our preceding work
\cite{Pagani:2025dtc}, where we have constructed the energy-momentum
(EM) tensor for the large $N$ limit of the O($N$) linear sigma model.
Associated with the conservation of energy, momentum, and angular
momentum, the EM tensor is a crucial quantity in quantum and
statistical field theory.  Moreover, the presence of either scale or
conformal symmetry can be established by studying the trace of the EM
tensor \cite{Polchinski:1987dy}.  Despite its relevance, the EM tensor
has been little studied within the exact renormalization group (ERG)
formalism; see, e.g.,
\cite{Sonoda:2015pva,Heller:2021wan,Pagani:2025dtc}.  The main aim of
this paper is to construct and study the EM tensor associated with a
fermionic theory within the ERG for the first time.  In the explicit
case studied in this work, we are able to show the presence of
conformal symmetry in the scaling limit of a non-trivial critical
point.

In this paper we consider a theory consisting of $N$ number of Dirac
fields in the Euclidean space of dimension $D$, where $2 < D < 4$.
The theory, with a scalar-scalar four-Fermi interaction, has been
known to be renormalizable for many years \cite{Wilson:1972cf}.  In
the large $N$ limit it is straightforward to calculate the
one-particle-irreducible (1PI) Wilson action.  In this paper we
construct the energy-momentum (EM) tensor as a functional of Dirac
fields.  In the zero cutoff limit the functional becomes the 1PI
generating functional of correlation functions with the insertion of a
single EM tensor carrying an arbitrary momentum.

In the case of scalar fields, the EM tensor can be made symmetric as a
consequence of the rotational invariance.  The construction of such an
EM tensor is originally due to Belinfante and Rosenfeld
\cite{Belinfante,Rosenfeld}.  In the case of Dirac fields, however,
the EM tensor cannot be made symmetric off-shell.  This is also a
consequence of the rotational invariance.  We shall see that the
intrinsic spin of the Dirac field introduces a well defined
antisymmetric part to the EM tensor.

In the large $N$ limit of a scalar theory, the trace of the EM tensor
is given by an equation-of-motion operator (to be explained in
Sec.~\ref{comp}), and this results in the Ward-Takahashi (WT) identity
for the special conformal invariance.  The implication for the
one-particle-irreducible vertex functions has recently been worked out
in detail in the paper by Cabrera et al.\cite{Cabrera:2026ggg} In the
present work we only derive the WT identity for special conformal
invariance of the fermionic theory at the critical point, leaving the
detailed study of the interaction vertices for a future work.

We denote the traceless gamma matrices by
$\gamma_\mu\,(\mu=1,2,\cdots,D)$, which satisfy the Dirac algebra
\begin{equation}
  \lb \gamma_\mu, \gamma_\nu \rb = 2 \delta_{\mu\nu} \mathbf{1},\quad
  \Tr \gamma_\mu = 0.
\end{equation} 
We use the convenient notation
\begin{equation}
  \gamma_\mu p_\mu = \fs{p},
\end{equation}
where the repeated $\mu$ is summed over $\mu = 1, \cdots, D$.  We also
write
\begin{equation}
  \int_p = \int \frac{d^D p}{(2 \pi)^D} 
\end{equation}
for the integrals in the momentum space.

\section{Review of the ERG formalism for large $N$\label{largeN}}

There are two ways of introducing a one-particle-irreducible (1PI)
Wilson action $\Gamma_\Lambda$.  One is to start from a Wilson action
$S_\Lambda$ where $\Lambda$ is an ultraviolet (UV) cutoff momentum.
The momentum modes larger than $\Lambda$ are incorporated into the
vertices of $\SL$, and the functional integral over the exponentiated
Wilson action $e^{S_\Lambda}$ does not involve integration over the
momentum modes beyond $\Lambda$.  The Wilson action is related to the
generating functional $W_\Lambda$ of the connected correlation
functions via change of field variables, and the 1PI Wilson action
$\Gamma_\Lambda$ is defined as the Legendre transform of $W_\Lambda$.
For both $W_\Lambda$ and $\Gamma_\Lambda$, the cutoff $\Lambda$ plays
the role of an infrared (IR) cutoff because the momentum modes below
$\Lambda$ have not been integrated out.  (See, for example
\cite{Igarashi:2009tj}, for this viewpoint.)  In the following we take
an alternative way, introducing $W_\Lambda$ directly in terms of a
functional integral over a bare action.  This is a less technical way
of introducing the 1PI Wilson action, and we believe it more friendly
to those who are not experts on the subject of the exact
renormalization group.  See
\cite{Wetterich:1992yh,Morris:1993qb,Ellwanger:1993mw,Reuter:1993kw}
for some early works and
\cite{Berges:2000ew,Pawlowski:2005xe,Delamotte:2007pf,Igarashi:2009tj,Metzner:2011cw,Dupuis:2020fhh}
for reviews.

Given a UV regularized bare action $\Sb [\psi, \bar{\psi}]$ of the
Dirac field $\psi$ and its conjugate $\bar{\psi}$, we define a
generating functional $W_\Lambda [J,\bar{J}]$ by
\begin{align}
  e^{W_\Lambda [J,\bar{J}]}
  &\equiv
  \int [d\psi d\bar{\psi}]\,\exp \left[
    \Sb [\psi,\bar{\psi}] - \int_p \RL (p)\, \bar{\psi} (-p) \psi
    (p) \right.\nt\\
  &\qquad\left.+ \int_p \left( \bar{J} (-p) \psi (p) + \bar{\psi} (-p) J(p)
    \right)\right],
\end{align}
where $J, \bar{J}$ are anticommuting sources.  The cutoff function
\begin{equation}
  \RL (p) \equiv \Lambda R (p/\Lambda)
\end{equation}
has the properties
\begin{equation}
  R (0) = 1,\quad R (p) \overset{p^2 \to +\infty}{\longrightarrow} 0.
\end{equation}
For example, we can take $R (p) = \exp \left( - p^2 \right)$.
$\RL (p)$ acts as a momentum dependent mass parameter that suppresses
the fluctuations of the momentum modes smaller than $\Lambda$.  Hence,
$\Lambda$ is to be considered as an IR cutoff.  In the limit
$\Lambda \to 0+$, the functional $W_\Lambda$ reduces to the generating
functional of connected correlation functions.

We then define the 1PI Wilson action $\Gamma_\Lambda$ by
\begin{equation}
  \Gamma_\Lambda [\Psi, \bar{\Psi}] - \int_p \RL (p) \bar{\Psi} (-p)
  \Psi (p) = W_\Lambda [J, \bar{J}] - \int_p \left( \bar{J} (-p) \Psi
    (p) + \bar{\Psi} (-p) J (p) \right),
\end{equation}
where we define
\begin{equation}
  \Psi (p) \equiv \Ld{\bar{J} (-p)} W_\Lambda [J,\bar{J}],\quad
  \bar{\Psi} (-p) \equiv W_\Lambda [J,\bar{J}] \Rd{J (p)}.
\end{equation}
This is a Legendre transformation.  The cutoff dependence of the
  1PI Wilson action is given by
\begin{equation}
  - \Lambda \partial_\Lambda \Gamma_\Lambda [\Psi, \bar{\Psi}]
  = - \int_p \Lambda \partial_\Lambda \RL (p) \,\Tr\, \G_{\Lambda;
    -p,p} [\Psi, \bar{\Psi}],
\end{equation}
where $\G_{\Lambda; -p, q} [\Psi, \bar{\Psi}]$ is defined by
\begin{subequations}
\begin{equation}
  \int_q \G_{\Lambda; -p, q} [\Psi, \bar{\Psi}] \left( \RL (q) \delta
    (q-r) - \Ld{\bar{\Psi} (-q)} \Gamma_\Lambda \Rd{\Psi (r)} \right)
  = \delta (p-r),
\end{equation}
or alternatively by
\begin{equation}
  \int_q \left( \RL (p) \delta (p-q) - \Ld{\bar{\Psi} (-p)}
    \Gamma_\Lambda \Rd{\Psi (q)} \right) \G_{\Lambda; -q, r} [\Psi,
  \bar{\Psi}] = \delta (p-r).
\end{equation}
\end{subequations}
In the limit $\Lambda \to 0+$, the 1PI Wilson action reduces to the
effective action
\begin{equation}
  \Gamma_\mathrm{eff} [\Psi, \bar{\Psi}]
  = \lim_{\Lambda \to 0+} \Gamma_\Lambda [\Psi, \bar{\Psi}].
\end{equation}
  
We now introduce $N$ number of Dirac fields
$\Psi^I (p)\, (I=1,\cdots,N)$ and their conjugates
$\bar{\Psi}^I (-p)$.  The 1PI Wilson action $\Gamma_\Lambda [\Psi^I,
\bar{\Psi}^I]$ now satisfies the ERG differential equation
\begin{equation}
  - \Lambda \partial_\Lambda \Gamma_\Lambda [\Psi^I, \bar{\Psi}^I]
  = - \int_p \Lambda \partial_\Lambda \RL (p)\, \Tr \G_{\Lambda; -p,
    p}^{II} ,
\end{equation}
where the repeated $I$ is summed over $I=1,\cdots, N$,
and $\G^{IJ}_{\Lambda; -p, q}$ is defined by
\begin{subequations}
\begin{equation}
  \int_q \G^{IJ}_{\Lambda; -p, q} \left( \RL (q) \delta^{JK} \delta (q-r) -
    \Ld{\bar{\Psi}^J (-q)} \Gamma_\Lambda \Rd{\Psi^K (r)} \right)
  = \delta^{IK} \delta (p-r),
\end{equation}
or alternatively by
\begin{equation}
  \int_q \left( \RL (p) \delta^{IJ} \delta (p-q) - \Ld{\bar{\Psi}^I
      (-p)} \Gamma_\Lambda \Rd{\Psi^J (q)} \right) \G^{JK}_{\Lambda;
    -q, r} = \delta^{IK} \delta (p-r).
\end{equation}
\end{subequations}
In the large $N$ approximation we take the 1PI
Wilson action in a particular form:
\begin{equation}
  \Gamma_\Lambda [\Psi^I, \bar{\Psi}^I]
  = \Gamma_G [\Psi^I, \bar{\Psi}^I] +  N \Gamma_{I\Lambda}
  [\varphi],\label{largeN-ansatz}
\end{equation}
where
\begin{equation}
\Gamma_G [\Psi^I, \bar{\Psi}^I] \equiv -  \int_p \bar{\Psi}^I (-p) i
\fs{p} \Psi^I (p)\label{largeN-GammaGauss}
\end{equation}
is the effective action of the Gaussian theory (free massless
fermions), and $\varphi$ is a bosonic scalar field defined by
\begin{equation}
  \varphi (p) \equiv \frac{1}{N} \int_q \bar{\Psi}^I (-q) \Psi^I (q+p)
  = \int d^D x\, e^{- i p x} \frac{1}{N} \bar{\Psi}^I (x) \Psi^I (x).
\end{equation}

Under the assumption (\ref{largeN-ansatz}), we can
  approximate
\begin{equation}
  \G_{\Lambda; -p, q}^{IJ} = \delta^{IJ} \G_{\Lambda; -p, q} [\varphi]
  + \cdots
\end{equation}
for large $N$, where the suppressed part does not contribute
  to the leading order.  $\G_{\Lambda; -p, q} [\varphi]$ is now a
  functional of $\varphi$, and defined by
\begin{subequations}
\begin{equation}
  \int_q \G_{\Lambda; -p, q} [\varphi] \lb \left( i \fs{q} + \RL
    (q)\right) \delta (q-r) - \frac{\delta \Gamma_{I\Lambda}}{\delta
    \varphi (r-q)} \rb = \delta (p-r),
\end{equation}
or alternatively by
\begin{equation}
  \int_q \lb \left( i \fs{p} + \RL (p)\right) \delta (p-q) -
  \frac{\delta \Gamma_{I\Lambda}}{\delta \varphi (-p+q)} \rb
  \G_{\Lambda; -q, r} [\varphi] = \delta (p-r).
\end{equation}
\end{subequations}
The cutoff dependence of the 1PI Wilson action is now given by
\begin{equation}
  - \Lambda \partial_\Lambda \Gamma_{I\Lambda} [\varphi]
  = - \int_p \Lambda \partial_\Lambda \RL (p)\, \Tr \G_{\Lambda; -p,p}
  [\varphi].\label{largeN-ERG}
\end{equation}

We can solve (\ref{largeN-ERG}) using a simple trick
\cite{Sonoda:2023ohb}\footnote{This trick was inspired by the
  well-known technique of solving a first order partial differential
  equations used in \cite{Morris:1997xj} for the effective potential.
  See \cite{DAttanasio:1997yph,Morris:1997xj,Blaizot:2005xy} for some
  early works on the large $N$ limit within ERG.}.  We introduce a
Legendre transformation of $\Gamma_{I\Lambda} [\varphi]$ by
\begin{equation}
  \Gamma_{I\Lambda} [\varphi] = F_\Lambda [\sigma] + \int_p \sigma (p)
  \varphi (-p),
\end{equation}
where
\begin{equation}
  \sigma (p) \equiv \frac{\delta \Gamma_{I\Lambda} [\varphi]}{\delta
    \varphi (-p)}.
\end{equation}
The inverse transformation gives
\begin{equation}
  \varphi (p) = - \frac{\delta F_\Lambda [\sigma]}{\delta \sigma
    (-p)}.\label{largeN-varphi-sigma}
\end{equation}
We can now regard $\G_{\Lambda; -p, q} [\varphi]$ as a functional of
$\sigma$ satisfying
\begin{subequations}
  \label{largeN-GhinvG}
  \begin{align}
    \int_q \G_{-p,q} [\sigma] \lb \left( i \fs{q} + \RL (q) \right)
    \delta (q-r) - \sigma (q-r) \rb &= \delta (p-r),\label{largeN-Ghinv}\\
    \int_q \lb \left( i \fs{p} + \RL (p) \right) \delta (p-q) - \sigma
    (p-q) \right) \G_{-q,r} [\sigma] &= \delta (p-r).\label{largeN-hinvG}
  \end{align}
\end{subequations}
(We suppress the $\Lambda$-dependence of $\G_{\Lambda; -p,q} [\sigma]$ from now
on to simplify the notation.)

We rewrite (\ref{largeN-ERG}) as
\begin{equation}
  - \Lambda \partial_\Lambda F_\Lambda [\sigma]
  = - \int_p \Lambda \partial_\Lambda \RL (p)\, \Tr \G_{-p,p}
  [\sigma].\label{largeN-ERG-F}
\end{equation}
To solve this ERG equation, we first expand $\G_{-p,q} [\sigma]$ in 
powers of $\sigma$:
\begin{align}
  &\G_{-p,q} [\sigma]
  = h_\Lambda (p) \delta (p-q)\nt\\
  &\quad + h_\Lambda (p) \cdot \sum_{n=1}^\infty
    \int_{p_1, \cdots, p_n} \sigma (p_1) \cdots \sigma (p_n)\, \delta
    \left(\sum_1^n p_i - p + q\right)\nt\\
  &\qquad \times h_\Lambda (p-p_1) h_\Lambda (p-p_1-p_2) \cdots
    h_\Lambda \left( p - p_1 - \cdots - p_{n-1}\right) \cdot h_\Lambda
    (q),
\end{align}
where
\begin{equation}
  h_\Lambda (p) \equiv \frac{1}{i \fs{p} + \RL (p)}
\end{equation}
is the high-momentum fermion propagator with $\RL (p)$ as a momentum
dependent mass.

The general solution to (\ref{largeN-ERG-F}) is
\begin{equation}
  F_\Lambda [\sigma] = \tilde{F} [\sigma] + I_\Lambda [\sigma],
\end{equation}
where $\tilde{F} [\sigma]$ is an arbitrary functional independent of
$\Lambda$.  A naive solution for $I_\Lambda [\sigma]$ is given by
\begin{align}
  I_\Lambda [\sigma]
  &= c_\Lambda \delta (0) - \sum_{n=1}^\infty \frac{1}{n} \int_{p_1,
    \cdots, p_n}  \sigma (p_1) \cdots \sigma (p_n)\, \delta
    \left(\sum_1^n p_i \right)\nt\\
  &\qquad \times \int_q \Tr \big[ h_\Lambda (q) h_\Lambda (q-p_1)
    \cdots h_\Lambda (q-p_1-\cdots-p_{n-1}) \big].\label{largeN-ILambda-naive}
\end{align}
The constant $c_\Lambda$ is
\begin{equation}
  c_\Lambda \equiv \Tr \mathbf{1} \int_q \frac{1}{2} \ln \frac{q^2 +
    \RL (q)^2}{q^2} = c\,\Lambda^D,\label{largeN-c}
\end{equation}
which is a finite integral, assuming a fast decay of $\RL (q)$ for
large $q^2$.  The $n=1$ term is UV finite, since
\begin{equation}
  c_{1\Lambda} 
  \equiv  -  \int_q \Tr h_\Lambda (q) = - \Tr \mathbf{1} \int_q
  \frac{\RL (q)}{q^2 + \RL(q)^2} = c_1 \Lambda^{D-1}.\label{largeN-c1}
\end{equation}
The $n=2$ term is UV divergent for $2 < D < 4$:
\begin{equation}
  - \int_q \Tr h_\Lambda (q) h_\Lambda (q-p) \longrightarrow \infty,
\end{equation}
where the integrand behaves as $1/q^2$ for large $q^2$.  This
must be replaced by a solution to
\begin{equation}
  - \Lambda \partial_\Lambda  \F (p) = - \int_q \Lambda
  \partial_\Lambda \RL (q) \Tr h_\Lambda (q) \left(h_\Lambda (q-p)
    + h_\Lambda (q+p) \right) h_\Lambda (q).\label{largeN-ERG-FLambda}
\end{equation}
Demanding locality (i.e., $\F (p)$ can be expanded in powers of $p^2$
at $p=0$), $\F (p)$ becomes unique since the addition of a constant
multiple of $(p^2)^{\frac{D-2}{2}}$ is not allowed by locality.
The unique solution to $\F (p)$ is given explicitly by
\begin{equation}
  \F (p) \equiv - \int_q \Tr \left[ h_\Lambda (q) \frac{1}{2}
    \left(h_\Lambda (q+p) + h_\Lambda (q-p)\right) + \frac{1}{q^2}
  \right],\label{largeN-FLambda}
\end{equation}
where the asymptotic part of the integrand is subtracted.
  
Hence, the correct definition of $I_\Lambda [\sigma]$ is
\begin{align}
I_\Lambda [\sigma]
  &= c_\Lambda \delta (0) + c_{1\Lambda} \sigma (0) + \frac{1}{2}
    \int_p \sigma (p) \sigma (-p) \F (p)\nt\\
  &\quad - \sum_{n=3}^\infty \frac{1}{n} \int_{p_1,
    \cdots, p_n}  \sigma (p_1) \cdots \sigma (p_n)\, \delta
    \left(\sum_1^n p_i \right)\nt\\
  &\qquad \times \int_q \Tr \left[ h_\Lambda (q) h_\Lambda (q-p_1)
    \cdots h_\Lambda (q-p_1-\cdots-p_{n-1}) \right].\label{largeN-ILambda}
\end{align}
For $3 \le D < 4$, the $n=3$ term is potentially UV divergent, but
thanks to
\begin{equation}
  \Tr \frac{1}{\fs{q}} \frac{1}{\fs{q}} \frac{1}{\fs{q}} =
  \frac{1}{q^4} \Tr \fs{q} = 0,
\end{equation}
it is UV finite.  Furthermore, let us note that $I_\Lambda [\sigma]$ defined naively by
(\ref{largeN-ILambda-naive}) satisfies
\begin{equation}
  \frac{\delta I_\Lambda [\sigma]}{\delta \sigma (-p)}
= - \int_q \Tr \G_{-q-p,q} [\sigma],\label{largeN-dIdsigma-naive}
\end{equation}
whose term of order $\sigma$ is divergent.  Using the correct definition
(\ref{largeN-ILambda}), we obtain instead
\begin{equation}
  \frac{\delta I_\Lambda [\sigma]}{\delta \sigma (-p)}
  = c_{1\Lambda} \delta (p) + \F (p) \sigma (p) - \int_q \Tr \lb
  \G_{-q-p,q} [\sigma] - h_\Lambda (q) \delta (p) - h_\Lambda (q+p)
  \sigma (p) h_\Lambda (q) \rb .\label{largeN-dIdsigma}
\end{equation}

The solution with $\tilde{F} = 0$ gives the scaling limit of a
critical theory which is scale invariant.  This is the fermionic
analog of the Wilson-Fisher point. In Sec.~\ref{conformal} we
  will show
\begin{enumerate}
\item that the theory with $\tilde{F} = 0$ is invariant under scale
  transformations,
\item that this scale invariant theory is also invariant under special
  conformal transformations.
\end{enumerate}

In the rest of this section we would like to describe the salient
  features of the large $N$ limit.  We keep the description concise
  giving more details in Appendix \ref{appendix-largeN}.
\begin{enumerate}  
\item \textbf{Parameters} $f_2, f_{21}$ --- Away from the critical point, we
  can introduce two parameters $f_2$ and $f_{21}$ so that
\begin{equation}
  \tilde{F} [\sigma] = \frac{1}{2}  \int_p (f_2 + f_{21} p^2) \sigma
  (p) \sigma (-p) .
  \label{largeN-Ftilde}
\end{equation}
We then obtain, from (\ref{largeN-varphi-sigma}),
\begin{align}
  \varphi (p)
  &= - \frac{\delta F_\Lambda [\sigma]}{\delta \sigma (-p)}\nt\\
  &= - \left( f_2 + f_{21} p^2\right) \sigma (p) - \frac{\delta
    I_\Lambda [\sigma]}{\delta \sigma
    (-p)}.\label{largeN-varphi-sigma2}
\end{align}
\item \textbf{Relation to a four-Fermi theory} --- 
The physical meaning of $f_2, f_{21}$ can be seen easily from its
Legendre transform:
\begin{equation}
  \tilde{\Gamma} [\varphi]
  = \tilde{F} [\sigma] + \int_p \sigma (p) \varphi (-p)
=  \frac{1}{2} \int_p \varphi (p) \frac{1}{- f_2 - f_{21} p^2} \varphi
(-p).
\label{largeN-four-Fermi}
\end{equation}
This is a four-Fermi interaction mediated by a scalar field coupled to
$\varphi$, where the propagator of the scalar field is given by
$1/(-f_2-f_{21} p^2)$. This is why we call our model a four-Fermi
  theory in the title of the paper.  The above argument may be too
  crude, however; in Appendix \ref{appendix-largeN}
we show how (\ref{largeN-Ftilde}) is obtained from a bare theory with
  a four-Fermi interaction in the limit of an infinite momentum cutoff.
\item \textbf{RG flow of} $f_2, f_{21}$ --- 
The parameter $f_2$ has the mass dimension $D-2 > 0$, and $f_{21}$ has
$D- 4 < 0$.  At the critical point, $f_2$ is relevant, but $f_{21}$ is
irrelevant.  As we take $f_{21} \to - \infty$, we obtain the Gaussian
theory
\begin{equation}
  \Gamma_{I\Lambda} [\varphi] = \mathrm{constant}.
\end{equation}
See Fig.~\ref{Fig1} for RG flows.  
\begin{figure}[ht]
  \centering
  \includegraphics{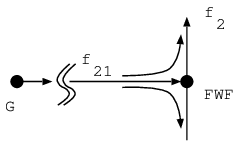}
  \caption{RG flows are sketched.  The theory is critical if
    $f_2 = 0$.  FWF, the fermionic analog of the Wilson-Fisher fixed
    point, is the scaling limit (i.e., $f_{21} =
    0$) of a critical point.  As $f_{21} \to
    -\infty$, the theory approaches the Gaussian theory (G).}
  \label{Fig1}
\end{figure}
\item $\mathbf{Z}_2$ \textbf{invariance} ---
In the limit $\Lambda \to 0+$, we find $I_\Lambda [\sigma]$ invariant
under the $\mathbf{Z}_2$ transformation
\begin{equation}
  \sigma (p) \longrightarrow - \sigma (-p).
\end{equation}
This comes from the invariance of the theory under
\begin{equation}
\mathbf{Z}_2:  \Psi^I (p) \longrightarrow \Psi^I (-p),\quad
  \bar{\Psi}^I (-p) \longrightarrow - \bar{\Psi}^I (-p).
\end{equation}
Though the invariance is explicitly broken by the mass term
proportional to the cutoff function $\RL (p)$, it is restored in the
limit $\Lambda \to 0+$.  The corresponding symmetry is exact for
$f_2 \le 0$, and spontaneously broken for $f_2 > 0$.  We give the
effective potential in Appendix \ref{appendix-largeN}.
\end{enumerate}

\section{Properties of the energy-momentum tensor\label{properties}}

The EM tensor $\Theta_{\mu\nu} (p)$ is a cutoff dependent functional
of $\Psi^I$ and its conjugate $\bar{\Psi}^I$.  In the limit
$\Lambda \to 0+$, the EM tensor becomes the effective action with the
insertion of a single EM tensor with momentum $p$.  The EM tensor is
what is called a composite operator whose cutoff dependence is
precisely controlled by the ERG differential equation to be discussed
in Sec.~\ref{comp}.  In practice, we shall see that the EM tensor, up
to Belinfante and Rosenfeld improvements, is identified by demanding
the WT identities associated with translations and rotations and by
requiring it be a composite operator.

Let us discuss briefly where the EM tensor comes from.
The change of the Dirac fields under infinitesimal translations is
given by
\begin{equation}
  D^T_\mu \Psi^I (p) = p_\mu \Psi^I (p),\quad
  D^T_\mu \bar{\Psi}^I (-p) = - p_\mu \bar{\Psi}^I (-p).\label{properties-DT}
\end{equation}
Similarly, the change under infinitesimal rotations is given by
\begin{subequations}
\begin{align}
  D^R_{\mu\nu} \Psi^I (p)
  &= \left( p_\mu \frac{\partial}{\partial
    p_\nu} - p_\nu \frac{\partial}{\partial p_\mu} + \Sigma_{\mu\nu} \right) \Psi^I (p),\\
  D^R_{\mu\nu} \bar{\Psi}^I (-p)
  &= \left( p_\mu \frac{\partial}{\partial
    p_\nu} - p_\nu \frac{\partial}{\partial p_\mu} \right)
    \bar{\Psi}^I (-p) - \bar{\Psi}^I (-p)  \Sigma_{\mu\nu},\label{properties-DR}
\end{align}
\end{subequations}
where the spin matrix is defined by
\begin{equation}
  \Sigma_{\mu\nu} \equiv \frac{1}{4} \left( \gamma_\mu \gamma_\nu -
    \gamma_\nu \gamma_\mu \right).\label{properties-spin-matrix}
\end{equation}
The 1PI Wilson action is invariant under both translations and
rotations as long as the cutoff function $\RL (p)$ is rotation
invariant:
\begin{align}
  & \int_q \left( \Gamma_\Lambda \Rd{\Psi^I (q)} D^T_\mu \Psi^I (q) +
    D^T_\mu \bar{\Psi}^I (-q) \Ld{\bar{\Psi}^I (-q)} \Gamma_\Lambda
    \right) = 0,\label{properties-translation}\\
  &\int_p \left( \Gamma_\Lambda \Rd{\Psi^I (q)} D^R_{\mu\nu} \Psi^I (q) +
    D^R_{\mu\nu} \bar{\Psi}^I (-q) \Ld{\bar{\Psi}^I (-q)} \Gamma_\Lambda
    \right) = 0.\label{properties-rotation}
\end{align}

Following a discussion summarized in Appendix \ref{appendix-BR},
we can show the existence of an EM tensor $\Theta_{\mu\nu} (p)$ with
the following two properties after a Belinfante-Rosenfeld type
improvement.  The first property is the Ward-Takahashi (WT) identity
for the translation invariance:
\begin{align}
  p_\mu \Theta_{\mu\nu} (p)
  &= \int_q \left(  \Gamma_\Lambda \Rd{\Psi^I (q)} (q+p)_\nu \Psi^I
    (q+p) + \left(-q+p\right)_\nu \bar{\Psi}^I (-q+p) \Ld{\bar{\Psi}^I (-q)}
    \Gamma_\Lambda \right)\nt\\
  &\quad -  N \int_q \RL (q) \Tr \lb (q+p)_\nu \G_{-q-p,q} +
    (-q+p)_\nu \G_{-q,q-p} \rb \nt\\
  &= \int_q \left(  \Gamma_G \Rd{\Psi^I (q)} (q+p)_\nu \Psi^I
    (q+p) + \left(-q+p\right)_\nu \bar{\Psi}^I (-q+p) \Ld{\bar{\Psi}^I (-q)}
    \Gamma_G \right)\nt\\
  &\quad + N \int_q 
    (q+p)_\nu \varphi (q+p) \frac{\delta \Gamma_{I\Lambda}
    [\varphi]}{\delta \varphi (q)} \nt\\
   &\quad -  N \int_q \RL (q) \Tr \lb (q+p)_\nu \G_{-q-p,q} +
    (-q+p)_\nu \G_{-q,q-p} \rb.\label{properties-WT}
\end{align}
The second property is the antisymmetric part of the EM tensor
\begin{align}
    \frac{1}{2} \left(\Theta_{\mu\nu} (p) - \Theta_{\nu\mu} (p)\right)
  &= A_{\mu\nu} (p)\nt\\  
  &\equiv \int_q \left( \Gamma_G \Rd{\Psi^I (q)} \frac{1}{2} \Sigma_{\mu\nu} \Psi^I
    (q+p) - \bar{\Psi}^I (-q+p) \frac{1}{2} \Sigma_{\mu\nu} \Ld{\bar{\Psi}^I (-q)}
    \Gamma_G \right)\nt\\
  &\quad + N \int_q \RL (q)\, \Tr \left(
    - \frac{1}{2} \Sigma_{\mu\nu} \G_{-q-p,q} + \G_{-q,q-p}
    \frac{1}{2} \Sigma_{\mu\nu}
    \right). \label{properties-antisym}
\end{align}
This is a consequence of the rotational invariance.  For scalar
theories, we can take $\Theta_{\mu\nu} (p)$ symmetric with respect to
$\mu, \nu$.  The right-hand sides of (\ref{properties-WT}) and
(\ref{properties-antisym}) are equation-of-motion composite operators
to be explained in Sec.~\ref{comp-subsec-eom}.

We find it convenient to first discuss the Gaussian part of the EM
tensor. Let $\Theta^G_{\mu\nu} (p)$ be the EM tensor of the Gaussian
theory.  (See Appendix \ref{appendix-free} for more details.)  It is
given by
\begin{equation}
  \Theta^G_{\mu\nu} (p)
  \equiv \int_q \bar{\Psi}^I (-q) t_{\mu\nu} (-q,q+p) \Psi^I (q+p),
  \label{properties-EMGauss}
\end{equation}
where
\begin{subequations}
  \label{properties-tmunu}
\begin{align}
  t_{\mu\nu} (-q,q+p)
  &\equiv s_{\mu\nu} (-q,q+p) + a_{\mu\nu} (-q,q+p),\\
s_{\mu\nu} (-q,q+p) &\equiv - \delta_{\mu\nu} \frac{1}{2} \left( i \fs{q} + i
    (\fs{q}+\fs{p})\right)
    + \frac{1}{4} \left( \gamma_\mu i (q + q+p)_\nu + \gamma_\nu i (q
    + q+p)_\mu \right),\label{properties-smunu}\\
  a_{\mu\nu} (-q,q+p) &\equiv - i \fs{q} \frac{1}{2} \Sigma_{\mu\nu} + \frac{1}{2}
  \Sigma_{\mu\nu} i \left(\fs{q}+\fs{p}\right).\label{properties-amunu}
\end{align}
\end{subequations}
We find
\begin{subequations}
  \label{properties-properties-tmunu}
\begin{align}
  p_\mu t_{\mu\nu} (-q,q+p)
  &= - (q+p)_\nu i \fs{q} + q_\nu i \left(\fs{q}+\fs{p}\right),\\
  \left( t_{\mu\nu}  - t_{\nu\mu} \right) (-q,q+p)
  &= 2 a_{\mu\nu} (-q,q+p).
\end{align}
\end{subequations}
In coordinate space we obtain
\begin{align}
  \Theta^G_{\mu\nu} (x)
  &\equiv \int_p e^{i p x} \Theta^G_{\mu\nu} (p)\nt\\
  &= - \delta_{\mu\nu} \frac{1}{2} \bar{\Psi}^I (x)
    \overleftrightarrow{\fs{\partial}} \Psi^I (x) + \frac{1}{4}
    \bar{\Psi}^I (x) \left( \gamma_\mu
    \overleftrightarrow{\partial}_\nu + \gamma_\nu
    \overleftrightarrow{\partial}_\mu \right) \Psi^I (x)\nt\\
  &\quad + \frac{1}{2} \bar{\Psi}^I (x) \left( \Sigma_{\mu\nu} \fs{\partial} +
    \overleftarrow{\fs{\partial}} \Sigma_{\mu\nu} \right) \Psi^I (x),
\end{align}
where
$f\overleftrightarrow{{\partial}}g \equiv f\partial g-\partial f \cdot
g$ and
$\bar{\Psi}^I (x) \overleftarrow{\fs{\partial}} \equiv \partial_\mu
\bar{\Psi}^I (x) \gamma_\mu$.  Eq.~(\ref{properties-properties-tmunu})
imply
\begin{subequations}
  \begin{align}
    p_\mu \Theta^G_{\mu\nu} (p)
    &= \int_q \left( \Gamma_G \Rd{\Psi^I (q)} (q+p)_\nu \Psi^I (q+p) +
      (-q+p)_\nu \bar{\Psi}^I (-q+p) \Ld{\bar{\Psi}^I (-q)} \Gamma_G
      \right),\\
    \Theta^G_{\mu\nu} (p) - \Theta^G_{\nu\mu} (p)
    &= \int_q \left( \Gamma_G \Rd{\Psi^I (q)} \Sigma_{\mu\nu} \Psi^I
      (q+p) - \bar{\Psi}^I (-q+p) \Sigma_{\mu\nu} \Ld{\bar{\Psi}^I
      (-q)} \Gamma_G \right).
  \end{align}
\end{subequations}
Hence, we obtain
\begin{align}
  p_\mu \frac{1}{N} \left( \Theta_{\mu\nu} (p) - \Theta^G_{\mu\nu} (p) \right)
  &= \int_q (q+p)_\nu \varphi (q+p) \, \frac{\delta
    \Gamma_{I\Lambda} [\varphi]}{\delta \varphi (q)}\nt\\
  &\quad -  \int_q \RL (q) \,\Tr \lb
    (q+p)_\nu \G_{-q-p,q}+ (-q+p)_\nu \G_{-q,q-p} \rb
    \label{properties-diff-WT}
\end{align}
from (\ref{properties-WT}), and
\begin{equation}
   \frac{1}{N} \lb \left(\Theta_{\mu\nu} - \Theta^G_{\mu\nu}\right)
    (p) - \left( \mu \leftrightarrow \nu \right) \rb
  = \int_q \RL (q) \, \Tr \left( - \Sigma_{\mu\nu} \G_{-q-p,q} + \G_{-q,q-p}
    \Sigma_{\mu\nu} \right)\label{properties-diff-antisym}
\end{equation}
from (\ref{properties-antisym}).

Using the $\sigma$ field, we can rewrite (\ref{properties-diff-WT}) as
\begin{align}
&  p_\mu \frac{1}{N} \left( \Theta_{\mu\nu} (p) - \Theta^G_{\mu\nu} (p) \right)\nt\\
  &= \int_q q_\nu \frac{\delta F_\Lambda [\sigma]}{\delta \sigma (q)}
    \sigma (q+p) - \int_q R_\Lambda (q) \,\Tr \lb (q+p)_\nu
    \G_{-q-p,q} + (-q+p)_\nu \G_{-q,q-p} \rb\nt\\
  &= \int_q q_\nu \left( f_2  + f_{21} q^2\right) \sigma (-q) \sigma
    (q+p) \nt\\
  &\quad + \int_q q_\nu \frac{\delta I_\Lambda [\sigma]}{\delta
    \sigma (q)} \sigma (q+p)
     - \int_q R_\Lambda (q) \,\Tr \lb (q+p)_\nu
    \G_{-q-p,q} + (-q+p)_\nu \G_{-q,q-p} \rb.
\end{align}
To solve this partially, we consider the $\Lambda$-independent part of
$F_\Lambda$ and introduce a symmetric tensor
\begin{align}
  \tilde{\Theta}_{\mu\nu} (p)
  &\equiv - \frac{1}{2} \int_q \sigma (q+p) \sigma (-q) \lb
    \delta_{\mu\nu} \left( f_2 + f_{21} \left(p^2 + q(q+p) \right)
    \right)\right.\nt\\
  &\left.\qquad\qquad + f_{21}\left( q_\mu (q+p)_\nu + q_\nu
    (q+p)_\mu\right) \rb\nt\\
  &\quad + \mathrm{const} \, f_{21} \left( p^2 \delta_{\mu\nu} - p_\mu p_\nu
    \right)  \frac{1}{2} \int_q \sigma (q+p) \sigma (-q)\label{properties-tildeTheta}
\end{align}
that satisfies
\begin{equation}
  p_\mu \tilde{\Theta}_{\mu\nu} (p)
  = \int_q q_\nu \left(f_2 + f_{21} q^2\right) \sigma (q+p) \sigma
  (-q) .
\end{equation}
Then, we can write
\begin{equation}
\Theta_{\mu\nu} (p) = \Theta^G_{\mu\nu} (p) + N
\left( \tilde{\Theta}_{\mu\nu} (p) +  \Theta'_{\mu\nu} (p)\right),
\end{equation}
so that $\Theta'_{\mu\nu} (p)$ satisfies the WT identity
\begin{equation}
  p_\mu \Theta'_{\mu\nu} (p) = \int_q q_\nu \frac{\delta I_\Lambda [\sigma]}{\delta
    \sigma (q)} \sigma (q+p)  - \int_q R_\Lambda (q) \,\Tr \lb 
  (q+p)_\nu \G_{-q-p,q} + (-q+p)_\nu \G_{-q,q-p} \rb.\label{properties-WTprime}
\end{equation}

Since the antisymmetric part of $\Theta'_{\mu\nu} (p)$ is determined
by (\ref{properties-diff-antisym}), it is more convenient to write
$\Theta'_{\mu\nu} (p)$ as the sum of the symmetric and
antisymmetric parts:
\begin{equation}
  \Theta'_{\mu\nu} (p) = S'_{\mu\nu} (p) + A'_{\mu\nu} (p),
\end{equation}
where 
\begin{equation}
  A'_{\mu\nu} (p) = \int_q \RL (q) \Tr \left( - \frac{1}{2}
    \Sigma_{\mu\nu} \G_{-q-p,q} + \G_{-q,q-p} \frac{1}{2}
    \Sigma_{\mu\nu} \right).
  \label{properties-antisymprime}
\end{equation}
Then, the symmetric part satisfies the WT identity
\begin{align}
  p_\mu S'_{\mu\nu} (p)
  &= \int_q q_\nu \frac{\delta I_\Lambda [\sigma]}{\delta
    \sigma (q)} \sigma (q+p)  - \int_q R_\Lambda (q) \,\Tr \lb 
    (q+p)_\nu \G_{-q-p,q} + (-q+p)_\nu \G_{-q,q-p} \rb\nt\\
  &\quad + \int_q \RL (q) \Tr \left[ \frac{1}{2} p_\mu \Sigma_{\mu\nu}
    \left( \G_{-q-p,q} - \G_{-q,q-p} \right) \right] .\label{properties-WTprime}
\end{align}
We are going to construct $S'_{\mu\nu} (p)$ that satisfies
(\ref{properties-WTprime}) in Sec.~\ref{construction}.

\section{Cutoff dependent composite operators in large $N$\label{comp}}

As a preparation for constructing $\Theta'_{\mu\nu} (p)$, we introduce
what we call composite operators that are functionals with specific
$\Lambda$-dependence.  This notion was first introduced by Becchi
\cite{Becchi:1996an}; see \cite{Igarashi:2009tj} for review.

Let $\Op_\Lambda$ be a functional of $\Psi^I, \bar{\Psi}^I$.  We call
it a composite operator if its cutoff dependence is the same as an
infinitesimal change $\Op_\Lambda$ of the 1PI Wilson action
$\Gamma_\Lambda$.  In the large $N$ limit this means
\begin{equation}
   - \Lambda \partial_\Lambda \Op_\Lambda
  = - \int_q \Lambda \partial_\Lambda \RL (q) \int_{r,s} \Tr
    \G_{-q,s} \Ld{\bar{\Psi}^I (-s)} \Op_\Lambda \Rd{\Psi^I
    (r)} \G_{-r,q}.\label{comp-ERG}
\end{equation}
The EM tensor $\Theta_{\mu\nu} (p)$ is a composite operator.  As
$\Gamma_\Lambda$ becomes the effective action in the limit
$\Lambda \to 0+$, the above $\Lambda$ dependence guarantees that
$\Theta_{\mu\nu} (p)$ becomes the 1PI generating functional with a single
insertion of the EM tensor.

\subsection{$\sigma$ as a composite operator\label{comp-subsec-sigma}}

Let us show that $\sigma (p)$, regarded as a functional of $\varphi$,
is a composite operator.  If $\Op_\Lambda$ is a functional of
$\varphi$, we find
\begin{equation}
  \Ld{\bar{\Psi}^I (-s)} \Op_\Lambda [\varphi] \Rd{\Psi^I (r)}
  = \frac{1}{N} \int_t \left(\Psi^I (t+s) \frac{\delta \Op_\Lambda [\varphi]}
    {\delta \varphi (t)}\right) \Rd{\Psi^I (r)}
= \frac{\delta \Op_\Lambda [\varphi]}{\delta \varphi (r-s)}
\end{equation}
in large $N$.  Hence, the ERG equation (\ref{comp-ERG}) becomes
\begin{equation}
  - \Lambda \partial_\Lambda \Op_\Lambda [\varphi] = - \int_q \Lambda
  \partial_\Lambda R_\Lambda (q) \int_{r,s} \Tr \G_{-q,s}
  \frac{\delta \Op_\Lambda [\varphi]}{\delta \varphi (r-s)} \G_{-r,q}
  .\label{comp-ERG-varphi} 
\end{equation}

Differentiating the ERG equation
\begin{equation}
  - \Lambda \partial_\Lambda \Gamma_{I\Lambda} [\varphi] = - \int_q
  \Lambda \partial_\Lambda R_\Lambda (q)\,\Tr \G_{-q, q}
  [\sigma]
\end{equation}
with respect to $\varphi$, we obtain
\begin{equation}
  - \Lambda \partial_\Lambda \frac{\delta
    \Gamma_{I\Lambda}[\varphi]}{\delta \varphi (-p)} = - \int_q
  \Lambda \partial_\Lambda R_\Lambda (q)\, \Tr \frac{\delta}{\delta
    \varphi (-p)} \G_{ -q,q} [\sigma],
\end{equation}
where $\varphi$ is fixed under the derivative with respect to
$\Lambda$.  Differentiating (\ref{largeN-Ghinv})
\begin{equation}
  \int_r \G_{ -q, r} [\sigma] \lb (i \fs{r} + R_\Lambda (r)) \delta
  (r-s) - \sigma (r-s)\rb = \delta (q-s)
\end{equation}
with respect to $\varphi (-p)$, we obtain
\begin{equation}
  \int_r \frac{\delta \G_{ -q,r}}{\delta \varphi (-p)} \lb
    ( i\fs{r} + R_\Lambda (r)) \delta (r-s) - \sigma (r-s) \rb =
  \int_r \G_{ -q, r} \frac{\delta \sigma (r-s)}{\delta \varphi
    (-p)}.
\end{equation}
This gives
\begin{equation}
  \frac{\delta \G_{ -q,r}}{\delta \varphi (-p)} = \int_{s,t}
  \G_{ -q,s} \frac{\delta \sigma (s-t)}{\delta \varphi (-p)}
  \G_{ -t, r}.
\end{equation}
Hence, regarding $\sigma$ as a functional of $\varphi$, we obtain
\begin{equation}
  - \Lambda \partial_\Lambda \sigma (p)\Big|_\varphi = - \int_{q,r,s}
  \Lambda \partial_\Lambda R_\Lambda (q) \,\Tr \left[ \G_{
      -q,r} \frac{\delta \sigma (r-s)}{\delta \varphi (-p)}
    \G_{ -s, q} \right].
\end{equation}
Using
\begin{equation}
  \frac{\delta \sigma (r-s)}{\delta \varphi (-p)} = \frac{\delta
    \sigma (p)}{\delta \varphi (s-r)},
\end{equation}
we obtain
\begin{equation}
  - \Lambda \partial_\Lambda \sigma (p)\Big|_\varphi = - \int_{q,r,s}
  \Lambda \partial_\Lambda R_\Lambda (q) \,\Tr \left[ \G_{
      -q,r} \frac{\delta \sigma (p)}{\delta \varphi (s-r)}
    \G_{ -s, q} \right].\label{comp-ERG-sigma}
\end{equation}
Comparing this with (\ref{comp-ERG-varphi}), we find that $\sigma (p)$
is a composite operator.  Products of $\sigma$'s with
$\Lambda$-independent coefficients are also composite operators.
Hence, $\tilde{\Theta}_{\mu\nu} (p)$ given by
(\ref{properties-tildeTheta}) is a composite operator.

Conversely, we can ask if a polynomial of $\sigma$
\begin{equation}
  \Op_\Lambda (p) = \int_{p_1, \cdots, p_n} \delta \left(\sum_1^n p_i
    - p \right) \, \sigma (p_1) \cdots \sigma (p_n)\, C_\Lambda
  (p_1,\cdots, p_n)
\end{equation}
is a composite operator, where $C_\Lambda (p_1, \cdots, p_n)$ is a
function of the momenta.  Since $\sigma$ is a functional of $\varphi$,
$\Op_\Lambda (p)$ is a functional of $\varphi$, and
$\Op_\Lambda (p)$ must satisfy
(\ref{comp-ERG-varphi}) for it to be a composite operator.   This gives
\begin{equation}
  - \Lambda \partial_\Lambda C_\Lambda (p_1, \cdots, p_n) = 0
\end{equation}
since each $\sigma$ satisfies (\ref{comp-ERG-varphi}).   Thus,
$C_\Lambda (p_1,\cdots,p_n)$ must be independent of $\Lambda$.

\subsection{Quadratic composite operators\label{comp-subsec-quadratic}}

We next consider a functional of the following type:
\begin{align}
  \Op_\Lambda (p)
  &= \int_{p_1, p_2} \delta (p_1+p_2 - p) \Tr \left[
    C (p_1, p_2) \left( -\Psi^I (p_1) \bar{\Psi}^I (p_2) - N \G_{
    -p_1, - p_2} [\sigma] \right) \right]\nt\\
  &= \int_{p_1, p_2} \delta (p_1+p_2 - p) \lb
    \bar{\Psi}^I (p_2) C(p_1, p_2) \Psi^I (p_1) - N \Tr \left[
    C(p_1, p_2) \G_{ -p_1, -p_2} [\sigma] \right] \rb,
\end{align}
where $C (p_1, p_2)$ is a $\Lambda$-independent polynomial in
  momenta.  We wish to show that this is a composite operator.  The
main part of the EM tensor, which is
$\Theta^G_{\mu\nu} + N \Theta'_{\mu\nu}$, is a composite operator of
this type.

We first compute the $\Lambda$-dependence:
\begin{equation}
  - \Lambda \partial_\Lambda \Op_\Lambda (p)\Big|_{\Psi, \bar{\Psi}}
  = - N \int_{p_1, p_2} \delta (p_1+p_2-p) \, \Tr C (p_1, p_2) \left(
    - \Lambda \partial_\Lambda \right) \G_{ -p_1, -p_2}
  [\sigma]\Big|_\varphi ,
\end{equation}
where
\begin{equation}
  - \Lambda \partial_\Lambda \G_{ -p_1, -p_2}
  [\sigma]\Big|_\varphi
  = - \Lambda \partial_\Lambda \G_{ -p_1, -p_2}
  [\sigma]\Big|_\sigma + \int_r \frac{\delta \G_{ -p_1,
      -p_2}}{\delta \sigma (r)} \left( - \Lambda \partial_\Lambda
    \sigma (r) \right)\Big|_\varphi .\label{comp-ERG-Gvarphi}
\end{equation}
Differentiating again (\ref{largeN-Ghinv})
\begin{equation}
  \int_q \G_{ -p,q} [\sigma] \lb (i\fs{q} + R_\Lambda (q)) \delta (q-r)
  - \sigma (q-r)\rb = \delta (p-r)
\end{equation}
with respect to $\Lambda$, we obtain
\begin{equation}
  - \Lambda \partial_\Lambda \G_{ -p_1, -p_2} [\sigma]\Big|_\sigma
  = \int_q \G_{ -p_1, q} \Lambda \partial_\Lambda R_\Lambda (q) 
  \G_{ -q, -p_2} .\label{comp-ERG-Gsigma}
\end{equation}
(This can also be obtained from the series expansion of
$\G_{-p_1,-p_2}$.)  Hence, from
(\ref{comp-ERG-sigma},\ref{comp-ERG-Gvarphi},\ref{comp-ERG-Gsigma}),
we obtain
\begin{align}
  - \Lambda \partial_\Lambda \G_{ -p_1. -p_2}
  [\sigma]\Big|_\varphi
  &= \int_q \Lambda \partial_\Lambda R_\Lambda (q) \, \G_{
    -p_1, q} \G_{ -q, -p_2} \nt\\
  &\quad - \int_r \frac{\delta \G_{ -p_1, -p_2}}{\delta \sigma
    (r)} \int_{q,s,t} \Lambda \partial_\Lambda R_\Lambda (q) \,
    \Tr \left[ \G_{ -q,s} \frac{\delta \sigma (r)}{\delta
    \varphi (t-s)} \G_{ -t,q} \right]\nt\\
  &= \int_q \Lambda \partial_\Lambda R_\Lambda (q) \, \left[ \G_{
    -p_1, q} \G_{ -q, -p_2} 
 - \int_{s,t} \frac{\delta \G_{ -p_1, -p_2}}{\delta
    \varphi (t-s)}  \Tr \left( \G_{ -q,s} \G_{ -t,q} \right) \right].
\end{align}
We thus obtain
\begin{align}
  - \Lambda \partial_\Lambda \Op_\Lambda (p)\Big|_{\Psi,\bar{\Psi}}
  &=  - N \int_{p_1, p_2} \delta (p_1+p_2-p) \, \Tr \Bigg[ C(p_1, p_2) \int_q
    \Lambda \partial_\Lambda R_\Lambda (q)\nt\\
  &\quad \times \lb \G_{ -p_1, q}
    \G_{ -q, -p_2} - \int_{s,t} \frac{\delta \G_{ -p_1,
    -p_2}}{\delta \varphi (t-s)} \,\Tr \left( \G_{ -q,s}
    \G_{ -t,q} \right) \rb \Bigg].
\end{align}
This agrees with the expected ERG equation:
\begin{align}
  - \Lambda \partial_\Lambda \Op_\Lambda (p)\Big|_{\Psi,\bar{\Psi}}
  &= - \int_q \Lambda \partial_\Lambda R_\Lambda (q) \int_{r,s} \Tr
    \G_{ -q,s} \Ld{\bar{\Psi}^I (-s)} \Op_\Lambda \Rd{\Psi^I
    (r)} \G_{ -r,q}\nt\\
  &=  - N \int_q \Lambda \partial_\Lambda R_\Lambda (q) \int_{p_1, p_2}
    \delta (p_1+p_2-p) \, \Tr \left[\G_{ -q, -p_2} C(p_1, p_2)
    \G_{ - p_1,q}\right]\nt\\
  &\quad + N \int_q \Lambda \partial_\Lambda R_\Lambda (q) \int_{p_1, p_2}
    \delta (p_1+p_2-p) \nt\\
  &\qquad \times \int_{r,s} \Tr \, \lb \G_{ -q,s} \G_{ -r, q}\rb
    \cdot \Tr \left[ C(p_1, p_2) \frac{\delta \G_{ -p_1, -p_2}}{\delta
    \varphi (r-s)}\right] .
\end{align}
Hence, $\Op_\Lambda (p)$ is a composite operator.

The only problem with $\Op_\Lambda (p)$ is its potential UV
divergence; the integral
\[
-  \int_{p_1, p_2} \delta (p_1+p_2-p) \,\Tr \left[ C(p_1, p_2) \,\G_{
    -p_1, -p_2} [\sigma] \right]
\]
may not converge.  For example, look at the terms up to quadratic in
$\sigma$:
\begin{align}
  &(\textrm{terms up to $\sigma^2$})
  = - \int_{p_1, p_2} \delta (p_1+p_2-p)\, \Tr C (p_1, p_2)\,
    \Bigg[ h_\Lambda (p_1) \delta (p_1+p_2)\nt\\
  &\quad + h_\Lambda (p_1) \lb
    \sigma (p_1+p_2) + \int_{q_1,q_2}  \sigma (q_1) \sigma (q_2) \,\delta
    (q_1+q_2-p_1-p_2)\, h_\Lambda (p_1-q_1)  \rb h_\Lambda (-p_2) \Bigg]\nt\\
  &= - \delta (p) \,\int_q \Tr C(q,-q) h_\Lambda (q)
    - \sigma (p) \int_q \Tr C (q+p, -q) h_\Lambda (q+p) h_\Lambda
    (q)\nt\\
  &\quad - \frac{1}{2} \int_r \sigma (r+p) \sigma (-r) \int_q \Tr C (q+p, -q)
    h_\Lambda (q+p) \left( h_\Lambda (q-r) + h_\Lambda (q+p+r) \right)
    h_\Lambda (q).
\end{align}
The integrals over $q$ may be UV divergent.  Suppose
\[
-  \int_q \Tr C(q,-q) h_\Lambda (q)
\]
is UV divergent.   We must replace it by a UV finite $A_\Lambda$ that satisfies
\begin{equation}
  - \Lambda \partial_\Lambda A_\Lambda = - \int_q \Tr C(q,-q) f_\Lambda
  (q),
\end{equation}
where
\begin{equation}
  f_\Lambda (q) \equiv - \Lambda \partial_\Lambda h_\Lambda (q)
  = \Lambda \partial_\Lambda \RL (q)\, h_\Lambda (q)^2.\label{comp-fLambda}
\end{equation}
Similarly, we replace the $q$-integrals by UV finite terms with the
same $\Lambda$-dependence.  We will consider concrete examples
$\N_\Lambda (p), A_{\mu\nu} (p)$ in Sec.~\ref{comp-subsec-eom}.

\subsection{Composite operator $\left[ \varphi (p)\right]$\label{comp-subsec-varphi}}

Let us recall (\ref{largeN-varphi-sigma2}):
\begin{align}
  \varphi (p)
  &= \frac{1}{N} \int_q \bar{\Psi}^I (-q) \Psi^I (q+p)
  = - \frac{\delta F_\Lambda [\sigma]}{\delta \sigma (-p)}
  = - \left( f_2 + f_{21} p^2 \right) \sigma (p) - \frac{\delta
    I_\Lambda [\sigma]}{\delta \sigma (-p)}\nt\\
  &= - \left( f_2 + f_{21} p^2 \right) \sigma (p) - c_{1\Lambda}
    \delta (0) - \F (p) \sigma (p) \nt\\
  &\quad + \int_q \Tr \left( \G_{ -q-p,q} [\sigma] - h_\Lambda
    (q) \delta (p) - h_\Lambda (q+p) \sigma (p) h_\Lambda (q) \right).
\end{align}
The second line is a quadratic composite operator discussed in
Sec.~\ref{comp-subsec-quadratic} above, corresponding to $C=1$:
\begin{align}
N  \left[ \varphi (p) \right]
  &\equiv \int_q \bar{\Psi}^I (-q) \Psi^I (q+p) + N c_{1\Lambda} \delta
    (0) + N \F (p) \sigma (p) \nt\\
  &\quad - N \int_q \Tr \left( \G_{ -q-p,q} [\sigma] - h_\Lambda
    (q) \delta (p) - h_\Lambda (q+p) \sigma (p) h_\Lambda (q)
    \right)\nt\\
  &= - N \left( f_2 + f_{21} p^2 \right) \sigma (p),
\end{align}
where
\[
  - \int_q \Tr h_\Lambda (q+p) h_\Lambda (q)
\]
is replaced by $\F (p)$ for UV finiteness.  At the critical point,
where $f_2 = f_{21} = 0$, we find
\begin{equation}
  \left[ \varphi (p) \right] = 0.\quad \left(\textrm{critical
      point}\right)
  \label{comp-vanishingvarphi}
\end{equation}

\subsection{Equation-of-motion composite operators\label{comp-subsec-eom}}

There is a special class of composite operators, called the
equation-of-motion composite operators
\cite{Becchi:1996an,Igarashi:2009tj}.  These are obtained as the
change of the Wilson action under infinitesimal changes $\delta
\Psi^I, \delta \bar{\Psi}^I$ of the Dirac fields.
For the 1PI Wilson action in the large $N$ limit, the
equation-of-motion composite operators are given in the form:
\begin{align}
  \E_\Lambda [\Psi^I, \bar{\Psi}^I]
  &\equiv \int_q \left( - \Gamma_\Lambda \Rd{\Psi^I (q)}
    \delta \Psi^I (q) - \delta \bar{\Psi}^I (-q) \Ld{\bar{\Psi}^I
    (-q)} \Gamma_\Lambda \right) \nt\\
  &\quad + \int_{q,r} \RL (q) \Tr \left[
    \delta \Psi^I (q) \Rd{\Psi^I (r)} \G_{-r,q} + \G_{-q,r}
    \Ld{\bar{\Psi}^I (-r)} \delta \bar{\Psi}^I (-q)\right].
\end{align}

The number operator $\N_\Lambda (p)$ which shifts the fields by
momentum $p$ corresponds to the choice
\begin{equation}
  \delta \Psi^I (q) = \Psi^I (q+p),\quad
  \delta \bar{\Psi}^I (-q) = \bar{\Psi}^I (-q+p).
\end{equation}
This gives the composite operator
\begin{align}
  \N_\Lambda (p)
  &\equiv  \int_q \left( -\Gamma_\Lambda \Rd{\Psi^I (q)} \Psi^I (q+p) -
    \bar{\Psi}^I (-q+p) \Ld{\bar{\Psi}^I (-q)} \Gamma_\Lambda \right)\nt\\
  &\quad + N \int_q \RL (q) \Tr \left[ \G_{-q-p, q} + \G_{-q, q-p}
    \right].
\end{align}
We call $\N_\Lambda (p)$ a number operator because if it is
  inserted into the correlation function of elementary fields, it
  replaces each elementary field $\Psi^I (q)$ (or
$\bar{\Psi}^I (-q)$ ) by $\Psi^I (q+p)$ (or
  $\bar{\Psi}^I (-q+p)$), and especially for $p=0$ the
  correlation function of $2 n$ elementary fields is multiplied by
  a factor $2n$.  Similarly, the spin operator $A_{\mu\nu} (p)$
corresponds to
\begin{equation}
  \delta \Psi^I (q) = - \frac{1}{2} \Sigma_{\mu\nu} \Psi^I (q+p),\quad
  \delta \bar{\Psi}^I (-q) =  \bar{\Psi}^I (-q+p) \frac{1}{2} \Sigma_{\mu\nu}.
\end{equation}
This gives
\begin{align}
  A_{\mu\nu} (p)
  &\equiv \int_q \left(  \Gamma_\Lambda \Rd{\Psi^I (q)} \frac{1}{2} \Sigma_{\mu\nu}
    \Psi^I (q+p) - \bar{\Psi}^I (-q+p) \frac{1}{2} \Sigma_{\mu\nu} \Ld{\bar{\Psi}^I
    (-q)} \Gamma_\Lambda \right)\nt\\
  &\quad + N \int_q \RL (q)\, \Tr \left[ - \frac{1}{2} \Sigma_{\mu\nu} \G_{-q-p,q} +
    \G_{-q,q-p} \frac{1}{2} \Sigma_{\mu\nu} \right].
\end{align}

Substituting (\ref{largeN-ansatz}, \ref{largeN-GammaGauss}) into the
above, we obtain
\begin{align}
  \N_\Lambda (p)
  &= \int_q \left( \bar{\Psi}^I (-q) i \fs{q} \Psi^I (q+p) + \bar{\Psi}^I
    (-q+p) i \fs{q} \Psi^I (q) \right) \nt\\
  &\quad - \int_r \frac{\delta
    \Gamma_{I\Lambda}[\varphi]}{\delta \varphi (r)} \int_q \left(
    \bar{\Psi}^I (r-q) \Psi^I (q+p) + \bar{\Psi}^I (-q+p) \Psi^I (r+q)
    \right)\nt\\
  &\quad + N \int_q \RL (q) \,\Tr \left( \G_{-q-p,q} + \G_{-q,q-p}
    \right)\nt\\
  &= \int_q \bar{\Psi}^I (-q) \lb i \fs{q} + i (\fs{q}+\fs{p}) \rb
    \Psi^I (q+p)\nt\\
  &\quad + N \int_q \lb
    \left( \RL (q) + \RL (q+p)\right) \,\Tr \G_{-q-p,q} + 2 \sigma
    (-q) \frac{\delta F_\Lambda [\sigma]}{\delta \sigma (-q-p)} \rb,
    \label{comp-NLambda}
\end{align}
and
\begin{align}
  A_{\mu\nu} (p)
  &= \int_q \left( - \bar{\Psi}^I (-q) i \fs{q} \frac{1}{2} \Sigma_{\mu\nu} \Psi^I
    (q+p) + \bar{\Psi}^I (-q+p) \frac{1}{2} \Sigma_{\mu\nu} i \fs{q} \Psi^I (q)
    \right)\nt\\
  &\quad + \int_r \frac{\delta \Gamma_{I\Lambda}[\varphi]}{\delta
    \varphi (r)} \int_q \left( \bar{\Psi}^I (r-q) \frac{1}{2} \Sigma_{\mu\nu}
    \Psi^I (q+p) - \bar{\Psi}^I (-q+p) \frac{1}{2} \Sigma_{\mu\nu} \Psi^I (r+q)
    \right)\nt\\
  &\quad + N \int_q \RL (q)\,\Tr \left[ \frac{1}{2} \Sigma_{\mu\nu} \left(
    - \G_{-q-p,q} + \G_{-q,q-p} \right)\right]\nt\\
  &= \int_q \bar{\Psi}^I (-q) \lb - i \fs{q} \frac{1}{2} \Sigma_{\mu\nu} +
    \frac{1}{2} \Sigma_{\mu\nu} i (\fs{q}+\fs{p}) \rb \Psi^I (q+p)\nt\\
  &\quad + N \int_q  \RL (q)  \Tr \left[
    \frac{1}{2} \Sigma_{\mu\nu} \left( - \G_{-q-p,q} + \G_{-q,q-p}\right)\right].
    \label{comp-Amunu}
\end{align}
Note that this $A_{\mu\nu} (p)$ is the same as
(\ref{properties-antisym}); hence we have used the same notation.  By
construction, both $\N_\Lambda (p)$ and $A_{\mu\nu} (p)$ are UV
finite.  The integrals over $q$ are UV finite thanks to the fast decay
of the cutoff function $\RL (q)$ for large $q^2$.

Notice that both $\N_\Lambda (p)$ and $A_{\mu\nu} (p)$ are quadratic
composite operators discussed in Sec.~\ref{comp-subsec-quadratic}.  We
wonder if they are the same as
\begin{align}
  \N^\mathrm{formal}_\Lambda (p)
  &\equiv \int_q \bar{\Psi}^I (-q) \lb i \fs{q} + i (\fs{q}+\fs{p})
    \rb \Psi^I (q+p) - N \int_q \Tr \left[ \lb i \fs{q} + i (\fs{q}+\fs{p})
    \rb  \G_{-q-p,q}\right] \nt\\
  &\quad + N \int_q 2 \sigma (-q) \frac{\delta \tilde{F}
    [\sigma]}{\delta \sigma (-q-p)},\label{comp-Nprime}\\
  A^\mathrm{formal}_{\mu\nu} (p)
  &\equiv \int_q \bar{\Psi}^I (-q) \lb - i\fs{q} \frac{1}{2} \Sigma_{\mu\nu} +
    \frac{1}{2} \Sigma_{\mu\nu} i (\fs{q}+\fs{p}) \rb \Psi^I (q+p) \nt\\
  &\quad - N \int_q \Tr \left[ \lb - i\fs{q} \frac{1}{2} \Sigma_{\mu\nu} +
    \frac{1}{2} \Sigma_{\mu\nu} i (\fs{q}+\fs{p}) \rb \G_{-q-p,q}\right] .\label{comp-Aprime}
\end{align}
Almost but not quite. $\N^\mathrm{formal}_\Lambda (p)$ and
$A^\mathrm{formal}_{\mu\nu} (p)$ have the same $\Lambda$-dependence as
$\N_\Lambda (p)$ and $A_{\mu\nu} (p)$, respectively, but they contain
UV divergences in the low order terms in $\sigma$.  Hence,
$\N_\Lambda (p)$ and $A_{\mu\nu} (p)$ should be regarded as the UV
finite improvement of $\N^\mathrm{formal}_\Lambda (p)$ and
$A^\mathrm{formal}_{\mu\nu} (p)$, respectively.  In Appendix
\ref{appendix-N} we discuss the relation between $\N_\Lambda (p)$ and
$\N^\mathrm{formal}_\Lambda (p)$, and show that $\N_\Lambda (p)$ is an
UV improvement of $\N^{\mathrm{formal}}_\Lambda (p)$.

We have explained that the antisymmetric part $A_{\mu\nu}
(p)$ of the EM tensor is an equation-of-motion composite operator.
Since the EM tensor itself is a composite operator, the right-hand
side of the WT identity (\ref{properties-WT}) is also a composite
operator.  It is indeed an equation-of-motion composite operator
corresponding to the infinitesimal translation
\begin{equation}
  \delta \Psi^I (q)
  = - (q+p)_\nu \Psi^I (q+p),\quad
  \delta \bar{\Psi}^I (-q)
  = - (-q+p)_\nu \bar{\Psi}^I (-q+p).\label{comp-deltaPsi}
\end{equation}
It may be more precise to say that $p_\mu \Theta_{\mu\nu} (p)$ is
determined by how the Dirac fields changes under the infinitesimal
translation as in (\ref{comp-deltaPsi}).

\section{Construction of the energy-momentum tensor\label{construction}}

In Sec.~\ref{properties} we have found the EM tensor given in the
following form:
\begin{equation}
  \Theta_{\mu\nu} (p) =
  \Theta^G_{\mu\nu} (p) + N \left( \tilde{\Theta}_{\mu\nu} (p) +
    S'_{\mu\nu} (p) + A'_{\mu\nu} (p) \right),\label{construction-EMtensor}
\end{equation}
where $\Theta^G_{\mu\nu} (p)$, given by (\ref{properties-EMGauss}), is
the EM tensor of the Gaussian theory, and the antisymmetric part
$A'_{\mu\nu} (p)$ is given by (\ref{properties-antisymprime}).  Since
$\tilde{\Theta}_{\mu\nu} (p)$, given by (\ref{properties-tildeTheta}),
is a composite operator on its own, it is easy to guess that the
remaining symmetric part is a quadratic composite operator introduced
in Sec.~\ref{comp-subsec-quadratic}.  Hence, we guess
\begin{equation}
S'_{\mu\nu} (p) = - \int_q \Tr s_{\mu\nu} (-q,q+p)\, \G_{-q-p,
    q},\label{construction-guess}
\end{equation}
where the coefficient $s_{\mu\nu}$ is given by (\ref{properties-smunu}).

As it stands, the expression of $S'_{\mu\nu} (p)$ contains UV
divergences up to the order quadratic in powers of $\sigma$.  First we
would like to show formally, ignoring the UV divergences, that
(\ref{construction-guess}) satisfies the WT identity
(\ref{properties-WTprime}).  This way we can explain better what is
involved.  We will then take care of the UV divergences.

\subsection{Check of the WT identity\label{construction-subsec-WT}}

We wish to check the WT identity (\ref{properties-WTprime}).  We find
\begin{align}
  p_\mu S'_{\mu\nu} (p)
  &= - \int_q \Tr p_\mu s_{\mu\nu} (-q,q+p)\, \G_{-q-p,q}\nt\\
  &= -  \int_q \Tr \left( - (q+p)_\nu i \fs{q} + q_\nu i (\fs{q}+\fs{p}) + i \fs{q}
    \frac{1}{2} p_\mu \Sigma_{\mu\nu} - \frac{1}{2} p_\mu \Sigma_{\mu\nu} i
    (\fs{q} + \fs{p}) \right) \G_{-q-p, q}\nt\\
  &= \int_q \Tr \Bigg[
    \left(i\fs{q}+\RL (q)\right) \lb (q+p)_\nu - \frac{1}{2} p_\mu
    \Sigma_{\mu\nu} \rb\nt\\
&\qquad\qquad + \left( - q_\nu + \frac{1}{2} p_\mu \Sigma_{\mu\nu} \right)
    \lb i (\fs{q}+\fs{p}) + \RL (q+p) \rb \nt\\
  &\quad + \RL (q) \left( - (q+p)_\nu + \frac{1}{2} p_\mu
    \Sigma_{\mu\nu} \right) + \RL (q+p) \left( q_\nu - \frac{1}{2}
    p_\mu \Sigma_{\mu\nu} \right) \Bigg] \G_{-q-p,q} .
\end{align}
Using (\ref{largeN-GhinvG}) we obtain
\begin{align}
  & \int_q \Tr \lb  \left( (q+p)_\nu - \frac{1}{2} p_\mu
    \Sigma_{\mu\nu} \right) \G_{-q-p,q} \frac{1}{h_\Lambda (q)} 
    + \left(  - q_\nu + \frac{1}{2} p_\mu \Sigma_{\mu\nu} \right)
    \frac{1}{h_\Lambda (q+p)} \G_{-q-p,q} \rb\nt\\ 
  &= \int_q \Tr \lb \left( (q+p)_\nu - \frac{1}{2} p_\mu
    \Sigma_{\mu\nu} \right) \left( \delta (p) + \int_r \G_{-q-p,r}
    \sigma (r-q) \right) \right.\nt\\
  &\qquad\qquad\left. + \left(- q_\nu + \frac{1}{2} p_\mu \Sigma_{\mu\nu}
    \right) \left( \delta (p) + \int_r \sigma
    (-r+q+p) \G_{-r,q} \right) \rb\nt\\
  &= \int_{q,r}  \Tr \lb \left((q+p)_\nu - \frac{1}{2} p_\mu
    \Sigma_{\mu\nu} \right)  \G_{-q-p, -r+q} \sigma (-r) + \left(
    -q_\nu + \frac{1}{2} p_\mu \Sigma_{\mu\nu} \right)  \sigma (-r) \G_{-r-q-p, q} \rb\nt\\
  &= \int_r \sigma (-r) (r+p)_\nu \int_q \Tr \G_{-q-r-p, q} .
\end{align}

Ignoring the UV divergence of the term linear in
$\sigma$, we obtain (\ref{largeN-dIdsigma-naive}):
\begin{equation}
  \frac{\delta I_\Lambda}{\delta \sigma (-p)}
  = - \int_q \Tr \G_{-q,q-p} = - \int_q \Tr \G_{-q-p,q} .
\end{equation}
Hence,
\begin{align}
 & \int_q \Tr \lb  \left( (q+p)_\nu - \frac{1}{2} p_\mu
    \Sigma_{\mu\nu} \right) \G_{-q-p,q} \frac{1}{h_\Lambda (q)} 
    + \left(  - q_\nu + \frac{1}{2} p_\mu \Sigma_{\mu\nu} \right)
   \frac{1}{h_\Lambda (q+p)} \G_{-q-p,q} \rb\nt\\
 & = - \int_r \sigma (-r) (r+p)_\nu \frac{\delta I_\Lambda}{\delta
   \sigma (-r-p)} .
\end{align}

We thus obtain
\begin{align}
  p_\mu  S'_{\mu\nu} (p)
&= - \int_q \sigma (-q) (q+p)_\nu \frac{\delta I_\Lambda}{\delta \sigma
  (-q-p)} \nt\\
  &\quad+ \int_q \Tr \lb \RL (q) \left(-(q+p)_\nu + \frac{1}{2} p_\mu
    \Sigma_{\mu\nu} \right) + \RL (q+p) \left( q_\nu -
    \frac{1}{2}p_\mu \Sigma_{\mu\nu}\right) \rb \G_{-q-p,q} \nt\\
  &= \int_q \frac{\delta I_\Lambda [\sigma]}{\delta \sigma (q)} q_\nu \sigma
    (q+p) - \int_q \RL (q) \Tr \lb (q+p)_\nu \G_{-q-p,q} + (-q+p)_\nu
    \G_{-q,q-p} \rb\nt\\
  &\quad + \int_q \RL (q) \Tr \left[ \frac{1}{2} p_\mu \Sigma_{\mu\nu}
    \left( \G_{-q-p,q} - \G_{-q,q-p} \right)\right] .
\end{align}
which is (\ref{properties-WTprime}).  Hence, $\Theta_{\mu\nu} (p)$ given by
(\ref{construction-EMtensor}) satisfies the WT identity (\ref{properties-WT}).

\subsection{Taking care of UV divergences\label{construction-subsec-UV}}

Let us discuss how to modify the symmetric $S'_{\mu\nu} (p)$, given by
(\ref{construction-guess}), to make it free from UV divergences.  We
find
\begin{align}
S'_{\mu\nu} (p)
  &= - \int_q \Tr s_{\mu\nu} (-q,q+p) \G_{-q-p,q} [\sigma]\nt\\
  &= - \int_q \Tr s_{\mu\nu} (-q,q+p) \Bigg[ h_\Lambda (q) \delta (p)
    + h_\Lambda (q+p) \sigma (p) h_\Lambda (q) \nt\\
  &\qquad + h_\Lambda (q+p) \int_{p_1, p_2} \sigma (p_1) \sigma (p_2)
    \,\delta (p_1+p_2 - p)\, h_\Lambda (q+p_2) h_\Lambda (q) \nt\\
  &\qquad + \textrm{higher order terms in $\sigma$} \Bigg],
\end{align}
where the coefficient $s_{\mu\nu} (-q,q+p)$ is given by
(\ref{properties-smunu}).  The higher order terms contribute terms of
order $1/q^3$ and higher to the $q$-integral.  Upon symmetrization
with respect to $q \to -q$, the integrals over $q$ are absolutely
convergent for $2 < D < 4$.

We modify the first three terms so that
\begin{align}
  &S'_{\mu\nu} (p)
  = A_\Lambda \delta_{\mu\nu}\,\delta (p) + B_{\Lambda; \mu\nu} (p)
    \sigma (p) \nt\\
  &\quad + \int_{p_1,p_2} \sigma (p_1) \sigma (p_2) \delta
    (p_1+p_2-p) \, C_{\Lambda; \mu\nu} (p_1, p_2)\nt\\
  &\quad - \int_q \Tr s_{\mu\nu} (-q,q+p) \big\lbrace \G_{-q-p,q} [\sigma]
    - h_\Lambda (q) \delta (p) - h_\Lambda (q+p) \sigma (p) h_\Lambda
    (q)\nt\\
  &\qquad - h_\Lambda (q+p) \int_{p_1, p_2} \sigma (p_1) \sigma
    (p_2) \delta (p_1+p_2-p)\, h_\Lambda (q+p_2) h_\Lambda (q)
    \big\rbrace ,
\end{align}
where the coefficients $A_\Lambda, B_{\Lambda;\mu\nu} (p)$, and $C_{\Lambda;
  \mu\nu} (p_1, p_2)$ are the solutions to
\begin{subequations}
  \label{construction-ERG-ABC}
  \begin{align}
    - \Lambda \partial_\Lambda A_\Lambda \delta_{\mu\nu}
    &= - \int_q \Tr s_{\mu\nu} (-q,q)\, f_\Lambda (q),\label{construction-ERG-A}\\
    - \Lambda \partial_\Lambda B_{\Lambda;\mu\nu} (p)
    &= - \int_q \Tr s_{\mu\nu} (-q,q+p) \left( f_\Lambda (q+p) h_\Lambda
      (q) + h_\Lambda (q+p) f_\Lambda (q) \right),\label{construction-ERG-B}\\
    - \Lambda \partial_\Lambda C_{\Lambda;\mu\nu} (p_1, p_2)
    &= - \int_q \Tr s_{\mu\nu} (-q,q+p) \left[
      f_\Lambda (q+p) \frac{1}{2} \left( h_\Lambda (q+p_1) + h_\Lambda
      (q+p_2) \right) h_\Lambda (q)\right.\nt\\
    &\qquad + h_\Lambda (q+p) \frac{1}{2} \left( f_\Lambda (q+p_1) + f_\Lambda
      (q+p_2) \right) h_\Lambda (q) \nt\\
    &\left.\qquad + h_\Lambda (q+p) \frac{1}{2} \left( h_\Lambda (q+p_1) + h_\Lambda
      (q+p_2) \right) f_\Lambda (q) \right],\label{construction-ERG-C}
  \end{align}
\end{subequations}
where $f_\Lambda (q)$ is defined by (\ref{comp-fLambda}).  These are
necessary for $S'_{\mu\nu} (p)$ to be a quadratic composite operator.
For the modified $S'_{\mu\nu} (p)$ to satisfy the WT identity
(\ref{properties-WTprime}), we expand the identity in powers of
$\sigma$ to obtain
\begin{subequations}
  \label{construction-WT-BC}
\begin{align}
&  p_\mu B_{\Lambda;\mu\nu} (p)
  = - \int_q  \Tr \lb (q+p)_\nu \RL (q) - q_\nu \RL (q+p) \rb
    h_\Lambda (q+p) h_\Lambda (q)\nt\\
  &\quad +  p_\mu \int_q \left(\RL (q) - \RL (q+p) \right) \Tr
    \frac{1}{2} \Sigma_{\mu\nu} h_\Lambda (q+p) h_\Lambda (q),
    \label{construction-WT-B}\\
&p_\mu C_{\Lambda;\mu\nu} (p_1,p_2)
  = - \frac{1}{2} \left( p_{1\nu} \F (p_1) + p_{2\nu} \F (p_2)
    \right)\nt\\
  &\quad - \int_q \Tr \lb (q+p)_\nu \RL (q) - q_\nu \RL (q+p)\rb
    h_\Lambda (q+p) \frac{1}{2} \left( h_\Lambda (q+p_1) + h_\Lambda
    (q+p_2) \right) h_\Lambda (q)\nt\\
  &\quad + p_\mu \int_q \left( \RL (q) - \RL (q+p)\right) \Tr
    \Sigma_{\mu\nu} h_\Lambda (q+p) \frac{1}{2} \left( h_\Lambda (q+p_1) + h_\Lambda
    (q+p_2) \right) h_\Lambda (q) .\label{construction-WT-C}
\end{align}
\end{subequations}
We discuss how (\ref{construction-ERG-ABC}, \ref{construction-WT-BC})
determine the three coefficients in Appendix \ref{appendix-ABC}.

Thus, to summarize this section, we have shown that the EM tensor is
given by (\ref{construction-EMtensor}) as
\begin{equation}
  \Theta_{\mu\nu} (p) = \Theta^G_{\mu\nu} (p) + N \left(
    \tilde{\Theta}_{\mu\nu} (p) + S'_{\mu\nu} (p) + A'_{\mu\nu} (p)
  \right),
  \label{construction-EM-result}
\end{equation}
where $\Theta^G_{\mu\nu} (p)$ is the EM tensor of the Gaussian theory
\begin{equation}
  \Theta^G_{\mu\nu} (p) = \int_q \bar{\Psi}^I (-q) t_{\mu\nu} (-q,q+p)
  \Psi^I (q+p),
\end{equation}
$\tilde{\Theta}_{\mu\nu} (p)$ is given by
\begin{align}
  \tilde{\Theta}_{\mu\nu} (p)
  &\equiv - \frac{1}{2} \int_q \sigma (q+p) \sigma (-q) \lb
    \delta_{\mu\nu} \left( f_2 + f_{21} \left(p^2 + q(q+p) \right)
    \right)\right.\nt\\
  &\left.\qquad\qquad + f_{21}\left( q_\mu (q+p)_\nu + q_\nu
    (q+p)_\mu\right) \rb\nt\\
  &\quad + \mathrm{const} \, f_{21} \left( p^2 \delta_{\mu\nu} - p_\mu p_\nu
    \right)  \frac{1}{2} \int_q \sigma (q+p) \sigma (-q),
\end{align}
where the parameters $f_2, f_{21}$ denote the deviations away from the
critical point, the symmetric part is given by
\begin{align}
S'_{\mu\nu} (p)
  &= A_\Lambda \delta_{\mu\nu}\,\delta (p) + B_{\Lambda; \mu\nu} (p)
    \sigma (p) \nt\\
  &\quad + \int_{p_1,p_2} \sigma (p_1) \sigma (p_2) \delta
    (p_1+p_2-p) \, C_{\Lambda; \mu\nu} (p_1, p_2)\nt\\
  &\quad - \int_q \Tr s_{\mu\nu} (-q,q+p) \big\lbrace \G_{-q-p,q} [\sigma]
    - h_\Lambda (q) \delta (p) - h_\Lambda (q+p) \sigma (p) h_\Lambda
    (q)\nt\\
  &\qquad - h_\Lambda (q+p) \int_{p_1, p_2} \sigma (p_1) \sigma
    (p_2) \delta (p_1+p_2-p)\, h_\Lambda (q+p_2) h_\Lambda (q)
    \big\rbrace ,
\end{align}
and finally the antisymmetric part is given by
\begin{equation}
  A'_{\mu\nu} (p) =  \int_q \RL (q)  \Tr \left[ \frac{1}{2}
  \Sigma_{\mu\nu} \left( - \G_{-q-p,q} + \G_{-q,q-p} \right) \right].
\end{equation}
The coefficients $t_{\mu\nu}, s_{\mu\nu}$ are given by
(\ref{properties-tmunu}).  Eq.~(\ref{construction-EM-result}),
together with the explicit expressions for each term, makes the
so-defined EM tensor completely explicit and constitutes one of the
main results of this paper.

\section{Trace\label{trace}}

The trace of the EM tensor is given by
\begin{equation}
  \Theta_\Lambda (p) \equiv \Theta_{\mu\mu} (p)
  = \Theta^G (p) + N \left( \tilde{\Theta} (p) + \Theta'_\Lambda (p) \right),
\end{equation}
where
\begin{align}
  \Theta^G (p)
  &= \frac{1-D}{2} \int_q \bar{\Psi}^I (-q) \left( 2 i
    \fs{q} + i \fs{p} \right) \Psi^I (q+p),\\
  \tilde{\Theta} (p)
  &= \frac{1}{2} \int_q \sigma (q+p) \sigma (-q) \lb
    - D f_2 + \left( (D-1) \mathrm{const} - D \right) f_{21} p^2 -
    (D+2) f_{21} q(q+p) \rb,\\
  \Theta'_\Lambda (p)
  &= D A_\Lambda \delta (p) + B_{\Lambda;\mu\mu} (p) \sigma (p)\nt\\
  &\quad + \int_{p_1, p_2} \sigma (p_1) \sigma (p_2) \delta
    (p_1+p_2-p)\,C_{\Lambda; \mu\mu} (p_1, p_2)\nt\\
  &\quad + \frac{D-1}{2} \int_q \Tr \Bigg[
    \left( 2 i \fs{q} + i \fs{p} \right) \Big\lbrace \G_{-q-p,q} - h_\Lambda
    (q) \delta (p) - h_\Lambda (q+p) \sigma (p) h_\Lambda (q) \nt\\
  &\qquad - h_\Lambda (q+p) \int_{p_1,p_2} \sigma (p_1) \sigma (p_2)
    \delta (p_1+p_2-p) \frac{1}{2} \left( h_\Lambda (q+p_1) +
    h_\Lambda (q+p_2) \right) h_\Lambda (q) \Big\rbrace \Bigg].
\end{align}

We now compare the trace with the number operator $\N_\Lambda (p)$
defined by (\ref{comp-NLambda}).  For the comparison it is more
convenient to use the expression of $\N_\Lambda (p)$ as a quadratic
composite operator given as (\ref{appendix-N-quadratic}) in Appendix
\ref{appendix-N}:
\begin{align}
  & \N_\Lambda (p) - N \int_q 2 \sigma (-q) \frac{\delta \tilde{F}
    [\sigma]}{\delta \sigma (-q-p)}\nt\\
  &= \int_q \bar{\Psi}^I (-q) \left( 2 i \fs{q} + i \fs{p}\right)
    \Psi^I (q+p)\nt\\
  &\quad - N \int_q \Tr \Bigg[ \left( 2 i \fs{q} + i \fs{p} \right)
    \Big\lbrace \G_{-q-p,q} - h_\Lambda (q) \delta (p) - h_\Lambda
    (q+p) \sigma (p) h_\Lambda (q)\nt\\
  &\qquad - h_\Lambda (q+p) \int_{p_1, p_2} \sigma (p_1) \sigma (p_2)
    \delta (p_1+p_2-p) h_\Lambda (q+p_2) h_\Lambda (q)
    \Big\rbrace\Bigg]\nt\\
  &\quad + N  \int_q 2 \RL (q) \Tr h_\Lambda (q)\, \delta (p)\nt\\
  &\quad + N \left( 2 c_{1\Lambda} + \int_q  \left( \RL (q) + \RL (q+p) \right)
    \Tr h_\Lambda (q+p) h_\Lambda (q) \right) \sigma (p) \nt\\
&\quad  + N \int_{p_1, p_2} \sigma (p_1)
    \sigma (p_2) \delta (p_1+p_2-p)\, \Bigg[\F (p_1) + \F (p_2)\nt\\
&\qquad + \int_q \left(\RL (q) + \RL (q+p)\right) \Tr \lb h_\Lambda (q+p)
  \frac{1}{2} \left( h_\Lambda (q+p_1) + h_\Lambda (q+p_2)\right)
  h_\Lambda (q) \rb \Bigg].
\end{align}
We find
\begin{align}
  \Theta_\Lambda (p)
  &= - \frac{D-1}{2} \N_\Lambda (p) + N \Bigg[ \lb (D-2) f_2
    + \left( (D-1)\,\mathrm{const}-\frac{D-2}{2} \right) f_{21} p^2 \rb\nt\\
  &\quad \times \frac{1}{2} \int_q \sigma (q+p) \sigma (-q)
   + (D-4) f_{21} \frac{1}{4} \int_q \left( q^2 + (q+p)^2
    \right) \sigma (-q) \sigma (q+p)\nt\\
  &\quad +
    D A_\Lambda \delta (p) + B_{\Lambda;\mu\mu} (p) \sigma (p) +
    \int_{p_1, p_2} \sigma (p_1) \sigma (p_2) \delta
    (p_1+p_2-p)\,C_{\Lambda;\mu\mu} (p_1, p_2) \nt\\
  &\quad + \frac{D-1}{2} \Bigg(
    2 \int_q \RL (q) \Tr h_\Lambda (q) \cdot \delta (p)\nt\\
  &\qquad + \lb 2 c_{1\Lambda} + \int_q \left(
    \RL (q) + \RL (q+p)\right) \Tr h_\Lambda (q) h_\Lambda (q+p) \rb
    \sigma (p)\nt\\
  &\qquad + \int_{p_1, p_2} \sigma (p_1) \sigma (p_2) \delta
    (p_1+p_2-p)\,\Big\lbrace \F (p_1) + \F (p_2)\nt\\
  &\qquad\qquad + \int_q \left(\RL (q) + \RL (q+p)\right)
    \Tr h_\Lambda (q+p) \frac{1}{2} \left( h_\Lambda (q+p_1) + h_\Lambda
    (q+p_2) \right) h_\Lambda (q) \Big\rbrace \Bigg)\Bigg].
\end{align}
Choosing
\begin{equation}
  \mathrm{const} = \frac{1}{2} \frac{D-2}{D-1},
\end{equation}
to remove the total derivative term proportional to
\[
  p^2 \int_{p_1,p_2} \sigma (p_1) \sigma (p_2) \delta (p_1+p_2-p),
\]
we obtain
\begin{align}
  \Theta_\Lambda (p)
  &= - \frac{D-1}{2} \N_\Lambda (p) \nt\\
  &\quad + N \left[
    (D-2) f_2 \frac{1}{2} \int_q \sigma (q+p) \sigma (-q) + (D-4) f_{21}
    \frac{1}{4} \int_q \left( q^2+(q+p)^2\right) \sigma (q+p) \sigma
    (-q) \right]\nt\\
  &\quad + N \left[ A\, \delta (p) + B (p)\,\sigma (p) + \int_{p_1, p_2} \sigma
    (p_1) \sigma (p_2)\delta (p_1+p_2-p)\, C(p_1, p_2) \right],
\end{align}
where $A, B(p), C(p_1, p_2)$ are defined by
\begin{subequations}
  \begin{align}
    A
    &\equiv D A_\Lambda + (D-1) \int_q \RL (q) \Tr h_\Lambda
      (q),\label{trace-A}\\
    B (p)
    &\equiv B_{\Lambda;\mu\mu} (p) + (D-1) \left(
      c_{1\Lambda} + \frac{1}{2} \int_q \left( \RL (q) + \RL
      (q+p)\right) \Tr h_\Lambda (q) h_\Lambda (q+p) \right),
    \label{trace-B}\\
    C(p_1,p_2)
    &\equiv C_{\Lambda;\mu\mu} (p_1,p_2) + \frac{D-1}{2} \Big\lbrace \F (p_1)
      + \F (p_2) \nt\\
    &\quad + \int_q \left( \RL (q) + \RL (q+p) \right) \Tr h_\Lambda
      (q) \frac{1}{2} \left( h_\Lambda (q+p_1) + h_\Lambda (q+p_2)
      \right) h_\Lambda (q+p) \Big\rbrace.\label{trace-C}
  \end{align}
\end{subequations}
Since $\Theta_\Lambda (p)$ is a composite operator,
\begin{equation}
  A\, \delta (p) + B (p)\,\sigma (p) + \int_{p_1, p_2} \sigma
  (p_1) \sigma (p_2)\delta (p_1+p_2-p)\, C(p_1, p_2)
\end{equation}
must also be a composite operator.  As we have pointed out at the
  end of Sec.~\ref{comp-subsec-sigma}, this means $A, B(p), C(p_1,p_2)$
are independent of $\Lambda$.

In Appendix \ref{appendix-ABC} we show
\begin{equation}
  A = B (p) = C (p_1,p_2) = 0.
\end{equation}
Hence, we obtain the trace formula of the EM tensor as
\begin{align}
&   \Theta_\Lambda (p)
= - \frac{D-1}{2} \N_\Lambda (p)\nt\\
&\quad + N \left[
    (D-2) f_2 \frac{1}{2} \int_q \sigma (q+p) \sigma (-q) + (D-4) f_{21}
    \frac{1}{4} \int_q \left( q^2+(q+p)^2\right) \sigma (q+p) \sigma
    (-q) \right],\label{trace-result}
\end{align}
where $D-2, D-4$ are the scale dimensions of the parameters
$f_2, f_{21}$ at the scaling limit of the critical point
$f_2 = f_{21}= 0$.  Especially, at the critical point, we obtain
\begin{equation}
  \Theta_\Lambda (p)
  = - \frac{D-1}{2} \N_\Lambda (p), \label{trace-result-critical}
\end{equation}
i.e., the trace vanishes up to an equation-of-motion composite
operator.

\section{Conformal invariance \label{conformal}}

In this last section before concluding the paper, we show the
conformal invariance of the critical point at $f_2=f_{21} = 0$.
First of all, let us note that the 1PI Wilson action
(\ref{largeN-ansatz}) is manifestly invariant under
rotations. Moreover, it is known that a 1PI Wilson action that
satisfies the fixed point equation automatically satisfies the
Ward-Takahashi identity associated with scale invariance.
We refer the reader to
\cite{Rosten:2014oja,Delamotte:2015aaa,Sonoda:2015pva,Rosten:2016zap,Sonoda:2017zgl}
for a detailed discussion related to scale invariance, conformal
invariance, and the ERG.
Therefore, in order to show conformal invariance, we shall check that
the Ward identity associated with special conformal transformations is
also satisfied.  To do so, we shall show that
Eq.~(\ref{trace-result-critical}) implies the Ward-Takahashi identity
associated with special conformal transformations.

We recapitulate what we know.  The energy-momentum tensor satisfies
the WT identity
\begin{align}
  &p_\mu \Theta_{\mu\nu} (p)
  = \int_q \left[ \Gamma_G \Rd{\Psi^I (q)} (q+p)_\nu \Psi^I (q+p)
    + \left(-q+p\right)_\nu \bar{\Psi}^I (-q+p) \Ld{\bar{\Psi}^I (-q)}
    \Gamma_G \right]\nt\\
  &\quad + N \int_q \left[ (q+p)_\nu \varphi (q+p) \sigma (-q)
    - \RL (q) \Tr \lb (q+p)_\nu \G_{-q-p,q} +
    (-q+p)_\nu \G_{-q,q-p} \rb\right],\label{conformal-WT}
\end{align}
and its antisymmetric part is given by
\begin{align}
  \Theta_{\mu\nu} (p) - \Theta_{\nu\mu} (p)
  &= \int_q \left[ \Gamma_G \Rd{\Psi^I (q)} \Sigma_{\mu\nu} \Psi^I
    (q+p) - \bar{\Psi}^I (-q+p) \Sigma_{\mu\nu} \Ld{\bar{\Psi}^I (-q)}
    \Gamma_G \right]\nt\\
  &\quad + \int_q R_\Lambda (q) \,\Tr \Sigma_{\mu\nu} \left( -
    \G_{-q-p,q} + \G_{-q,q-p} \right).\label{conformal-antisym}
\end{align}
The trace is given by
\begin{align}
  \Theta_\Lambda (p)
  &= - \frac{D-1}{2} \N_\Lambda (p)\nt\\
  &= \frac{D-1}{2} \Bigg[ \int_q \lb \Gamma_G \Rd{\Psi^I (q)}
    \Psi^I (q+p) + \bar{\Psi}^I (-q+p) \Ld{\bar{\Psi}^I (-q)} \Gamma_G
    \rb\nt\\ 
  &\quad + N \int_q
    \lb  2 \sigma (-q) \varphi (q+p) - \RL (q) \Tr \left( \G_{-q-p,q} + \G_{-q,q-p}
    \right) \rb \Bigg]. \label{conformal-trace}
\end{align}

\subsection{Scale invariance}

Before we derive the WT identity for special conformal invariance,
  let us derive the WT identity for scale invariance.  We follow
\cite{Sonoda:2015pva} and calculate
\begin{equation}
  \lim_{p \to 0} \frac{\partial}{\partial p_\nu} \left( p_\mu
    \Theta_{\mu\nu} (p) \right) = - \int_x x_\nu \partial_\mu \Theta_{\mu\nu} (x)
\end{equation}
in two ways.  Using (\ref{conformal-WT}), we obtain
\begin{align}
&\lim_{p \to 0} \frac{\partial}{\partial p_\nu} \left( p_\mu
  \Theta_{\mu\nu} (p) \right)   \nt\\
  &= \int_q \left[ \Gamma_G \Rd{\Psi^I (q)} \left( q \cdot \partial_q
     + D \right) \Psi^I (q) + \left(
    q \cdot \partial_q + D \right) \bar{\Psi}^I
    (-q) \Ld{\bar{\Psi}^I (-q)} \Gamma_G \right]\nt\\
  &\quad + N \int_q \left( q \cdot \partial_q + D \right) \varphi
    (q)\,\sigma (-q)
   - N \int_q \RL (q) \left(q \cdot \partial_q + 2 D\right)
    \,\Tr \G_{-q,q}.
\end{align}
Alternatively, using (\ref{conformal-trace}), we obtain
\begin{align}
  &\lim_{p \to 0} \frac{\partial}{\partial p_\nu} \left( p_\mu
    \Theta_{\mu\nu} (p) \right)  = \Theta_\Lambda (0) = \int_x
    \Theta_\Lambda (x)\nt\\
  &= \frac{D-1}{2} \Bigg[ \int_q \lb \Gamma_G \Rd{\Psi^I (q)}
    \Psi^I (q) + \bar{\Psi}^I (-q) \Ld{\bar{\Psi}^I (-q)} \Gamma_G
    \rb\nt\\ 
  &\quad + N \int_q
    \lb  2 \varphi (q) \sigma (-q) - 2 \RL (q) \Tr \G_{-q,q} \rb
    \Bigg].
\end{align}
Combining the two, we obtain the WT identity for scale invariance
\begin{align}
&  \int_q \left[ \Gamma_G \Rd{\Psi^I (q)} \left( q \cdot \partial_q +
  \frac{D+1}{2} \right) \Psi^I (q) + \left(q \cdot \partial_q +
  \frac{D+1}{2}\right) \bar{\Psi}^I (-q) \Ld{\bar{\Psi}^I (-q)}
  \Gamma_G \right]\nt\\
  &\quad + N \int_q \left( q \cdot \partial_q + 1 \right) \varphi
    (q) \frac{\delta \Gamma_{I\Lambda} [\varphi]}{\delta \varphi
    (q)} - N \int_q \RL (q) \left( q \cdot \partial_q + D+1 \right)
    \Tr \G_{-q,q}  = 0.
\end{align}
The infinitesimal transformation of the Dirac field is given by
\begin{equation}
  D^S \Psi^I (q) \equiv \left( q \cdot \partial_q + \frac{D+1}{2}
  \right) \Psi^I (q),\quad
  D^S \bar{\Psi}^I (-q) \equiv \left( q \cdot \partial_q +
    \frac{D+1}{2} \right) \bar{\Psi}^I (-q).
\end{equation}
This implies
\begin{equation}
  D^S \varphi (q) = \frac{1}{N} \int_p D^S \left(\bar{\Psi}^I (-p)
    \Psi^I (p+q) \right) = \left( q \cdot \partial_q + 1 \right) \varphi (q).
\end{equation}
Hence, we can rewrite the WT identity for scale invariance as
\begin{align}
&  \int_q \left[ \Gamma_G \Rd{\Psi^I (q)} D^S \Psi^I (q)
+ D^S \bar{\Psi}^I (-q) \Ld{\bar{\Psi}^I (-q)}
  \Gamma_G \right] +  \int_q D^S  \varphi (q) \, N \frac{\delta
  \Gamma_{I\Lambda} [\varphi]}{\delta \varphi(q)} \nt\\
  &\quad - N \int_q \RL (q) \left( q \cdot \partial_q + D+1 \right)
    \Tr \G_{-q,q}  = 0.\label{conformal-scale-invariance}
\end{align}

\subsection{Special conformal invariance}

Similarly, we can derive the WT identity for special conformal
  invariance.  We follow \cite{Sonoda:2015pva} and compute
  \begin{equation}
    \lim_{p \to 0} \left( \frac{\partial^2}{\partial p_\alpha
    \partial p_\nu} -   \frac{1}{2} \delta_{\alpha\nu} \frac{\partial^2}{\partial p_\beta
    \partial p_\beta} \right) \frac{1}{N} \left(p_\mu \Theta_{\mu\nu} (p)\right)
\end{equation}
in two ways.  Calculating this we ignore the contribution of the Gaussian 1PI
action, since we already know its conformal invariance (see Appendix
\ref{appendix-free}):
\begin{equation}
  \int_q \left[ \Gamma_G \Rd{\Psi^I (q)} D_\nu^K \Phi^I (q) + D_\nu^K
    \bar{\Psi}^I (-q) \Ld{\bar{\Psi}^I (-q)} \Gamma_G \right] = 0,
\end{equation}
where the infinitesimal special conformal transformation is
  defined by
\begin{subequations}
  \label{conformal-transformation}
  \begin{align}
    D_\nu^K \Psi^I (q)
    &\equiv \lb - q_\alpha \frac{\partial^2}{\partial q_\alpha
      \partial q_\nu} + \frac{1}{2} q_\nu \frac{\partial^2}{\partial
      q_\alpha \partial q_\alpha} - \frac{D+1}{2}
      \frac{\partial}{\partial q_\nu} + \Sigma_{\nu\alpha}
      \frac{\partial}{\partial q_\alpha} \rb \Psi^I (q),\\
    D_\nu^K \bar{\Psi}^I (-q)
    &\equiv \lb   q_\alpha \frac{\partial^2}{\partial q_\alpha
      \partial q_\nu} - \frac{1}{2} q_\nu \frac{\partial^2}{\partial
      q_\alpha \partial q_\alpha} + \frac{D+1}{2}
      \frac{\partial}{\partial q_\nu} \rb \bar{\Psi}^I (-q) +
      \frac{\partial}{\partial q_\alpha} \bar{\Psi}^I (-q)
      \Sigma_{\nu\alpha} .
  \end{align}
\end{subequations}

Using (\ref{conformal-WT}), we obtain
\begin{align}
&  \lim_{p \to 0} \left( \frac{\partial^2}{\partial p_\alpha
    \partial p_\nu} - 
    \frac{1}{2} \delta_{\alpha\nu} \frac{\partial^2}{\partial p_\beta
    \partial p_\beta} \right) \frac{1}{N} \left(p_\mu \Theta_{\mu\nu}
  (p)\right)\nt\\
&= \int_q \sigma (-q) \lb q_\nu \frac{\partial^2}{\partial
  q_\alpha \partial q_\nu} - \frac{1}{2} q_\alpha
  \frac{\partial^2}{\partial q_\beta \partial q_\beta} + D
  \frac{\partial}{\partial q_\alpha} \rb \varphi (q)\nt\\
  &\quad - \int_q \RL (q) \lb q_\nu \frac{\partial^2}{\partial
  q_\alpha \partial q_\nu} - \frac{1}{2} q_\alpha
  \frac{\partial^2}{\partial q_\beta \partial q_\beta} + D
  \frac{\partial}{\partial q_\alpha} \rb \Tr \left( \G_{-q,p} +
    \G_{p,-q} \right)\Big|_{p=q}. \label{conformal-first}
\end{align}
Alternatively, we can calculate this as
\begin{align}
&   \lim_{p \to 0} \left( \frac{\partial^2}{\partial p_\alpha
    \partial p_\nu} - 
    \frac{1}{2} \delta_{\alpha\nu} \frac{\partial^2}{\partial p_\beta
    \partial p_\beta} \right) \frac{1}{N} \left(p_\mu \Theta_{\mu\nu}
  (p)\right)\nt\\
&= \lim_{p \to 0} \Bigg[ p_\mu \left( \frac{\partial^2}{\partial
    p_\alpha \partial p_\nu} - 
    \frac{1}{2} \delta_{\alpha\nu} \frac{\partial^2}{\partial p_\beta
  \partial p_\beta} \right) \frac{1}{N} \Theta_{\mu\nu} (p)
 + \frac{\partial}{\partial p_\alpha} \frac{1}{N} \Theta_\Lambda (p) -
    \frac{\partial}{\partial p_\nu} \frac{1}{N} \left( \Theta_{\nu\alpha} -
    \Theta_{\alpha\nu}\right) (p) \Bigg].
\end{align}
The first term vanishes.  The second term is obtained
using (\ref{conformal-trace}) as
\begin{align}
&  \frac{1}{N} \lim_{p \to 0} \frac{\partial}{\partial p_\alpha} \Theta
  (p)\nt\\
&= \frac{D-1}{2} \int_q \lb 2 \sigma (-q) \frac{\partial}{\partial
  q_\alpha} \varphi (q) - \RL (q) \frac{\partial}{\partial q_\alpha}
  \Tr  \left( \G_{-q,p} + \G_{p,-q} \right)\Big|_{p=q} \rb.\label{conformal-second-a}
\end{align}
The third term is obtained using (\ref{conformal-antisym}) as
\begin{equation}
  \frac{1}{N} \lim_{p\to 0} (-) \frac{\partial}{\partial p_\nu} \left(
    \Theta_{\nu\alpha} - \Theta_{\alpha\nu}\right) (p)
  = \int_q \RL (q) \Tr \Sigma_{\nu\alpha} \frac{\partial}{\partial
    q_\nu} \left(  \G_{-q,p} - \G_{p,-q} \right)\Big|_{p=q}.\label{conformal-second-b}
\end{equation}

Hence, equating (\ref{conformal-first}) with the sum of
(\ref{conformal-second-a}) and (\ref{conformal-second-b}), we obtain
the WT identity as
\begin{align}
&  \int_q \sigma (-q) D_\alpha^K \varphi (q)\nt\\
  &\quad + \int_q \RL (q) \lb 
q_\nu \frac{\partial^2}{\partial
  q_\alpha \partial q_\nu} - \frac{1}{2} q_\alpha
  \frac{\partial^2}{\partial q_\beta \partial q_\beta} + \frac{D+1}{2}
  \frac{\partial}{\partial q_\alpha} \rb \Tr \left( \G_{-q,p} +
    \G_{p,-q}\right)\Big|_{p=q}\nt\\
  &\quad + \int_q \RL (q) \,\Tr \Sigma_{\nu\alpha}
    \frac{\partial}{\partial q_\nu} \left(  \G_{-q,p} -
    \G_{p,-q}\right)\Big|_{p=q} = 0,\label{conformal-WT1}
\end{align}
where we define the special conformal transformation of $\varphi$ by
\begin{align}
  D_\alpha^K \varphi (q)
  &\equiv \frac{1}{N}  \int_p \left(D_\alpha^K \bar{\Psi}^I (p+q) \cdot
    \Psi^I (-p) + \bar{\Psi}^I (p+q) D_\alpha^K \Psi^I (-p) \right)\nt\\
&=  \left( - q_\nu
    \frac{\partial^2}{\partial q_\alpha \partial q_\nu} -
    \frac{\partial}{\partial q_\alpha} + \frac{1}{2} q_\alpha
    \frac{\partial^2}{\partial q_\beta \partial q_\beta} \right)
  \varphi (q),
\end{align}
using (\ref{conformal-transformation}) for the Dirac fields.
Eq.~(\ref{conformal-WT1}) entails the invariance of the critical point
1PI Wilson action under the special conformal transformation
(\ref{conformal-transformation}) as a Ward-Takahashi identity.

Finally, by using the identity\footnote{This identity has been used as
  (D.4) in \cite{Sonoda:2017zgl}.  In Sec.~II of \cite{Cabrera:2026ggg} this is
  given as
  $\delta_\mathrm{conf} R_k (x,y) = - (x+y)_\mu \partial_t R_k (x,y)$.
This identity was in fact first derived in \cite{Delamotte:2015aaa}.}
\begin{align}
&  \int_q \RL (q) 
\lb q_\nu \frac{\partial^2}{\partial
    q_\alpha \partial q_\nu}  - \frac{1}{2} q_\alpha \frac{\partial^2}{\partial
    q_\beta \partial q_\beta} \rb \Tr \left(\G_{-q,p}+\G_{p,-q}\right)
  \Big|_{p=q}\nt\\
& = - \frac{1}{2} \int_q \left(D + q \cdot \partial_q \right) \RL (q)
  \cdot \frac{\partial}{\partial q_\alpha} \Tr \left(\G_{-q,p} +
  \G_{p,-q}\right) \Big|_{p=q},\label{conformal-identity}
\end{align}
which is valid if $\RL (q)$ is a function of $q^2$, we can rewrite the WT
identity as
\begin{align}
  &  \int_q \sigma (-q) D_\alpha^K \varphi (q)
    + \frac{1}{2} \int_q \left(1 - q \cdot \partial_q\right) \RL (q) 
    \frac{\partial}{\partial q_\alpha}  \Tr \left( \G_{-q,p} +
    \G_{p,-q}\right)\Big|_{p=q}\nt\\
  &\quad + \int_q \RL (q) \,\Tr \Sigma_{\nu\alpha}
    \frac{\partial}{\partial q_\nu} \left(  \G_{-q,p} -
    \G_{p,-q}\right)\Big|_{p=q} = 0.\label{conformal-WT2}
\end{align}
Eqs.~(\ref{conformal-WT1}) and (\ref{conformal-WT2}) are equivalent,
and they imply the invariance of the critical point under the special
conformal transformation (\ref{conformal-transformation}).

A few comments are in order.  Field theoretic approaches to conformal
symmetry and associated techniques have been applied to models such as
the one studied in this paper in the past
\cite{Vasiliev:1992_CFT-GN,Derkachov:1993uw}.  To the best of our
knowledge, however, this is the first time that conformal symmetry is
studied in a fermionic theory within the ERG framework.
This confirms that the ERG is a powerful conceptual framework to study
fundamental aspects such as the realization of a symmetry.
As for more concrete applications, with conformal symmetry being often
broken by the standard approximation schemes, it can be employed as a
criterion for the choice of a good regulator \cite{Balog:2020fyt}.

Finally, let us note that fermionic theories have been and are an
interesting study case for rigorous RG approaches; see
e.g.~\cite{Gawedzki:1985jn,deCalan:1991km,Giuliani:2020aot,Giuliani:2024ach}.
In this setting, it has been noted that correlation functions are
scale invariant up to a remainder term that decays at distances larger
than the inverse of the cutoff \cite{Giuliani:2024ach}.  Within the
ERG approach followed in this paper, we note that it is possible to
avoid observing such breaking of scale invariance provided that one
employs modified correlation functions, which are obtained from the
naive correlation functions by multiplying suitable cutoff
dependent factors and by employing a modified two-point function; see
e.g.~\cite{Igarashi:2009tj,Sonoda:2015pva,Pagani:2024gdt}.
In order to get a feeling of the idea, let us consider a Wilson action
$S_\Lambda$ whose correlation functions are $\Lambda$-dependent. For
instance, the two-point function of the scalar free theory may look
like $K\left(p/\Lambda\right)/\left(p^2+m^2\right)$, where
$K\left(p/\Lambda\right)$ decays exponentially for large values of the
momentum.  However, the associated modified correlation function gives
a power law at criticality \cite{Sonoda:2015pva}.
It would be interesting if the notion of modified correlation
functions could be introduced also in the rigorous approaches to the
ERG.

\section{Conclusions}

In this paper we have considered the large $N$ limit of a four-Fermi
theory with $N$ Dirac fields defined in the Euclidean space of
dimension $2 < D < 4$, where the theory is renormalizable.  We have
constructed the energy-momentum (EM) tensor with an infrared cutoff
$\Lambda$ as a functional of the Dirac fields.  In the limit
$\Lambda \to 0+$, the functional reduces to the effective action with
the insertion of a single EM tensor.  We have constructed the EM
tensor by solving the Ward-Takahashi identities (\ref{properties-WT},
\ref{properties-antisym}) for the translation and rotation invariance.
The cutoff dependence of the EM tensor, given by the ERG differential
equation discussed in Sec.~\ref{comp}, has played an essential role
for the construction.  Our final result is given by
(\ref{construction-EM-result}).  We then obtained a trace formula
(\ref{trace-result}) in Sec.~\ref{trace}.  Finally, in
Sec.~\ref{conformal}, by combining all the results, we have derived
the Ward-Takahashi identity for special conformal invariance of the
critical point, thereby confirming the ERG as a powerful tool for
understanding the realization of symmetries.
In this sense, we would like to call the reader's attention that,
within the ERG framework, it has been possible to show that conformal
invariance of the three dimensional Ising model is implied by its
scale invariance \cite{Delamotte:2015aaa}.

We believe that there are two directions we can proceed in the near
future.  One is to consider multiple products of the EM tensor; we
refer the reader to
\cite{Pagani:2017tdr,Pagani:2020ejb,Sonoda:2020gqc,Rose:2021zdk} for
some examples of operator products in the ERG context. The other is to
consider $1/N$ corrections to the results of this paper.  We believe
it important to show that the conformal invariance of the critical
point extends beyond the leading order in large $N$ approximations.

\appendix

\section{More about the large $N$ limit\label{appendix-largeN}}

The summary of the salient features of the model given in
  Sec.~\ref{largeN} is sketchy at best, and we would like to
  provide more details in this appendix.
  
\subsection{Renormalizability of a four-Fermi theory\label{appendix-largeN-subsec-4Fermi}}

Renormalizability of a four-Fermi theory ($D < 4$) was originally
shown by K.~Wilson \cite{Wilson:1972cf} in large $N$.  In this
appendix we would like to explain how the theory parametrized by
$f_2, f_{21}$ is related to a four-Fermi theory.

We recall, from Sec.~\ref{largeN}, that the interaction part of the
1PI Wilson action is given by
\begin{equation}
  \Gamma_{I\Lambda} [\varphi]
  = F_\Lambda [\sigma] + \int_p \sigma (p) \varphi (-p)
  \end{equation}
where
\begin{equation}
  \varphi (p) \equiv \frac{1}{N} \int_x e^{- i p x}\, \bar{\Psi}^I (x)
  \Psi^I (x),\quad
  \sigma (p) \equiv \frac{\delta \Gamma_{I\Lambda} [\varphi]}{\delta
    \varphi (-p)},
\end{equation}
and
\begin{align}
  F_\Lambda [\sigma]
  &= \tilde{F} [\sigma] + I_\Lambda [\sigma],\\
  I_\Lambda [\sigma]
  &\equiv \Lambda^D c\, \delta (0) + \Lambda^{D-1} c_1\, \sigma (0) +
    \frac{1}{2} \int_p \F (p) \sigma (p) \sigma (-p)\nt\\
  &\quad + \sum_{n=3}^\infty \frac{1}{n} \int_{p_1, \cdots, p_n}
    \delta \left( \sum_1^n p_i \right)\, \prod_{i=1}^n \sigma (p_i)
    \times I_{n\Lambda} (p_1,\cdots,  p_n),\\
  \F (p)
  &\equiv - \frac{1}{2} \int_q \Tr \left[ h_\Lambda (q) \left(
    h_\Lambda (q+p) + h_\Lambda (q-p) \right) + \frac{2}{q^2}
    \right]\nt\\
  &= \Lambda^{D-2} \frac{-1}{2} \int_q \Tr \left[ h(q) \lb h
    (q+p/\Lambda) + h (q-p/\Lambda) \rb + \frac{2}{q^2} \right]\nt\\
&= \Lambda^{D-2} \mathcal{F} (p/\Lambda), \\
  I_{n\Lambda} (p_1,\cdots,p_n)
  &\equiv - \int_q \Tr \left[ h_\Lambda (q) h_\Lambda (q-p_1) \cdots
    h_\Lambda (q-p_1-\cdots-p_{n-1}) \right].
\end{align}
The high momentum propagator is given by
\begin{equation}
  h_\Lambda (q) \equiv \frac{1}{i \fs{q} + \RL (q)} =
  \frac{1}{\Lambda} \frac{1}{i \frac{\fs{q}}{\Lambda} + R (q/\Lambda) }
= \frac{1}{\Lambda} h (p/\Lambda).
\end{equation}

For the $\Lambda$ independent functional $\tilde{F} [\sigma]$, we have
chosen
\begin{equation}
  \tilde{F} [\sigma] = \frac{1}{2} \int_p \left( f_2 + p^2 f_{21}
  \right) \sigma (p) \sigma (-p).\label{appendix-largeN-ourchoice}
\end{equation}
We would like to show that this is obtained as a scaling limit, i.e.,
the $\Lambda_0 \to +\infty$ limit of a bare
$\tilde{F}_{\Lambda_0} [\sigma]$ defined at a UV cutoff scale
$\Lambda_0$.  $\tilde{F}_{\Lambda_0} [\sigma]$ is the Legendre
transform of a bare four-Fermi interaction, and this is why we call
$\tilde{F} [\sigma]$ a four-Fermi theory.  The short but inaccurate
explanation given in Sec.~\ref{largeN} is to be elaborated.

In order to find such a bare theory, we first observe the following
asymptotic behavior of $I_\Lambda [\sigma]$ as $\Lambda \to +\infty$:
\begin{align}
  I_\Lambda [\sigma]
  &\overset{\Lambda \to +\infty}{\longrightarrow}
    c \Lambda^D \delta (0) + c_1 \Lambda^{D-1} \sigma (0) +
    \Lambda^{D-2} \mathcal{F} (0) \frac{1}{2} \int_p \sigma (p) \sigma
    (-p)\nt\\
  &\qquad + \Lambda^{D-3} I_3 (0,0,0) \frac{1}{3} \int_{p_1, p_2}
    \sigma (p_1) \sigma (p_2) \sigma (-p_1-p_2),
\end{align}
where
\begin{subequations}
  \begin{align}
    \mathcal{F} (0)
    &= - \int_q \Tr \left( h (q)^2 + \frac{1}{q^2} \right) < 0,\\
    I_3 (0,0,0)
    &\equiv - \int_q \Tr h (q)^3,\label{appendix-largeN-I3}
  \end{align}
\end{subequations}
and we need to keep the cubic term only for $3 \le D < 4$.

We now introduce an alternative choice of $\tilde{F} [\sigma]$:
\begin{align}
  \tilde{F}_{\Lambda_0} [\sigma]
  &\equiv  \frac{1}{2} \int_p \left( f_2 + p^2 f_{21} \right) \sigma
    (p) \sigma (-p) - I_{\Lambda_0} [\sigma]\nt\\
  &\quad + c \Lambda_0^D \delta (0) + c_1 \Lambda_0^{D-1} \sigma (0) +
    \Lambda_0^{D-2} \mathcal{F} (0) \frac{1}{2} \int_p \sigma (p)
    \sigma (-p)\nt\\
  &\quad + \Lambda_0^{D-3} I_3 (0,0,0) \frac{1}{3} \int_{p_1, p_2}
    \sigma (p_1) \sigma (p_2) \sigma (-p_1-p_2).
\end{align}
This has been chosen so that
\begin{equation}
  \lim_{\Lambda_0 \to +\infty}  \tilde{F}_{\Lambda_0} [\sigma]
  = \frac{1}{2} \int_p \left( f_2 + p^2 f_{21} \right) \sigma (p)
  \sigma (-p).\label{appendix-largeN-scalinglimit}
\end{equation}

The new solution
\begin{equation}
  F_{\Lambda, \Lambda_0} [\sigma]
  \equiv \tilde{F}_{\Lambda_0} [\sigma] + I_\Lambda [\sigma]
\end{equation}
has the property that at $\Lambda = \Lambda_0$ we find
\begin{align}
  F_{\Lambda_0, \Lambda_0} [\sigma]
  &=  \frac{1}{2} \int_p \left( f_2 + p^2 f_{21} \right) \sigma
    (p) \sigma (-p) \nt\\
  &\quad + c \Lambda_0^D \delta (0) + c_1 \Lambda_0^{D-1} \sigma (0) +
    \Lambda_0^{D-2} \mathcal{F} (0) \frac{1}{2} \int_p \sigma (p)
    \sigma (-p)\nt\\
  &\quad + \Lambda_0^{D-3} I_3 (0,0,0) \frac{1}{3} \int_{p_1, p_2}
    \sigma (p_1) \sigma (p_2) \sigma (-p_1-p_2).
\end{align}
For $2 < D < 3$, the cubic term is absent, and the inverse Legendre
transformation gives
\begin{align}
  &\Gamma_{I\Lambda_0} [\varphi]
  = F_{\Lambda_0, \Lambda_0} [\sigma] + \int_p \sigma (p) \varphi
    (-p)\nt = c \Lambda_0^D \delta (0) \nt\\
&\quad  - \frac{1}{2} \int_p \frac{1}{f_2 + p^2
    f_{21} + \Lambda_0^{D-2} \mathcal{F} (0) } \left( \varphi (p) +
    c_1 \Lambda_0^{D-1} \delta (p) \right) \left( \varphi (-p) + c_1
    \Lambda_0^{D-1} \delta (p) \right).
\end{align}
This is a four-Fermi theory.  Hence,  (\ref{appendix-largeN-scalinglimit})
implies that our choice (\ref{appendix-largeN-ourchoice}) is the
scaling limit (i.e., $\Lambda_0 \to +\infty$) of a four-Fermi theory.

For $3 \le D < 4$ there is a cubic term in
$F_{\Lambda_0, \Lambda_0} [\sigma]$.  The corresponding
$\Gamma_{I\Lambda_0} [\varphi]$ is an infinite series with terms cubic
in $\varphi$ and higher.  This is mainly due to our choice of the
cutoff function.  Without going into details, we merely mention that
we could replace the momentum dependent mass $\RL (p)$ by a kinetic
term such as
\begin{equation}
  \RL (p) \longrightarrow i \fs{p} \frac{e^{-
      \frac{p^2}{\Lambda^2}}}{1 - e^{- \frac{p^2}{\Lambda^2}}}.
\end{equation}
This would give a high momentum propagator
\begin{equation}
  h_\Lambda (p) = \frac{1}{i \fs{p}} \frac{1}{1 + \frac{e^{-
        \frac{p^2}{\Lambda^2}}}{1 - e^{- \frac{p^2}{\Lambda^2}}}}
  = \frac{1}{i \fs{p}} \left( 1 - e^{- \frac{p^2}{\Lambda^2}} \right)
\end{equation}
which is suppressed at low momenta.  This would give
\begin{equation}
  I_3 (0,0,0) = - \int_p \Tr h (p)^3
  = - \int_p \left( 1 - e^{-p^2}\right)^3 \Tr \frac{1}{\left( i
      \fs{p}\right)^3} = 0,
\end{equation}
and $\tilde{F}_{\Lambda_0, \Lambda_0} [\sigma]$ would be free from the
cubic term.\footnote{$c_1$ would also vanish.}  Hence, the bare theory
$\Gamma_{I\Lambda_0} [\varphi]$ would be a four-Fermi theory also for
$3 \le D < 4$.

\subsection{Construction of $\Gamma_{I\Lambda} [\varphi]$\label{appendix-largeN-subsec-Gamma}}

We would like to show that our theory is reduced to the Gaussian
theory in the limit $f_{21} \to - \infty$.  For this purpose we
calculate $\Gamma_{I\Lambda} [\varphi]$ and show that it reduces to a
field independent constant in the limit.

We first write
\begin{align}
  F_\Lambda [\sigma]
  &= \frac{1}{2} \int_p \left( f_2 + p^2 f_{21}\right) \sigma (p)
    \sigma (-p) + I_\Lambda [\sigma]\nt\\
  &= c \Lambda^D \delta (0) + c_1 \Lambda^{D-1} \sigma (0) 
   + \frac{1}{2} \int_p \left( f_2 + p^2 f_{21} + \F (p)
    \right) \sigma (p) \sigma (-p)\nt\\
  &\quad + \sum_{n=3}^\infty \frac{1}{n!} \int_{p_1, \cdots, p_n}
    \sigma (p_1) \cdots \sigma (p_n)\, \delta \left(\sum_1^n
    p_i\right)\, \cdot c_{n\Lambda} (p_1, \cdots, p_n),
\end{align}
where $c_{n\Lambda}$ is the sum of  symmetrized $I_{n\Lambda}$ such as
\begin{equation}
  c_{3\Lambda} (p_1, p_2, p_3)
  = I_{3\Lambda} (p_1, p_2, p_3) + I_{3\Lambda} (p_1, p_3, p_2).
\end{equation}
The inverse Legendre transformation gives
\begin{align}
  \Gamma_{I\Lambda} [\varphi]
  &= F_\Lambda [\sigma] + \int_p \sigma (p) \varphi (-p)\nt\\
  &= c \Lambda^D \delta (0)  \nt\\
  &\quad + \sum_{n=2}^\infty \frac{1}{n!} \int_{p_1, \cdots, p_n}
    \delta \left(\sum_1^n p_i\right)\, 
    \prod_{i=1}^n \left( \varphi (p_i) + c_1 \Lambda^{D-1} \delta
    (p_i) \right)\, \cdot d_{n\Lambda} (p_1, \cdots, p_n),
\end{align}
where
\begin{subequations}
  \begin{align}
    d_{2\Lambda} (p)
    &\equiv d_{2\Lambda} (p,-p) = \frac{1}{- f_2 - p^2 f_{21} - \F (p)},\\
    d_{3\Lambda} (p_1, p_2, p_3)
    &= \prod_{i=1}^3 d_{2\Lambda} (p_i) \cdot c_{3\Lambda} (p_1, p_2, p_3),\\
    d_{4\Lambda} (p_1, p_2, p_3, p_4)
    &= \prod_{i=1}^4 d_{2\Lambda} (p_i)\cdot \Big[ c_{4\Lambda} (p_1, p_2, p_3,
      p_4)\nt\\
    &\quad + c_{3\Lambda} (p_1, p_2, p_3+p_4) d_{2\Lambda} (p_1+p_2)
      c_{3\Lambda} (p_1+p_2, p_3, p_4)\nt\\
    &\quad + c_{3\Lambda} (p_1, p_3, p_2+p_4) d_{2\Lambda} (p_1+p_3)
      c_{3\Lambda}  (p_1+p_3, p_2, p_4)\nt\\
    &\quad + c_{3\Lambda} (p_1, p_4, p_2+p_3) d_{2\Lambda} (p_1+p_4)
      c_{3\Lambda} (p_1+p_4, p_2, p_3) \Big],
  \end{align}
\end{subequations}
and so on.\footnote{Strictly speaking this expansion is valid only for
  $f_2 < 0$ (symmetric phase).  For $f_2 > 0$ (broken phase) we need
  to shift $\varphi (p)$.}

In the limit $f_{21} \to - \infty$, we find
\begin{equation}
  \lim_{f_{21} \to - \infty} d_2 (p) = 0.
\end{equation}
Hence, we obtain
\begin{equation}
  \lim_{f_{21} \to -\infty} \Gamma_{I\Lambda} [\varphi] = c \Lambda^D
  \delta (0),
\end{equation}
and the theory is reduced to the massless Gaussian theory.
\begin{equation}
  \lim_{f_{21} \to -\infty} \Gamma_\Lambda [\Phi^I, \bar{\Phi}^I] =
  \Gamma_G [\Phi^I, \bar{\Phi}^I] + N c \Lambda^D \delta (0).
\end{equation}

\subsection{Calculation of $I_\Lambda (\sigma)$ for constant $\sigma$\label{appendix-largeN-subsec-ILambda}}

For the constant field $\sigma (p) = \sigma \delta (p)$, we obtain
\begin{equation}
  I_\Lambda [\sigma]
  = I_\Lambda (\sigma)\,\delta (0),
\end{equation}
where
\begin{align}
  I_\Lambda (\sigma)
  &= \frac{1}{2} \Tr \mathbf{1} \int_p \ln \frac{p^2 + \RL (p)^2}{p^2}
    - \sigma\,\Tr \mathbf{1} \int_p \frac{\RL (p)}{p^2 + \RL
    (p)^2} - \frac{1}{2} \sigma^2 \int_p \Tr \left[ h_\Lambda (p)^2 +
    \frac{1}{p^2}\right]\nt\\
&\quad - \sum_{n=3}^\infty \frac{1}{n} \sigma^n \int_p \Tr h_\Lambda  (p)^n
\end{align}
We can calculate
\begin{align}
&  - \sum_{n=3}^\infty \frac{1}{n} \sigma^n \int_p \Tr h_\Lambda
    (p)^n
  = \int_p \Tr \left[ \ln \left(1 - \sigma h_\Lambda
    (p)\right) + \sigma h_\Lambda (p) + \frac{1}{2} \sigma^2 h_\Lambda
    (p)^2 \right]\nt\\
  &= \int_p \Tr \left[ \ln \frac{i \fs{p} + \RL (p) - \sigma}{i
    \fs{p} + \RL (p)} + \sigma \frac{- i \fs{p} + \RL (p)}{p^2 + \RL
    (p)^2} + \frac{1}{2} \sigma^2 h_\Lambda (p)^2 \right]\nt\\
  &=  \Tr \mathbf{1} \, \int_p \left[ \frac{1}{2} \ln
    \frac{p^2 + \left(-\sigma + \RL (p)\right)^2}{p^2 + \RL (p)^2}
    + \frac{\sigma \RL (p)}{p^2 + \RL (p)^2} + \frac{1}{2} \sigma^2
    h_\Lambda (p)^2 \right].
\end{align}
Hence, we obtain
\begin{equation}
  I_\Lambda (\sigma)
  \equiv \Tr \mathbf{1} \int_p \left[ \frac{1}{2} \ln \frac{p^2 + \left(-
        \sigma + \RL (p)\right)^2}{p^2} - \frac{1}{2} \sigma^2
    \frac{1}{p^2} \right].
\end{equation}
In the limit $\Lambda \to 0+$, we obtain
\begin{equation}
  I_\mathrm{eff} (\sigma) = \lim_{\Lambda \to 0+} I_\Lambda (\sigma)
  = \Tr \mathbf{1} \cdot \frac{1}{2} \int_p \left[ \ln \frac{p^2 +
      \sigma^2}{p^2} - \frac{\sigma^2}{p^2} \right]
=  - k_D (\sigma^2)^{\frac{D}{2}}, 
\end{equation}
where the positive constant $k_D$ is given by
\begin{equation}
  k_D \equiv \Tr \mathbf{1} \cdot \frac{1}{(4 \pi)^{\frac{D}{2}}}
  \frac{1}{D} (-) \Gamma \left(\frac{2-D}{2}\right) > 0.\quad (2 < D <
  4)
\end{equation}
Hence, we obtain
\begin{equation}
  F_\mathrm{eff}  (\sigma)
  \equiv \lim_{\Lambda \to 0+} F_\Lambda (\sigma)
  = f_2 \frac{1}{2} \sigma^2 - k_D (\sigma^2)^{\frac{D}{2}}.
\end{equation}

To see the physical meaning of $F_\mathrm{eff} (\sigma)$, we observe
that in the limit $\Lambda \to 0+$ the effective action is given by
\begin{equation}
  \Gamma_\mathrm{eff} [\Psi^I, \bar{\Psi}^I]
  = - \int_p \bar{\Psi}^I (-p) i \fs{p} \Psi^I (p) + N F_\mathrm{eff}
  [\sigma] + \int_p \sigma (p) \int_q \bar{\Psi}^I (-q) \Psi^I (q+p),
  \label{appendix-largeN-Gammaeff}
\end{equation}
where $\sigma (p)$ is determined by
\begin{equation}
\frac{1}{N}  \int_q \bar{\Psi}^I (-q) \Psi^I (q+p) + \frac{\delta
  F_\mathrm{eff} [\sigma]}{\delta \sigma (-p)} = 0.\label{appendix-largeN-sigma}
\end{equation}
We can regard (\ref{appendix-largeN-Gammaeff}) as an effective action
of a theory with the Dirac field $\Psi^I, \bar{\Psi}^I$ interacting
with a real scalar field $\sigma$ through a Yukawa coupling.  We can
obtain (\ref{appendix-largeN-sigma}) by integrating out the scalar
field to leading order in large $N$.  Setting the Dirac field to zero,
we obtain the effective action of the scalar field as
$- N F_\mathrm{eff} [\sigma]$.  It is in this sense that we can regard
\begin{equation}
  V_\mathrm{eff} (\sigma) \equiv - F_\mathrm{eff} (\sigma)
  = - f_2 \frac{1}{2} \sigma^2 + k_D (\sigma^2)^{\frac{D}{2}}
\end{equation}
as the effective potential of the scalar field $\sigma$ coupled to
$\varphi$.  The effective potential is invariant under
$\mathbf{Z}_2: \sigma \to - \sigma$.  For $f_2 < 0$, the symmetric is
intact, and for $f_2 > 0$, the symmetry is broken spontaneously.

\subsection{Fixed-point, relevant and irrelevant parameters\label{appendix-largeN-subsec-fixedpoint}}

At the end of Sec.~\ref{largeN} we have called the critical theory a
fixed point theory, and have mentioned the relevance of the parameter
$f_2$ and the irrelevance of $f_{21}$.  We owe an explanation.

To see how the critical theory at $f_2 = f_{21} = 0$, corresponding to the
choice $\tilde{F} [\sigma] = 0$, is related to a fixed point, we need
to introduce a dimensionless convention in which we measure
dimensional physical quantities in units of appropriate powers of the
momentum cutoff $\Lambda$.  We introduce the dimensionless fields
$\tilde{\Psi}, \bar{\tilde{\Psi}}$ by
\begin{equation}
  \Psi^I (p) = \Lambda^{- \frac{D+1}{2}} \tilde{\Psi}^I (p/\Lambda),\quad
  \bar{\Psi}^I (-p) = \Lambda^{-\frac{D+1}{2}} \bar{\tilde{\Psi}}^I
  (-p/\Lambda).
\end{equation}
Let
\begin{equation}
  \Gamma_\Lambda [\Psi^I, \bar{\Psi}^I]
  = \tilde{\Gamma}_t [\tilde{\Psi}^I, \bar{\tilde{\Psi}}^I ],
\end{equation}
where $t$ is a logarithmic scale parameter
\begin{equation}
  \Lambda = \mu\,e^{-t}
\end{equation}
defined with an arbitrary reference mass scale $\mu$.  Then, the ERG
differential equation for $\Gamma_\Lambda$
\begin{equation}
  - \Lambda \partial_\Lambda \Gamma_\Lambda [\Psi^I, \bar{\Psi}^I]
  = - \int_p \Lambda \partial_\Lambda \RL (p)\, \Tr \G_{\Lambda; p,
    -p}^{II}
\end{equation}
is rewritten as
\begin{align}
&  \partial_t \tilde{\Gamma}_t [\tilde{\Psi}^I,
  \bar{\tilde{\Psi}}^I]
  = \int_p \left[
    \tilde{\Gamma}_t \Rd{\tilde{\Psi}^I (p)} \left( p \cdot \partial_p
    + \frac{D+1}{2}\right) \tilde{\Psi}^I (p) \right.\nt\\
  &\quad\left. + \left(p \cdot \partial_p +
    \frac{D+1}{2} \right) \bar{\tilde{\Psi}}^I (-p)
    \Ld{\bar{\tilde{\Psi}}^I (-p)} \tilde{\Gamma}_t 
 - \left(1 - p \cdot
    \partial_p\right) R (p)\, \Tr \tilde{\G}^{II}_{t; -p,p} \right],\label{appendix-largeN-dimlessERG}
\end{align}
where
\begin{equation}
  \tilde{\G}^{IJ}_{t; -p,q} \equiv \Lambda^{D+1} \G^{IJ}_{\Lambda; -\Lambda p, \Lambda q}
\end{equation}
satisfies
\begin{equation}
  \int_q \tilde{\G}^{IJ}_{t; -p,q} \left( R (q) \delta^{JK} \delta (q-r)
    - \Ld{\bar{\tilde{\Psi}} (-q)} \tilde{\Gamma}_t \Rd{\tilde{\Psi}^K
      (r)} \right) = \delta^{IK} \delta (p-r).
\end{equation}
Hence, if there is a fixed point $\tilde{\Gamma}_*$ with no anomalous
dimension for the Dirac field, it satisfies
\begin{align}
&  \int_p \Bigg[
    \tilde{\Gamma}_* \Rd{\tilde{\Psi}^I (p)} \left( p \cdot \partial_p
    + \frac{D+1}{2}\right) \tilde{\Psi}^I (p)  + \left(p \cdot \partial_p +
    \frac{D+1}{2} \right) \bar{\tilde{\Psi}}^I (-p)
    \Ld{\bar{\tilde{\Psi}}^I (-p)} \tilde{\Gamma}_*\nt\\
&\quad  - \left(1 - p \cdot
    \partial_p\right) R (p)\, \Tr \tilde{\G}^{II}_{*; -p,p} \Bigg] =
  0. \label{appendix-largeN-ERG-fixedpoint}
\end{align}

As we show in Sec.~\ref{conformal} the theory at $f_2 = f_{21} = 0$,
i.e., $\tilde{F} [\sigma] = 0$, has scale invariance
(\ref{conformal-scale-invariance}).  Rewriting it in the dimensionless
convention, we get (\ref{appendix-largeN-ERG-fixedpoint}).  Similarly,
the dimensionless 1PI Wilson action of the Gaussian critical point,
satisfying (\ref{appendix-free-scale}), satisfies
(\ref{appendix-largeN-ERG-fixedpoint}).

We have parametrized $\Gamma_\Lambda$ near the critical point in terms
of two parameters $f_2, f_{21}$ independent of $\Lambda$.  In the
dimensionless convention $\tilde{\Gamma}_t$ is parametrized by
\begin{equation}
  \tilde{f}_2 \equiv \Lambda^{2-D} f_2,\quad
  \tilde{f}_{21} \equiv \Lambda^{4-D} f_{21}.
\end{equation}
Since $\Lambda$ is $t$-dependent, we find the RG flow
\begin{equation}
  \frac{d}{dt} \tilde{f}_2 = (D-2) \tilde{f}_2,\quad
  \frac{d}{dt} \tilde{f}_{21} = (D-4) \tilde{f}_{21}.
\end{equation}
Since $D-2 > 0$, $\tilde{f}_2$ is a relevant parameter whose magnitude
grows as $t$ grows.  Analogously, $\tilde{f}_{21}$ is an irrelevant
parameter whose magnitude decreases as $t$ grows.  It is this RG flow
that we have given in Fig.~\ref{Fig1}.  The fixed point has an
infinite number of irrelevant parameters corresponding to the higher
order powers of $\sigma$ in $\tilde{F} [\sigma]$.  We have
particularly chosen $\tilde{f}_{21}$ because it connects the Gaussian
fixed point ($t \to - \infty$) and the non-trivial fixed point
($t \to +\infty$) as we have shown in Appendix~\ref{appendix-largeN-subsec-Gamma}.

Any critical system that flows into our non-trivial fixed point
belongs to the universality class of this fixed point.  Hence, those
critical systems possess special conformal invariance at long
distances.

\section{The Belinfante-Rosenfeld construction of the EM tensor\label{appendix-BR}}

In Sec.~\ref{properties} we have stated the existence of an
energy-momentum (EM) composite operator $\Theta_{\mu\nu} (p)$ with two
properties (\ref{properties-WT}) and (\ref{properties-antisym}).  We
would like to outline an argument that incorporates the improvement
introduced by Belinfante \cite{Belinfante} and Rosenfeld
\cite{Rosenfeld}.  We follow mostly \cite{Sonoda:2015pva} with a
little change made for the spin of the Dirac fields.

To start with we assume the invariance of the 1PI Wilson action
$\Gamma_\Lambda [\Psi^I, \bar{\Psi}^I]$ under translations and
rotations:
  \begin{align}
  & \int_q \left( \Gamma_\Lambda \Rd{\Psi^I (q)} D^T_\mu \Psi^I (q) +
    D^T_\mu \bar{\Psi}^I (-q) \Ld{\bar{\Psi}^I (-q)} \Gamma_\Lambda
    \right) = 0,\label{appendix-BR-translation}\\
  &\int_p \left( \Gamma_\Lambda \Rd{\Psi^I (q)} D^R_{\mu\nu} \Psi^I (q) +
    D^R_{\mu\nu} \bar{\Psi}^I (-q) \Ld{\bar{\Psi}^I (-q)} \Gamma_\Lambda
    \right) = 0,\label{appendix-BR-rotation}
  \end{align}
  where
\begin{equation}
  D^T_\mu \Psi^I (p) = p_\mu \Psi^I (p),\quad
  D^T_\mu \bar{\Psi}^I (-p) = - p_\mu \bar{\Psi}^I (-p), 
\end{equation}
is the infinitesimal translation, and
\begin{subequations}
\begin{align}
  D^R_{\mu\nu} \Psi^I (p)
  &= \left( p_\mu \frac{\partial}{\partial
    p_\nu} - p_\nu \frac{\partial}{\partial p_\mu} + \Sigma_{\mu\nu} \right) \Psi^I (p),\\
  D^R_{\mu\nu} \bar{\Psi}^I (-p)
  &= \left( p_\mu \frac{\partial}{\partial
    p_\nu} - p_\nu \frac{\partial}{\partial p_\mu} \right)
    \bar{\Psi}^I (-p) - \bar{\Psi}^I (-p)  \Sigma_{\mu\nu}.
\end{align}
\end{subequations}
is the infinitesimal rotation.  $\Sigma_{\mu\nu}$ is the spin matrix
defined by (\ref{properties-spin-matrix}).

We first introduce an equation-of-motion composite operator by
\begin{align}
    J_\nu (p)
    &\equiv
      \int_q \left(  \Gamma_\Lambda \Rd{\Psi^I (q)} (q+p)_\nu \Psi^I
      (q+p) + \left(-q+p\right)_\nu \bar{\Psi}^I (-q+p) \Ld{\bar{\Psi}^I (-q)}
      \Gamma_\Lambda \right)\nt\\
    &\quad -  \int_q \RL (q) \Tr \lb (q+p)_\nu \G^{II}_{-q-p,q} +
      (-q+p)_\nu \G^{II}_{-q,q-p} \rb ,\label{appendix-BR-Jnu}
\end{align}
where
\begin{equation}
  \int_q \G^{IJ}_{-p,q} \left( \RL (q)  \delta^{JK}
  \delta (q-r) - \Ld{\bar{\Psi}^J (-q)} \Gamma_{\Lambda} \Rd{\Psi^K
    (r)} \right) = \delta^{IK} \delta (p-r).
\end{equation}
At $p=0$ the second line of (\ref{appendix-BR-Jnu}) vanishes, and
thanks to the translation invariance (\ref{appendix-BR-translation}) we
obtain
\begin{equation}
  J_\nu (0) = 0.
\end{equation}
This implies that $J_\nu (p)$ is a derivative of a composite
  operator that we call the EM tensor
\begin{equation}
  J_\nu (p) = p_\mu \Theta_{\mu\nu} (p),
\end{equation}
and we obtain the Ward-Takahashi (WT) identity
\begin{align}
  p_\mu \Theta_{\mu\nu} (p)
  &= \int_q \left(  \Gamma_\Lambda \Rd{\Psi^I (q)} (q+p)_\nu \Psi^I
    (q+p) + \left(-q+p\right)_\nu \bar{\Psi}^I (-q+p) \Ld{\bar{\Psi}^I (-q)}
    \Gamma_\Lambda \right)\nt\\
  &\quad -   \int_q \RL (q) \Tr \lb (q+p)_\nu \G^{II}_{-q-p,q} +
    (-q+p)_\nu \G^{II}_{-q,q-p} \rb .
        \label{appendix-BR-WT}
\end{align}
This is (\ref{properties-WT}).

Now, differentiating the WT identity with respect to $p_\mu$,
antisymmetrizing the result with respect to $\mu$ and $\nu$, and
setting $p=0$, we obtain
\begin{align}
  \Theta_{\mu\nu} (0) - \Theta_{\nu\mu} (0)
  &= \int_q \left[ \Gamma_\Lambda \Rd{\Psi^I (q)} \left(q_\nu
    \frac{\partial}{\partial q_\mu} - q_\mu \frac{\partial}{\partial
    q_\nu} \right) \Psi^I (q) \right.\nt\\
  &\left.\quad + \left(q_\nu
    \frac{\partial}{\partial q_\mu} - q_\mu \frac{\partial}{\partial
    q_\nu} \right) \bar{\Psi}^I (-q)\, \Ld{\bar{\Psi}^I (-q)} 
    \Gamma_\Lambda \right],
\end{align}
where we have used the rotation invariance of the cutoff function
\begin{equation}
  \left(q_\nu
    \frac{\partial}{\partial q_\mu} - q_\mu \frac{\partial}{\partial
      q_\nu} \right) \RL (q) = 0.
\end{equation}
Using (\ref{appendix-BR-rotation}), we can write the above as
\begin{equation}
  \Theta_{\mu\nu} (0) - \Theta_{\nu\mu} (0)
  = \int_q \left( \Gamma_\Lambda \Rd{\Psi^I (q)} \Sigma_{\mu\nu}
    \Psi^I (q) - \bar{\Psi}^I (-q) \Sigma_{\mu\nu} \Ld{\bar{\Psi}^I
      (-q)} \Gamma_\Lambda\right).\label{appendix-BR-intermediate}
\end{equation}
Let us introduce another equation-of-motion composite operator
\begin{align}
  A_{\mu\nu} (p)
  &\equiv \int_q \left( \Gamma_\Lambda \Rd{\Psi^I (q)} \frac{1}{2} \Sigma_{\mu\nu} \Psi^I
    (q+p) - \bar{\Psi}^I (-q+p) \frac{1}{2} \Sigma_{\mu\nu} \Ld{\bar{\Psi}^I (-q)}
    \Gamma_\Lambda \right)\nt\\
  &\quad +  \int_q \RL (q)\, \Tr \left(
    - \frac{1}{2} \Sigma_{\mu\nu} \G^{II}_{-q-p,q} + \G^{II}_{-q,q-p}
    \frac{1}{2} \Sigma_{\mu\nu} \right) \label{appendix-BR-Amunu}
\end{align}
which is antisymmetric $A_{\mu\nu} (p) = - A_{\nu\mu} (p)$.  We can
now write (\ref{appendix-BR-intermediate}) as
\begin{equation}
  \Theta_{\mu\nu} (0) - \Theta_{\nu\mu} (0) = 2 A_{\mu\nu} (0).
\end{equation}
This implies that there exists a local composite operator
$\tau_{\alpha,\mu\nu} (p)$ satisfying
\begin{equation}
  \Theta_{\mu\nu} (p) - \Theta_{\nu\mu} (p) - 2 A_{\mu\nu} (p)
  = p_\alpha \tau_{\alpha,\mu\nu} (p),\label{appendix-BR-tau}
\end{equation}
where
\begin{equation}
  \tau_{\alpha,\mu\nu} (p) = - \tau_{\alpha,\nu\mu} (p).
\end{equation}

We are now ready to follow the construction of the improved
energy-momentum tensor due to Belinfante \cite{Belinfante} and
Rosenfeld \cite{Rosenfeld}.  We introduce
\begin{equation}
  B_{\alpha\mu\nu} (p) \equiv \frac{1}{2} \left( \tau_{\alpha,\mu\nu}
    (p) + \tau_{\mu,\nu\alpha} (p) - \tau_{\nu,\alpha\mu} (p) \right),
\end{equation}
which satisfies
\begin{equation}
  B_{\alpha\mu\nu} (p) = - B_{\mu\alpha\nu} (p),\label{appendix-BR-Bantisym}
\end{equation}
and
\begin{equation}
  B_{\alpha\mu\nu} (p) - B_{\alpha\nu\mu} (p) = \tau_{\alpha,\mu\nu}
  (p).
\end{equation}
Define the improved EM tensor by
\begin{equation}
  \Theta'_{\mu\nu} (p) \equiv \Theta_{\mu\nu} (p) - p_\alpha
  B_{\alpha\mu\nu} (p).
\end{equation}
This satisfies
\begin{equation}
  p_\mu \Theta'_{\mu\nu} (p) = p_\mu \Theta_{\mu\nu} (p) - p_\alpha
  p_\mu B_{\alpha\mu\nu} (p) = p_\mu \Theta_{\mu\nu} (p),
\end{equation}
where we have used (\ref{appendix-BR-Bantisym}).  Hence,
$\Theta'_{\mu\nu} (p)$ satisfies the same WT identity
(\ref{appendix-BR-WT}) as $\Theta_{\mu\nu} (p)$.  Moreover, we obtain
\begin{align}
  \frac{1}{2} \left( \Theta'_{\mu\nu} (p) - \Theta'_{\nu\mu}
  (p)\right)
  &= \frac{1}{2} \left( \Theta_{\mu\nu} (p) - \Theta_{\nu\mu} (p)
    \right) - \frac{1}{2} p_\alpha \left( B_{\alpha\mu\nu} (p) -
    B_{\alpha\nu\mu} (p) \right)\nt\\
  &= \frac{1}{2} \left( \Theta_{\mu\nu} (p) - \Theta_{\nu\mu} (p)
    \right) - \frac{1}{2} p_\alpha \tau_{\alpha,\mu\nu} (p)\nt\\
  &= A_{\mu\nu} (p),
\end{align}
where we have used (\ref{appendix-BR-tau}).  This gives
(\ref{properties-antisym}) in the large $N$ limit, using the rotation
invariance of the field $\varphi$.

\section{The energy-momentum tensor of the free
  theory\label{appendix-free}}

The 1PI Wilson action of the free fermionic theory is given by
\begin{equation}
  \Gamma_\Lambda [\Psi, \bar{\Psi}]
  = - \int_p \bar{\Psi} (-p) \left( i \fs{p} + m \right) \Psi (p) +
  c_{F\Lambda} \delta (0),
\end{equation}
where $c_{F\Lambda}$, satisfying
\begin{equation}
  - \Lambda \partial_\Lambda c_{F\Lambda} = - \int_p \Lambda
  \partial_\Lambda \RL (p)\, \Tr \frac{1}{i \fs{p} + m + \RL
    (p)},\label{appendix-free-cFLambda}
\end{equation}
is minus the vacuum energy density with an infrared cutoff $\Lambda$.
Solving (\ref{appendix-free-cFLambda}) we obtain \cite{Pagani:2024gdt}
 \begin{equation}
   c_{F\Lambda} =  \frac{1}{D} \Tr \mathbf{1} \int_p \left( \frac{(m +
       \RL (p))^2}{p^2 + (m+\RL(p))^2} - \frac{m^2}{p^2} \right).
 \end{equation}
 
 The EM tensor is given by
 \begin{equation}
   \Theta_{\mu\nu} (p)
   \equiv \int_q \bar{\Psi} (-q) \,t_{\mu\nu}^\mathrm{free} (-q, q+p) \Psi
   (q+p) + c_\Lambda\,\delta_{\mu\nu} \delta (p),
 \end{equation}
 where the coefficient
 \begin{align}
   t_{\mu\nu}^\mathrm{free} (-q,q+p)
   &\equiv  - \delta_{\mu\nu} \frac{1}{2} \left( i \fs{q} + i
    (\fs{q}+\fs{p}) + 2 m \right)
    + \frac{1}{4} \left( \gamma_\mu i (2q +p)_\nu + \gamma_\nu i (2q
    +p)_\mu \right) \nt\\
&\quad - i \fs{q} \frac{1}{2} \Sigma_{\mu\nu} + \frac{1}{2}
  \Sigma_{\mu\nu} i \left(\fs{q}+\fs{p}\right)
 \end{align}
satisfies
 \begin{subequations}
   \begin{align}
     p_\mu t^\mathrm{free}_{\mu\nu} (-q,q+p)
     &= - (q+p)_\nu i \fs{q} + q_\nu i \left(\fs{q} + \fs{p}\right) -
       m p_\nu ,\\
     t^\mathrm{free}_{\mu\nu} (-q,q+p) - \left(\mu \leftrightarrow \nu\right)
     &= - i \fs{q} \Sigma_{\mu\nu} + \Sigma_{\mu\nu} i \left(\fs{q} + \fs{p}\right).
   \end{align}
 \end{subequations}
 Hence, the EM tensor satisfies the WT identity
\begin{equation}
   p_\mu \Theta_{\mu\nu} (p)
   = \int_q \left( \Gamma_\Lambda \Rd{\Psi (q)} (q+p)_\nu \Psi (q+p)
     + \left(-q+p\right)_\nu \bar{\Psi} (-q+p) \Ld{\bar{\Psi} (-q)}
     \Gamma_\Lambda \right),\label{appendix-free-WT}
 \end{equation}
 and its antisymmetric part is given by
\begin{equation}
\Theta_{\mu\nu} (p) - \Theta_{\nu\mu} (p)
   = \int_q \left( \Gamma_\Lambda \Rd{\Psi (q)} \Sigma_{\mu\nu} \Psi (q+p) -
     \bar{\Psi} (-q+p) \Sigma_{\mu\nu} \Ld{\bar{\Psi} (-q)}
     \Gamma_\Lambda \right).\label{appendix-free-antisym}
 \end{equation}
 
 The trace is given by
 \begin{equation}
   \Theta (p) = - \frac{D-1}{2} \N_\Lambda (p) + m \Op_m (p),
   \label{appendix-free-trace}
 \end{equation}
 where $\N_\Lambda (p)$ is the number operator
 \begin{align}
   \N_\Lambda (p)
   &\equiv \int_q \left[
     -  \Gamma_\Lambda \Rd{\Psi (q)} \Psi (q+p) - \bar{\Psi} (-q+p)
     \Ld{\bar{\Psi} (-q)} \Gamma_\Lambda \right] \nt\\
   &\quad +
     \delta (p)\, 2 \int_q \RL (q)\, \Tr \frac{1}{\RL (q) + i
     \fs{q} + m} \nt\\
   &=  \int_q \bar{\Psi} (-q) \left( i \left(\fs{q} + \fs{q} +
     \fs{p}\right) + 2m \right) \Psi (q+p) + \delta (p)\, \Tr \mathbf{1}
     \cdot 2 \int_q \frac{\RL (q) \left( m + \RL (q) \right)}{q^2 + (m
     + \RL (q))^2}, 
 \end{align}
 and
 \begin{equation}
   \Op_m (p) \equiv \int_q \bar{\Psi} (-q) \Psi (q+p) - \partial_m
   c_{F\Lambda} \delta (p)
 \end{equation}
 is the operator conjugate to the mass parameter.  We note
 \begin{equation}
   \Op_m (p=0) = - \partial_m \Gamma_\Lambda.
 \end{equation}
 The field independent part of (\ref{appendix-free-trace}) gives
 \begin{align}
   D c_\Lambda
   &= - (D-1)  \Tr \mathbf{1} \int_p \frac{\RL
     (p) \left( m + \RL (p) \right)}{p^2 + (m + \RL (p))^2} \nt\\
   &\quad + \frac{2}{D} \Tr \mathbf{1} \int_p \lb \frac{m (m+\RL
     (p)) p^2}{\left( p^2 + (m+\RL (p))^2\right)^2}  - \frac{m^2}{p^2} \rb.
 \end{align}

 At $m=0$, we can derive the WT identities for the invariance under
 scale transformations and special conformal transformations as
 follows \cite{Sonoda:2015pva}.  For $m=0$ we denote $\Gamma_\Lambda$
 as $\Gamma_G$.

 \subsection{Scale invariance}

 We calculate
 \begin{equation}
   \lim_{p \to 0} \frac{\partial}{\partial p_\nu} \left(p_\mu
     \Theta_{\mu\nu} (p) \right)
 \end{equation}
 in two ways.  Using (\ref{appendix-free-WT}), we obtain
 \begin{equation}
   \int_q \left( \Gamma_G \Rd{\Psi (q)} \left( q_\nu
       \frac{\partial}{\partial q_\nu} + D \right) \Psi (q) + \left( q_\nu
       \frac{\partial}{\partial q_\nu} + D \right) \bar{\Psi} (-q)
     \cdot \Ld{\bar{\Psi} (-q)} \Gamma_G\right).
 \end{equation}
 Alternatively, we obtain\footnote{$0 = \partial_\nu \left(p_\nu
     \delta (p)\right)
   = D \delta (p) + p_\nu \partial_\nu \delta (p)$ gives $- p \cdot
   \partial_p \delta (p) = D \delta (p)$.}
 \begin{align}
&   \lim_{p\to 0} p_\mu \frac{\partial}{\partial p_\nu} \Theta_{\mu\nu}
  (p) + \Theta (0)
   = \lim_{p \to 0} p_\mu \frac{\partial}{\partial p_\nu} c_\Lambda
     \delta_{\mu\nu} \delta (p) + \Theta (0)\nt\\
&= - D c_\Lambda \delta (0) + \Theta (0)
  = - (D-1) \int_q \bar{\Psi} (-q) i \fs{q} \Psi (q)\nt\\
   &= \frac{D-1}{2} \int_q \left( \Gamma_G \Rd{\Psi (q)} \Psi (q) +
     \bar{\Psi} (-q) \Ld{\bar{\Psi} (-q)} \Gamma_G \right).
 \end{align}
 Equating the two, we obtain the WT identity for scale invariance:
 \begin{equation}
   \int_q \left( \Gamma_G \Rd{\Psi (q)} D^S \Psi (q) + D^S \bar{\Psi} (-q)
     \cdot \Ld{\bar{\Psi} (-q)} \Gamma_G \right) = 0,\label{appendix-free-scale}
 \end{equation}
 where
 \begin{subequations}
   \begin{align}
     D^S \Psi (q)
     &\equiv   - \left( q_\nu
    \frac{\partial}{\partial q_\nu} + \frac{D+1}{2} \right) \Psi
       (q),\\
     D^S \bar{\Psi} (-q)
     &\equiv   - \left( q_\nu
       \frac{\partial}{\partial q_\nu} + \frac{D+1}{2} \right)
       \bar{\Psi} (-q).
   \end{align}
 \end{subequations}
Finally, let us note that the Gaussian fixed point equation for
$\Gamma_G$ can be brought to the form (\ref{appendix-free-scale}).

 \subsection{Special conformal invariance}

 We calculate
 \begin{equation}
\lim_{p\to 0}   \left[  \left(\frac{\partial^2}{\partial p_\alpha \partial p_\nu} -
     \frac{1}{2} \delta_{\alpha\nu} \frac{\partial^2}{\partial p_\beta
       \partial p_\beta} \right)  p_\mu \Theta_{\mu\nu} (p) \right]
\end{equation}
in two ways.

Using (\ref{appendix-free-WT}) we obtain
\begin{align}
& \int_q \left[ \Gamma_G \Rd{\Psi (q)} \left( q_\nu
  \frac{\partial^2}{\partial q_\alpha 
  \partial q_\nu} - \frac{1}{2} q_\alpha \frac{\partial^2}{\partial
  q_\beta \partial q_\beta} + D \frac{\partial}{\partial q_\alpha}
  \right) \Psi (q) \right.\nt\\
  &\quad \left. - \left( q_\nu
  \frac{\partial^2}{\partial q_\alpha 
  \partial q_\nu} - \frac{1}{2} q_\alpha \frac{\partial^2}{\partial
  q_\beta \partial q_\beta} + D \frac{\partial}{\partial q_\alpha}
  \right) \bar{\Psi} (-q) \cdot \Ld{\bar{\Psi} (-q)} \Gamma_G
    \right].
\end{align}
Alternatively, we obtain\footnote{$0 = \left( \partial_\alpha
    \partial_\nu - \frac{1}{2} \delta_{\alpha\nu} \partial^2 \right)
  p_\nu \delta (p) = p_\nu  \left( \partial_\alpha
    \partial_\nu - \frac{1}{2} \delta_{\alpha\nu} \partial^2 \right)
  \delta (p) + D \partial_\alpha \delta (p)$ gives
  $p_\nu  \left( \partial_\alpha
    \partial_\nu - \frac{1}{2} \delta_{\alpha\nu} \partial^2 \right)
  \delta (p) = - D \partial_\alpha \delta (p)$.}
\begin{align}
  & \lim_{p\to 0} p_\mu \left( \frac{\partial^2}{\partial p_\alpha
    \partial p_\nu} - \frac{1}{2} \delta_{\alpha\nu}
    \frac{\partial^2}{\partial p_\beta \partial p_\beta} \right)
    \delta_{\mu\nu} c_\Lambda \delta (p)
 + \frac{\partial}{\partial p_\alpha} \Theta (0) -
    \frac{\partial}{\partial p_\nu} \left( \Theta_{\nu\alpha} -
    \Theta_{\alpha\nu}\right) (0)\nt\\
  &= \frac{\partial}{\partial p_\alpha} \left( \Theta (p) - D
    c_\Lambda \delta (p) \right) \Big|_{p=0}
 -  \frac{\partial}{\partial p_\nu} \left( \Theta_{\nu\alpha} -
    \Theta_{\alpha\nu}\right) (0)\nt\\
  &= \int_q \left[ \Gamma_G \Rd{\Psi (q)} \left( \frac{D-1}{2}
    \frac{\partial}{\partial q_\alpha}  - \Sigma_{\nu\alpha}
    \frac{\partial}{\partial q_\nu} \right) \Psi (q)\right.\nt\\
&\left.\qquad    - \left( \frac{\partial}{\partial q_\nu} \bar{\Psi} (-q)
    \Sigma_{\nu\alpha} + \frac{D-1}{2} \frac{\partial}{\partial
    q_\alpha} \bar{\Psi} (-q) \right) \Ld{\bar{\Psi} (-q)}
  \Gamma_G\right] .
\end{align}
Equating the two, we obtain the WT identity for special conformal
invariance:
\begin{equation}
  \int_q \left( \Gamma_G \Rd{\Psi (q)} D^K_\alpha \Psi (q) +
    D^K_\alpha \bar{\Psi} (-q) \Ld{\bar{\Psi} (-q)} \Gamma_G
  \right) = 0,
\end{equation}
where
\begin{subequations}
  \begin{align}
    D^K_\alpha \Psi (q)
    &\equiv \lb - q_\nu \frac{\partial^2}{\partial q_\alpha \partial
      q_\nu} + \frac{1}{2} q_\alpha \frac{\partial^2}{\partial
      q_\nu\partial q_\nu} - \frac{D+1}{2} \frac{\partial}{\partial
      q_\alpha} - \Sigma_{\nu\alpha} \frac{\partial}{\partial q_\nu}
      \rb \Psi (q),\\
    D^K_\alpha \bar{\Psi} (-q)
    &\equiv \lb  q_\nu \frac{\partial^2}{\partial q_\alpha \partial
      q_\nu} - \frac{1}{2} q_\alpha \frac{\partial^2}{\partial
      q_\nu\partial q_\nu} + \frac{D+1}{2} \frac{\partial}{\partial
      q_\alpha} \rb \bar{\Psi} (-q) - \frac{\partial}{\partial
      q_\nu} \bar{\Psi} (-q) \Sigma_{\nu\alpha} .
  \end{align}
\end{subequations}

\section{The number operator $\N_\Lambda (p)$ as a quadratic composite
  operator\label{appendix-N}}

In this appendix we would like to rewrite $\N_\Lambda (p)$
to make it more obvious that this is a quadratic
composite operator explained in Sec.~\ref{comp-subsec-quadratic}.

The number operator is defined by (\ref{comp-NLambda}) as an
equation-of-motion composite operator:
\begin{align}
  \N_\Lambda (p)
  &= \int_q \bar{\Psi}^I (-q) \lb i \fs{q} + i (\fs{q}+\fs{p}) \rb
    \Psi^I (q+p)\nt\\
  &\quad + N \int_q \lb
    \left( \RL (q) + \RL (q+p)\right) \,\Tr \G_{-q-p,q} + 2 \sigma
    (-q) \frac{\delta F_\Lambda [\sigma]}{\delta \sigma (-q-p)}
    \rb\nt\\
  &= N \int_q 2 \sigma (-q) \frac{\delta \tilde{F} [\sigma]}{\delta
    \sigma (-q-p)} + \int_q \bar{\Psi}^I (-q) \left( 2 i \fs{q} + i \fs{p} \right)
    \Psi^I (q+p)\nt\\
  &\quad + N \int_q \lb \left( \RL (q) + \RL (q+p) \right) \Tr
    \G_{-q-p,q} + 2 \sigma (-q) \frac{\delta I_\Lambda
    [\sigma]}{\delta \sigma (-q-p)} \rb .
\end{align}
The first term is quadratic in $\sigma$, and is a composite operator
on its own.  We would like to show that the remaining part is a
quadratic composite operator.

Using
\begin{align}
  & \int_q \sigma (-q) \frac{\delta}{\delta \sigma (-q-p)}
    \left( I_\Lambda [\sigma] - c_{1\Lambda} \sigma (0) - \frac{1}{2}
    \int_r \sigma (r) \sigma (-r) \F (r) \right)\nt\\
  &= - \int_q \Tr \Bigg[ \frac{1}{h_\Lambda (q+p)} \Big\lbrace\G_{-q-p,q}
    - h_\Lambda (q) \delta (p) - h_\Lambda (q+p) \sigma (p) h_\Lambda
    (q)\nt\\
&\qquad    - h_\Lambda (q+p) \int_{p_1,p_2} \sigma (p_1) \sigma (p_2)\,\delta
  (p_1+p_2-p) h_\Lambda (q+p_2) h_\Lambda (q) \Big\rbrace\Bigg]\nt\\
  &= - \int_q \Tr \Bigg[ \Big\lbrace \G_{-q-p,q}  - h_\Lambda (q) \delta
    (p) - h_\Lambda (q+p) \sigma (p) h_\Lambda    (q)\nt\\  
&\qquad    - h_\Lambda (q+p) \int_{p_1,p_2} \sigma (p_1) \sigma (p_2)\,\delta
  (p_1+p_2-p) h_\Lambda (q+p_2) h_\Lambda (q) \Big\rbrace
  \frac{1}{h_\Lambda (q)} \Bigg],
\end{align}
we obtain
\begin{align}
& \int_q \lb \left( \RL (q) + \RL (q+p) \right) \Tr
    \G_{-q-p,q} + 2 \sigma (-q) \frac{\delta I_\Lambda
    [\sigma]}{\delta \sigma (-q-p)} \rb \nt\\
&= \int_q  \left( \RL (q) + \RL (q+p) \right)
  \Tr \Bigg[ h_\Lambda (q) \delta (p) + h_\Lambda (q+p) \sigma (p)
  h_\Lambda (q)\nt\\
  &\qquad + h_\Lambda (q+p) \int_{p_1, p_2} \sigma (p_1) \sigma (p_2)
    \delta (p_1+p_2-p) h_\Lambda (q+p_2) h_\Lambda (q) \Bigg]\nt\\
  &\quad - \int_q \Tr \Bigg[ \left( i \fs{q} + i (\fs{q}+\fs{p})\right)
    \Big\lbrace \G_{-q-p,q} - h_\Lambda (q) \delta (p) - h_\Lambda
    (q+p) \sigma (p) h_\Lambda (q)\nt\\
  &\qquad - h_\Lambda (q+p) \int_{p_1, p_2} \sigma (p_1) \sigma (p_2)
    \delta (p_1+p_2-p) h_\Lambda (q+p_2) h_\Lambda (q)
    \Big\rbrace\Bigg]\nt\\
  &\quad + 2 c_{1\Lambda} \sigma (p) + \int_{p_1, p_2} \sigma (p_1)
    \sigma (p_2) \delta (p_1+p_2-p)\, \left( \F (p_1) + \F (p_2)\right).
\end{align}
Hence, we obtain
\begin{align}
  & \N_\Lambda (p) - N \int_q 2 \sigma (-q) \frac{\delta \tilde{F}
    [\sigma]}{\delta \sigma (-q-p)}\nt\\
  &= \int_q \bar{\Psi}^I (-q) \left( 2 i \fs{q} + i \fs{p}\right)
    \Psi^I (q+p)\nt\\
  &\quad - N \int_q \Tr \Bigg[ \left( 2 i \fs{q} + i \fs{p} \right)
    \Big\lbrace \G_{-q-p,q} - h_\Lambda (q) \delta (p) - h_\Lambda
    (q+p) \sigma (p) h_\Lambda (q)\nt\\
  &\qquad - h_\Lambda (q+p) \int_{p_1, p_2} \sigma (p_1) \sigma (p_2)
    \delta (p_1+p_2-p) h_\Lambda (q+p_2) h_\Lambda (q)
    \Big\rbrace\Bigg]\nt\\
  &\quad + N  \int_q 2 \RL (q) \Tr h_\Lambda (q)\, \delta (p)\nt\\
  &\quad + N \left( 2 c_{1\Lambda} + \int_q  \left( \RL (q) + \RL (q+p) \right)
    \Tr h_\Lambda (q+p) h_\Lambda (q) \right) \sigma (p) \nt\\
&\quad  + N \int_{p_1, p_2} \sigma (p_1)
    \sigma (p_2) \delta (p_1+p_2-p)\, \Big\lbrace \F (p_1) + \F (p_2)\nt\\
&\qquad + \int_q \left(\RL (q) + \RL (q+p)\right) \Tr \left[h_\Lambda (q+p)
  \frac{1}{2} \left( h_\Lambda (q+p_1) + h_\Lambda (q+p_2)\right)
  h_\Lambda (q) \right] \Big\rbrace .\label{appendix-N-quadratic}
\end{align}
It is straightforward to check that the $\Lambda$-dependence of the
last three terms is given by
\begin{subequations}
\begin{align}
  & - \Lambda \partial_\Lambda \int_q 2 \RL (q) h_\Lambda (q) = \int_q
    2 (- i \fs{q}) (-\Lambda \partial_\Lambda) h_\Lambda (q),\\
  &- \Lambda \partial_\Lambda \int_q \left( 2 c_{1\Lambda} + \int_q
    \left( \RL (q) + \RL (q+p)\right) \Tr h_\Lambda (q) h_\Lambda
    (q+p) \right)\nt\\
  &\quad = - \int_q \Tr \left[ (2 i \fs{q} + i \fs{p}) (-\Lambda
    \partial_\Lambda) \left(  h_\Lambda (q+p) h_\Lambda (q) \right) \right],\\
  &  - \Lambda \partial_\Lambda
    \Big\lbrace \F (p_1) + \F (p_2)\nt\\
 &\qquad + \int_q \left(\RL (q) + \RL (q+p)\right) \Tr \left[h_\Lambda (q+p)
  \frac{1}{2} \left( h_\Lambda (q+p_1) + h_\Lambda (q+p_2)\right)
    h_\Lambda (q) \right] \Big\rbrace\nt\\
  &\quad = - \int_q \Tr  \left[
    \left( 2 i \fs{q} + i \fs{p}\right) \left(- \Lambda
    \partial_\Lambda\right)
    \left( h_\Lambda (q+p) \frac{1}{2} \left( h_\Lambda (q+p_1) +
    h_\Lambda (q+p_2) \right) h_\Lambda (q) \right) \right].
\end{align}
\end{subequations}
Hence, (\ref{appendix-N-quadratic}) is a quadratic composite operator
corresponding to
\begin{equation}
  C (-q,q+p) = 2 i \fs{q} + i \fs{p}.
\end{equation}

\section{Solving the ERG equations to determine
  $A_\Lambda, B_{\Lambda;\mu\nu},
  C_{\Lambda;\mu\nu}$\label{appendix-ABC}}

\subsection{$A_\Lambda$}

The coefficient $A_\Lambda$ is a solution to
(\ref{construction-ERG-A}):
\begin{align}
  - \Lambda \partial_\Lambda A_\Lambda \delta_{\mu\nu}
  &= - \int_q \Tr s_{\mu\nu} (-q,q) f_\Lambda (q)\nt\\
  &= \int_q \Tr \left( \delta_{\mu\nu} i \fs{q} - \frac{i}{2}
    \left(\gamma_\mu q_\nu + \gamma_\nu q_\mu\right) \right) f_\Lambda
    (q).
\end{align}
The solution is unique, and it is given by an integral with a
  UV subtraction
\begin{align}
  A_\Lambda \delta_{\mu\nu}
  &= \int_q \Tr \left( \delta_{\mu\nu} i \fs{q} - \frac{i}{2}
    \left(\gamma_\mu q_\nu + \gamma_\nu q_\mu\right) \right) \left(
    h_\Lambda (q) - \frac{1}{i \fs{q}} \right)\nt\\
  &= - \int_q \RL (q) \Tr \left( \delta_{\mu\nu} i \fs{q} - \frac{i}{2}
    \left(\gamma_\mu q_\nu + \gamma_\nu q_\mu\right) \right)
    \frac{1}{i\fs{q} \left(i\fs{q}+\RL (q)\right)}\nt\\
  &= - \delta_{\mu\nu} \left( 1 - \frac{1}{D}\right) 
    \int_q \RL (q) \Tr h_\Lambda (q) .
\end{align}
Hence, $A$ defined by (\ref{trace-A}) vanishes:
\begin{equation}
  A \equiv D A_\Lambda + (D-1) \int_q \RL (q) \Tr h_\Lambda (q) = 0.
\end{equation}

\subsection{$B_{\Lambda;\mu\nu} (p)$}

The coefficient $B_{\Lambda;\mu\nu} (p)$ is determined by 
(\ref{construction-ERG-B})
\begin{equation}
  - \Lambda \partial_\Lambda B_{\Lambda;\mu\nu} (p)
  = - \int_q \Tr s_{\mu\nu} (-q,q+p) (- \Lambda \partial_\Lambda)
  \left( h_\Lambda (q+p) h_\Lambda (q) \right),\label{appendix-ABC-ERG-B}
\end{equation}
where $s_{\mu\nu}$ is given by (\ref{properties-smunu}), and the
the WT identity (\ref{construction-WT-B})
\begin{align}
  p_\mu B_{\Lambda;\mu\nu} (p)
  &= - \int_q  \Tr \lb (q+p)_\nu \RL (q) - q_\nu \RL (q+p) \rb
    h_\Lambda (q+p) h_\Lambda (q)\nt\\
  &\quad +  p_\mu \int_q \left(\RL (q) - \RL (q+p) \right) \Tr
    \frac{1}{2} \Sigma_{\mu\nu} h_\Lambda (q+p) h_\Lambda (q).
    \label{appendix-ABC-WT-B}
\end{align}
In principle we can solve (\ref{appendix-ABC-ERG-B}) by an
  integral with a UV subtraction such as
\begin{equation}
  B_{\Lambda;\mu\nu} (p)
  = - \int_q \Tr \left[
    s_{\mu\nu} (-q,q+p) h_\Lambda (q+p) h_\Lambda (q) - \cdots
  \right],
\end{equation}
where the dots subtract the unintegrable terms of order $1$
and $\frac{1}{q^2}$.  With the UV subtraction, however, we
  expect it to be cumbersome to check (\ref{appendix-ABC-WT-B}).
For the purpose of determining $B (p)$, it suffices to
  construct $B_{\mu\nu} (p)$ indirectly using
(\ref{appendix-ABC-ERG-B}) and (\ref{appendix-ABC-WT-B}).

To check the consistency of the above two equations, we compare $p_\mu
(- \Lambda \partial_\Lambda B_{\Lambda;\mu\nu})$ and $- \Lambda
\partial_\Lambda (p_\mu B_{\Lambda;\mu\nu})$ and show them equal.  We
first rewrite (\ref{appendix-ABC-WT-B}) as
\begin{align}
  p_\mu     B_{\Lambda;\mu\nu} (p)
  &= \int_q \Tr p_\mu \Sigma_{\mu\nu} \lb h_\Lambda (q+p) h_\Lambda
    (q) \left( i \fs{q} + \RL (q) - i\fs{q}\right)\right.\nt\\
  &\quad\left. - \left( i(\fs{q}+\fs{p}) + \RL (q+p) - i
    (\fs{q}+\fs{p})\right) h_\Lambda (q+p) h_\Lambda (q) \rb\nt\\
  &\quad - \int_q \Big\lbrace (p+q)_\nu h_\Lambda (q+p) h_\Lambda (q)
    \left( i \fs{q} + \RL (q) - i \fs{q}\right)\nt\\
  &\qquad - q_\nu \left( i(\fs{q}+\fs{p}) + \RL (q+p) - i
    (\fs{q}+\fs{p}) \right) h_\Lambda (q+p) h_\Lambda (q)
    \Big\rbrace\nt\\
  &= \int_q \Tr \Bigg[ \left(p_\mu \Sigma_{\mu\nu} - (q+p)_\nu \right)
    h_\Lambda (q+p) + \left(- p_\mu \Sigma_{\mu\nu} + q_\nu \right)
    h_\Lambda (q)\nt\\
  &\qquad + \lb
    - i\fs{q} p_\mu \Sigma_{\mu\nu} +
    p_\mu \Sigma_{\mu\nu} i (\fs{q}+\fs{p})
    + (q+p)_\nu i \fs{q} - q_\nu i(\fs{q}+\fs{p}) \rb h_\Lambda (q+p)
    h_\Lambda (q) \Bigg].
\end{align}
Since
\begin{equation}
  \int_q \Tr \left[
    \left(p_\mu \Sigma_{\mu\nu} - (q+p)_\nu \right) f_\Lambda (q+p)
    + \left(- p_\mu \Sigma_{\mu\nu} + q_\nu \right) f_\Lambda (q)
  \right] = 0,
\end{equation}
we obtain
\begin{align}
  (- \Lambda \partial_\Lambda) p_\mu B_{\Lambda;\mu\nu} (p)
  &= - \int_q \Tr p_\mu s_{\mu\nu} (-q,q+p) \left( - \Lambda
    \partial_\Lambda \right)\left( h_\Lambda (q+p) h_\Lambda
    (q)\right)\nt\\
  &= p_\mu \left( - \Lambda \partial_\Lambda B_{\Lambda;\mu\nu} (p)
    \right)
\end{align}
from (\ref{appendix-ABC-ERG-B}).  Hence, (\ref{appendix-ABC-ERG-B}) and
(\ref{appendix-ABC-WT-B}) are consistent.

We now take the trace of (\ref{appendix-ABC-ERG-B}):
\begin{equation}
  - \Lambda \partial_\Lambda B_{\Lambda; \mu\mu} (p)
  = \frac{D-1}{2} \int_q \Tr \left(2 i \fs{q} + i \fs{p}\right)
  (- \Lambda \partial_\Lambda ) \left( h_\Lambda (q+p) h_\Lambda (q)
  \right).
\end{equation}
Then, $B(p)$ defined by (\ref{trace-B}) satisfies
\begin{align}
  - \Lambda \partial_\Lambda B (p)
  &= \frac{D-1}{2} \Bigg[\int_q \Tr \left(2 i \fs{q} + i \fs{p}\right)
  (- \Lambda \partial_\Lambda ) \left( h_\Lambda (q+p) h_\Lambda (q)
    \right) \nt\\
  &\quad - \Lambda \partial_\Lambda \left(
    2 c_{1\Lambda} +  \int_q \left( \RL (q)+ \RL (q+p)
    \right) \Tr h_\Lambda (q) h_\Lambda (q+p) \right) \Bigg]\nt\\
  &= \frac{D-1}{2} \Bigg[\int_q \Tr \left(2 i \fs{q} + i \fs{p}\right)
  (- \Lambda \partial_\Lambda ) \left( h_\Lambda (q+p) h_\Lambda (q)
    \right) \nt\\
  &\quad + \int_q \Tr \Bigg\lbrace - f_\Lambda (q) - f_\Lambda (q+p) -
     \left(\Lambda \partial_\Lambda \RL (q) +
    \Lambda \partial_\Lambda \RL (q+p) \right) \Tr h_\Lambda (q)
    h_\Lambda (q+p)\nt\\
  &\qquad +  \left( \RL (q)+ \RL (q+p) \right)
    (-\Lambda \partial_\Lambda) \left(h_\Lambda (q) h_\Lambda (q+p)
    \right) \Bigg\rbrace\Bigg].
\end{align}
Since
\begin{align}
  & \int_q \Tr \left[
    \left(\Lambda \partial_\Lambda \RL (q) + \Lambda \partial_\Lambda
    \RL (q+p) \right) h_\Lambda (q) h_\Lambda (q+p) \right]\nt\\
  &= \int_q \Tr \left[ f_\Lambda (q) \left( i \fs{q} + \RL (q)\right)
    h_\Lambda (q+p) + f_\Lambda (q+p) \left(i (\fs{q}+\fs{p}) + \RL
    (q+p)
    \right) h_\Lambda (q) \right]
\end{align}
we obtain
\begin{align}
  & \int_q \Tr \left[ f_\Lambda (q) + f_\Lambda (q+p)
    + \left(\Lambda \partial_\Lambda \RL (q) + \Lambda \partial_\Lambda
    \RL (q+p) \right) h_\Lambda (q) h_\Lambda (q+p) \right]\nt\\
  &= \int_q \Tr \left[ f_\Lambda (q) \left( 2 i \fs{q} + i \fs{p} +
    \RL (q) + \RL (q+p) \right) h_\Lambda (q+p)\right.\nt\\
  &\left.\qquad  + f_\Lambda (q+p) \left( 2 i \fs{q} + i \fs{p} + \RL (q) + \RL
    (q+p) \right) h_\Lambda (q) \right]\nt\\
  &= \int_q \Tr \left(  2 i \fs{q} + i \fs{p} + \RL (q) + \RL
    (q+p) \right) (-\Lambda \partial_\Lambda) \left( h_\Lambda (q)
    h_\Lambda (q+p) \right).
\end{align}
Hence, we obtain
\begin{equation}
  - \Lambda \partial_\Lambda B (p) = 0,
\end{equation}
and $B(p)$ is independent of $\Lambda$ as expected.

Now, $B (p)$ has the mass dimension $D-1$, and we expect it to be
proportional to $(p^2)^{\frac{D-1}{2}}$.  For $2 < D < 4$, this is not
local except for $D=3$.  Hence, for $D \ne 3$, we find
\begin{equation}
  B (p) = 0.\label{trace-B-zero}
\end{equation}
For $D=3$, (\ref{appendix-ABC-ERG-B}) and (\ref{appendix-ABC-WT-B})
leave $B_{\Lambda;\mu\nu} (p)$ undetermined up to the addition of a
constant multiple of $p^2 \delta_{\mu\nu} - p_\mu p_\nu$.  The
constant is determined by demanding (\ref{trace-B-zero}).

We have thus obtained the trace
\begin{align}
  B_{\Lambda;\mu\mu} (p)
  &= - \frac{D-1}{2} \int_q \Tr \left[\left( \frac{1}{h_\Lambda (q)} +
    \frac{1}{h_\Lambda (q+p)} - \RL (q) - \RL (q+p)\right)
 h_\Lambda (q+p) h_\Lambda (q) \right]\nt\\
  &= - \frac{D-1}{2} \int_q \Tr \left[ h_\Lambda (q) + h_\Lambda (q+p)
    - \left(\RL (q) + \RL (q+p)\right) h_\Lambda (q) h_\Lambda (q+p)
    \right]\nt\\
  &= (D-1) \left[ - \int_q \Tr h_\Lambda (q) + \frac{1}{2} \int_q \left( \RL
    (q) + \RL (q+p) \right) \Tr h_\Lambda (q) h_\Lambda (q+p) \right].
\end{align}

\subsection{$C_{\Lambda;\mu\nu} (p_1,p_2)$}

The coefficient $C_{\Lambda;\mu\nu} (p_1, p_2)$ is determined by
(\ref{construction-ERG-C})
\begin{align}
 - \Lambda \partial_\Lambda C_{\Lambda;\mu\nu} (p_1, p_2)
    &= - \int_q \Tr s_{\mu\nu} (-q,q+p) \left[
      f_\Lambda (q+p) \frac{1}{2} \left( h_\Lambda (q+p_1) + h_\Lambda
      (q+p_2) \right) h_\Lambda (q)\right.\nt\\
    &\qquad + h_\Lambda (q+p) \frac{1}{2} \left( f_\Lambda (q+p_1) + f_\Lambda
      (q+p_2) \right) h_\Lambda (q) \nt\\
    &\left.\qquad + h_\Lambda (q+p) \frac{1}{2} \left( h_\Lambda (q+p_1) + h_\Lambda
      (q+p_2) \right) f_\Lambda (q) \right], \label{appendix-ABC-ERG-C}
\end{align}
and the WT identity (\ref{construction-WT-C})
\begin{align}
  &p_\mu C_{\Lambda;\mu\nu} (p_1,p_2)
  = - \frac{1}{2} \left( p_{1\nu} \F (p_1) + p_{2\nu} \F (p_2)
    \right)\nt\\
  &\quad - \int_q \Tr \lb (q+p)_\nu \RL (q) - q_\nu \RL (q+p)\rb
    h_\Lambda (q+p) \frac{1}{2} \left( h_\Lambda (q+p_1) + h_\Lambda
    (q+p_2) \right) h_\Lambda (q)\nt\\
  &\quad + p_\mu \int_q \left( \RL (q) - \RL (q+p)\right) \Tr
    \Sigma_{\mu\nu} h_\Lambda (q+p) \frac{1}{2} \left( h_\Lambda (q+p_1) + h_\Lambda
    (q+p_2) \right) h_\Lambda (q) .\label{appendix-ABC-WT-C}
\end{align}
For the same reason as for $B_{\Lambda;\mu\nu} (p)$, we do not seek to
obtain $C_{\Lambda;\mu\nu} (p_1, p_2)$ as an integral with a UV
subtraction.

We rewrite (\ref{appendix-ABC-WT-C}) as
\begin{align}
&  p_\mu C_{\Lambda;\mu\nu} (p_1,p_2)
  = - \frac{1}{2} \left( p_{1\nu} \F (p_1) + p_{2\nu} \F (p_2)
    \right)\nt\\
  &\quad - \int_q \Tr \Bigg[
    \lb (q+p)_\nu \left( i \fs{q} + \RL (q) - i \fs{q}\right)
    - q_\nu \left( i (\fs{q}+\fs{p}) + \RL (q+p) - i (\fs{q}+\fs{p})
    \right)\rb \nt\\
  &\qquad \times h_\Lambda (q+p) \frac{1}{2} \left( h_\Lambda (q+p_1) +
    h_\Lambda (q+p_2) \right) h_\Lambda (q)\Bigg]\nt\\
  &\quad + \int_q \Tr \Bigg[
    \lb \left(  i \fs{q} + \RL (q) - i \fs{q} \right) p_\mu
    \Sigma_{\mu\nu} + p_\mu \Sigma_{\mu\nu} \left( - i (\fs{q}+\fs{p}
    + \RL (q+p)) + i (\fs{q}+\fs{p}) \right) \rb\nt\\
  &\qquad \times h_\Lambda (q+p) \frac{1}{2} \left( h_\Lambda (q+p_1) +
    h_\Lambda (q+p_2) \right) h_\Lambda (q)\Bigg] \nt\\
  &= - \frac{1}{2} \left( p_{1\nu} \F (p_1) + p_{2\nu} \F (p_2)
    \right)\nt\\
  &\quad + \int_q \Tr \Bigg[ \left( q_\nu - p_\mu \Sigma_{\mu\nu}
    \right)  \frac{1}{2} \left( h_\Lambda (q+p_1) +
    h_\Lambda (q+p_2) \right) h_\Lambda (q)\nt\\
  &\qquad + \left( -(q+p)_\nu + p_\mu \Sigma_{\mu\nu} \right)
    h_\Lambda (q+p) \frac{1}{2} \left( h_\Lambda (q+p_1) + 
    h_\Lambda (q+p_2) \right)\nt\\
  &\qquad + \lb (q+p)_\nu i \fs{q} - q_\nu i (\fs{q}+\fs{p}) - i\fs{q}
    p_\mu \Sigma_{\mu\nu} + p_\mu \Sigma_{\mu\nu} i (\fs{q}+\fs{p})
    \rb\nt\\
  &\qquad \times h_\Lambda (q+p) \frac{1}{2} \left( h_\Lambda (q+p_1) +
    h_\Lambda (q+p_2) \right) h_\Lambda (q)\Bigg].
\end{align}

Using (\ref{largeN-ERG-FLambda}) we obtain
\begin{align}
  &     (-\Lambda\partial_\Lambda) \left( p_{1\nu} \F (p_1) +
    p_{2\nu} \F (p_2) \right)\nt\\
  &= - p_{1\nu} \int_q \Tr f_\Lambda (q) \left( h_\Lambda (q-p_1) +
    h_\Lambda (q+p_1) \right) - p_{2\nu} \int_q \Tr f_\Lambda (q)
    \left( h_\Lambda (q-p_2) + h_\Lambda (q+p_2) \right)\nt\\
  &= - p_{1\nu} \int_q \Tr \lb f_\Lambda (q) h_\Lambda (q+p_1) +
    h_\Lambda (q) f_\Lambda (q+p_1) \rb\nt\\
 &\quad   - p_{2\nu} \int_q \Tr \lb f_\Lambda (q) h_\Lambda (q+p_2) +
    h_\Lambda (q) f_\Lambda (q+p_2) \rb.
\end{align}
This gives  
\begin{align}
  & (-\Lambda\partial_\Lambda) \frac{-1}{2} \left( p_{1\nu} \F (p_1) +
    p_{2\nu} \F (p_2) \right) \nt\\
  &+ \int_q \Tr \Bigg[
    \left(q_\nu - p_\mu \Sigma_{\mu\nu}\right) \frac{1}{2} \lb
    \left(f_\Lambda (q+p_1) + f_\Lambda (q+p_2) \right) h_\Lambda (q)
    + \left(h_\Lambda (q+p_1) + h_\Lambda (q+p_2)\right) f_\Lambda (q)
    \rb\nt\\
  &\quad + \left( - (q+p)_\nu + p_\mu \Sigma_{\mu\nu}\right)
    \frac{1}{2} \lb f_\Lambda (q+p) \left(h_\Lambda (q+p_1) + h_\Lambda
    (q+p_2)\right) \right.\nt\\
  &\left.\qquad + h_\Lambda (q+p) \left(f_\Lambda (q+p_1) +
    f_\Lambda (q+p_2) \right) \rb \Bigg] = 0.
\end{align}
We thus obtain
\begin{align}
  -\Lambda \partial_\Lambda \left( p_\mu C_{\Lambda;\mu\nu} (p_1, p_2)
  \right)
  &= - \int_q \Tr \Bigg[ p_\mu s_{\mu\nu} (-q,q+p)\nt\\
  &\quad\times    (- \Lambda \partial_\Lambda) \left( h_\Lambda (q+p) \frac{1}{2}
    \left( h_\Lambda (q+p_1) +  h_\Lambda (q+p_2) \right) h_\Lambda
    (q) \right)\Bigg] \nt\\
  &= p_\mu \left(- \Lambda \partial_\Lambda C_{\Lambda;\mu\nu} (p_1, p_2) \right).
\end{align}
Hence, (\ref{appendix-ABC-ERG-C}) and (\ref{appendix-ABC-WT-C}) are
consistent.

Now, taking the trace of (\ref{appendix-ABC-ERG-C}), we obtain
\begin{align}
&  - \Lambda \partial_\Lambda C_{\Lambda;\mu\mu} (p_1, p_2)\nt\\
&  = \frac{D-1}{2} \int_q \Tr \left(2 i \fs{q} + i \fs{p}\right)
  (- \Lambda \partial_\Lambda) \left( h_\Lambda (q+p) \frac{1}{2}
    \left( h_\Lambda (q+p_1) + h_\Lambda (q+p_2) \right) h_\Lambda (q)
  \right).
\end{align}
Since
\begin{align}
&  - \Lambda \partial_\Lambda 
 \Bigg[ \F (p_1) + \F (p_2)\nt\\
&\qquad    + \int_q \left(\RL (q) + \RL (q+p)\right) \Tr \left(h_\Lambda (q+p) 
\frac{1}{2}
    \left( h_\Lambda (q+p_1) + h_\Lambda (q+p_2) \right) h_\Lambda (q)\right)
  \Bigg] \nt\\
&= \int_q \left(- \Lambda \partial_\Lambda\right) \Tr \Bigg[
  - \frac{1}{2} h_\Lambda (q) \left(h_\Lambda (q+p_1) + h_\Lambda
  (q-p_1) + h_\Lambda (q+p_2) + h_\Lambda (q-p_2)\right)\nt\\
&\quad + 
  \left( i \fs{q} + \RL (q) + i (\fs{q}+\fs{p}) + \RL (q+p) - 2 i
  \fs{q} - i \fs{p} \right) \nt\\
  &\qquad\qquad \times h_\Lambda (q+p) \frac{1}{2} \left(
  h_\Lambda (q+p_1) + h_\Lambda (q+p_2)\right) h_\Lambda (q)
  \Bigg]\nt\\
&= - \int_q \left(- \Lambda \partial_\Lambda \right) \,\Tr \left[
  \left(2 i \fs{q} + i \fs{p}\right)  h_\Lambda (q+p) \frac{1}{2} \left(
  h_\Lambda (q+p_1) + h_\Lambda (q+p_2)\right) h_\Lambda (q)\right],
\end{align}
we obtain
\begin{equation}
  - \Lambda \partial_\Lambda C (p_1, p_2) = 0,
\end{equation}
where $C (p_1,p_2)$ is defined by (\ref{trace-C}).

Now $C (p_1,p_2)$ is a scalar, local, and has the mass dimension $D-2$.
Hence, it has to vanish:
\begin{equation}
  C (p_1, p_2) = 0.
\end{equation}
We thus obtain
\begin{align}
  &C_{\Lambda;\mu\mu} (p_1,p_2)
  = - \frac{D-1}{2} \Bigg[ \F (p_1) + \F (p_2) \nt\\
  &\quad + \int_q \left(\RL (q) + \RL (q+p)\right) \Tr
    \lb h_\Lambda (q+p) \frac{1}{2} \left( h_\Lambda (q+p_1) + h_\Lambda
    (q+p_2) \right) h_\Lambda (q) \rb \Bigg].
\end{align}

\section{Special conformal invariance of a scalar theory\label{appendix-scalar}}

The present paper is a straightforward extension of the preceding
paper \cite{Pagani:2025dtc}, where we have constructed the EM tensor
of the large $N$ limit of the $\mathrm{O} (N)$ linear sigma model in
dimension $2 < D < 4$.  Even though we pointed out the invariance of
the critical theory under special conformal transformations, we did
not derive the corresponding WT identity.  Let us do it here by
following the recipe given in \cite{Sonoda:2015pva}.  The derivation
is very much the same as in Sec.~\ref{conformal}, and the resulting WT
identity is somewhat simpler since the EM tensor is symmetric.

In the large $N$ limit, the 1PI Wilson action of the critical theory
is given by
\begin{equation}
  \Gamma_\Lambda [\Phi^I] = \Gamma_G [\Phi^I] + N \Gamma_{I\Lambda} [\rho],
\end{equation}
where the Gaussian 1PI action is
\begin{equation}
  \Gamma_G [\Phi^I] \equiv - \frac{1}{2} \int_p p^2 \Phi^I (p) \Phi^I
  (-p),\label{appendix-scalar-Gauss}
\end{equation}
and
\begin{equation}
  \rho (p) \equiv \frac{1}{N} \int_q \Phi^I (q+p) \Phi^I (-q).
\end{equation}
We define
\begin{equation}
  \sigma (p) \equiv \frac{\delta \Gamma_{I\Lambda} [\rho]}{\delta \rho (-p)}.
\end{equation}
Though we use the same symbols, such as $\sigma (p)$, $\RL (p)$, and
$\G_{-q,p}$, their definitions are a little different for the scalar
theory.  See \cite{Pagani:2025dtc} for details.

The EM tensor of the critical theory satisfies three properties.  The first is 
the WT identity for translation invariance:
\begin{equation}
  p_\mu \Theta_{\mu\nu} (p)
  = \int_q (q+p)_\nu \Phi^I (q+p) \frac{\delta \Gamma_G}{\delta \Phi^I
    (q)}
  + N \int_q (q+p)_\nu \lb \RL (q) \G_{-q-p,q} [\sigma] + \rho (q+p)
  \sigma (-q) \rb,
  \label{appendix-scalar-WT}
\end{equation}
where $\G_{-q,p} [\sigma] = \G_{p,-q} [\sigma]$ is the solution to
\begin{equation}
  \int_q \G_{p,-q} [\sigma] \lb \left(q^2 + \RL (q)\right) \delta
  (q-r) - \sigma (r-q) \rb = \delta (p-r).
\end{equation}
The second is its symmetry
\begin{equation}
  \Theta_{\mu\nu} (p) = \Theta_{\nu\mu} (p) \label{appendix-scalar-sym}
\end{equation}
as a consequence of the rotation invariance.
The third is the trace formula:
\begin{equation}
  \Theta_\Lambda (p)
  = - \frac{D-2}{2} \N_\Lambda (p), \label{appendix-scalar-trace}
\end{equation}
where the number operator is given by
\begin{equation}
  \N_\Lambda (p)
  \equiv - \int_q \frac{\delta \Gamma_G}{\delta \Phi^I (q)} \Phi^I
    (q+p) - N \int_q \lb \RL (q) \G_{-q-p,q} [\sigma] + 2 \rho (q+p)
    \sigma (-q) \rb.
\label{appendix-scalar-number}
\end{equation}

We first derive the WT identity for scale invariance, and then the one
for special conformal invariance.  We use the same derivation as in
Sec.~\ref{conformal} and in Appendix \ref{appendix-free}.

\subsection{Scale invariance}

We compute
\begin{equation}
\lim_{p\to 0}  \frac{\partial}{\partial p_\mu} \left( p_\nu \Theta_{\nu\mu}
  (p)\right)
\end{equation}
in two ways.  Using (\ref{appendix-scalar-WT}), we obtain
\begin{align}
&  \lim_{p\to 0}  \frac{\partial}{\partial p_\mu} \left( p_\nu \Theta_{\nu\mu}
  (p)\right)\nt\\
&=  \int_q \left( q \cdot \partial_q + D \right) \Phi^I (q) \frac{\delta
  \Gamma_G}{\delta \Phi^I (q)} + N \int_q \left( q \cdot \partial_q +
  D \right) \rho (q) \cdot \sigma (-q)\nt\\
&\quad  + N \int_q \RL (q) \left( \frac{1}{2} q \cdot \partial_q + D \right) \G_{-q,q}.
\end{align}
Alternatively, using (\ref{appendix-scalar-trace}), we find
\begin{align}
  &  \lim_{p\to 0}  \frac{\partial}{\partial p_\mu} \left( p_\nu \Theta_{\nu\mu}
    (p)\right)\nt = \Theta_\Lambda (0) = - \frac{D-2}{2} \N_\Lambda (0)\nt\\
  &=  \frac{D-2}{2} \int_q \Phi^I \frac{\delta \Gamma_G}{\delta \Phi^I
    (q)} + N (D-2) \int_q \rho (q) \sigma (-q) + N \frac{D-2}{2}
    \int_q \RL (q) \G_{-q,q} .
\end{align}
Equating the two, we obtain the WT identity for scale invariance
\begin{equation}
  \int_q D^S \Phi^I (q) \frac{\delta \Gamma_G}{\delta \Phi^I (q)}
  + N \int_q D^S \rho (q)\, \frac{\delta \Gamma_{I\Lambda}
    [\varphi]}{\delta \varphi (q)} + N \frac{1}{2} \int_q \RL (q)
  \left( q \cdot \partial_q + \frac{D+2}{2} \right) \G_{-q,q} = 0,
\end{equation}
where the infinitesimal scale transformation of $\Phi^I$ and
  $\rho$ is given by
\begin{subequations}
\begin{align}
  D^S \Phi^I (q)
  &\equiv \left( q \cdot \partial_q + \frac{D+2}{2} \right) \Phi^I
    (q),\\
  D^S \rho (q)
  &\equiv \frac{1}{2 N} \int_p \left( D^S \Phi^I (-p) \cdot \Phi^I
    (q+p) + \Phi^I (-p) \cdot D^S \Phi^I (q+p) \right) \nt\\
  &= \left( q \cdot \partial_q + 2 \right) \rho (q) .
\end{align}
\end{subequations}
Since $\Gamma_G$ itself is scale invariant
\begin{equation}
  \int_q  D^S \Phi^I (q) \frac{\delta \Gamma_G}{\delta \Phi^I (q)} = 0,
\end{equation}
the WT identity reduces to
\begin{equation}
  \int_q D^S \rho (q)\, \frac{\delta \Gamma_{I\Lambda}
    [\varphi]}{\delta \varphi (q)} +  \frac{1}{2} \int_q \RL (q)
  \left( q \cdot \partial_q + \frac{D+2}{2} \right) \G_{-q,q} = 0.
\end{equation}

\subsection{Special conformal invariance}

We now compute
\begin{equation}
  \lim_{p \to 0} \left( \frac{\partial^2}{\partial p_\nu
    \partial p_\alpha} - 
  \frac{1}{2} \delta_{\nu\alpha} \frac{\partial^2}{\partial p_\beta
    \partial p_\beta} \right) \left(p_\mu \Theta_{\mu\nu} (p)\right)
\end{equation}
in two ways.  Using (\ref{appendix-scalar-WT}), we obtain
\begin{align}
  & \lim_{p\to 0} \left(\frac{\partial^2}{\partial p_\nu
    \partial p_\alpha} - 
  \frac{1}{2} \delta_{\nu\alpha} \frac{\partial^2}{\partial p_\beta
    \partial p_\beta} \right) \left( p_\mu \Theta_{\mu\nu} (p)
    \right)\nt\\
  &= \int_q \lb q_\nu \frac{\partial^2}{\partial q_\alpha \partial
    q_\nu} + D \frac{\partial}{\partial q_\alpha} - \frac{1}{2}
    q_\alpha \frac{\partial^2}{\partial q_\beta \partial q_\beta} \rb
    \Phi^I (q) \cdot \frac{\delta \Gamma_G}{\delta \Phi^I (q)}\nt\\
  &\quad + N \int_q \RL (q) \lb q_\nu \frac{\partial^2}{\partial q_\alpha \partial
    q_\nu} + D \frac{\partial}{\partial q_\alpha} - \frac{1}{2}
    q_\alpha \frac{\partial^2}{\partial q_\beta \partial q_\beta} \rb
    \G_{-q,p} \Big|_{p=q}\nt\\
  &\quad + N \int_q \sigma (-q) \lb q_\nu \frac{\partial^2}{\partial q_\alpha \partial
    q_\nu} + D \frac{\partial}{\partial q_\alpha} - \frac{1}{2}
    q_\alpha \frac{\partial^2}{\partial q_\beta \partial q_\beta} \rb
    \rho (q).
\end{align}
Alternatively, we can calculate
\begin{align}
& \lim_{p\to 0} \left(\frac{\partial^2}{\partial p_\nu
    \partial p_\alpha} - 
  \frac{1}{2} \delta_{\nu\alpha} \frac{\partial^2}{\partial p_\beta
    \partial p_\beta} \right) \left( p_\mu \Theta_{\mu\nu} (p)
  \right)\nt\\
&= \lim_{p\to 0} \left[ p_\mu \left(\frac{\partial^2}{\partial p_\nu
    \partial p_\alpha} - 
  \frac{1}{2} \delta_{\nu\alpha} \frac{\partial^2}{\partial p_\beta
    \partial p_\beta} \right)  \Theta_{\mu\nu} (p)
  + \frac{\partial}{\partial p_\alpha} \Theta_\Lambda (p) \right],
\end{align}
where we have used (\ref{appendix-scalar-sym}).  The first term
vanishes, and we obtain
\begin{align}
    \frac{\partial}{\partial p_\alpha} \Theta (p) \Big|_{p=0}
    &= - \frac{D-2}{2} \frac{\partial}{\partial p_\alpha} \N_\Lambda (p)
    \Big|_{p=0}\nt\\
  &= \frac{D-2}{2} \Bigg[
    \int_q \frac{\partial}{\partial q_\alpha} \Phi^I (q) \cdot
    \frac{\delta \Gamma_G}{\delta \Phi^I (q)}\nt\\
  &\quad - N \int_q \lb \RL (q) \frac{\partial}{\partial q_\alpha}
    \G_{-q,p}\Big|_{p=q} + 2 \frac{\partial}{\partial q_\alpha} \rho (q) \cdot
    \sigma (-q) \rb \Bigg].
\end{align}

Equating the two, we obtain the WT identity for special conformal invariance
\begin{align}
  & \int_q \frac{\delta \Gamma_G}{\delta \Phi^I (q)} D_\alpha^K
    \Phi^I (q)  + N \int_q \frac{\delta \Gamma_{I\Lambda}
    [\rho]}{\delta \rho (q)} D_\alpha^K \rho (q)\nt\\
  &\quad + N \int_q \RL (q) \lb - q_\nu \frac{\partial^2}{\partial
    q_\alpha \partial q_\nu} - \frac{D+2}{2} \frac{\partial}{\partial
    q_\alpha} + \frac{1}{2} q_\alpha \frac{\partial^2}{\partial
    q_\beta \partial q_\beta} \rb \G_{-q,p}\Big|_{p=q} = 0,
\end{align}
where the infinitesimal special conformal transformation of
$\Phi^I (q)$ and $\rho (q)$ is given by
\begin{subequations}
  \begin{align}
    D_\alpha^K \Phi^I (q)
    &\equiv \lb - q_\nu \frac{\partial^2}{\partial q_\alpha \partial
      q_\nu} - \frac{D+2}{2} \frac{\partial}{\partial q_\alpha} +
      \frac{1}{2} q_\alpha \frac{\partial^2}{\partial q_\beta \partial
      q_\beta} \rb \Phi^I (q),\\
    D_\alpha^K \rho (q)
    &\equiv \frac{1}{2N} \int_p \left( D_\alpha^K \Phi^I (-p) \cdot
      \Phi^I (q+p) + \Phi^I (-p) D_\alpha^K \Phi^I (q+p) \right)\nt\\
&=  \lb - q_\nu \frac{\partial^2}{\partial q_\alpha \partial
      q_\nu} - 2 \frac{\partial}{\partial q_\alpha} +
      \frac{1}{2} q_\alpha \frac{\partial^2}{\partial q_\beta \partial
      q_\beta} \rb \rho (q).
  \end{align}
\end{subequations}
Using the conformal invariance of the Gaussian action
\begin{equation}
  \int_q \frac{\delta \Gamma_G}{\delta \Phi^I (q)} D_\alpha^K \Phi^I
  (q) = 0,
\end{equation}
the WT identity for the special conformal invariance is reduced to
\begin{equation}
  \int_q \frac{\delta \Gamma_{I\Lambda}}{\delta \rho (q)} D_\alpha^K \rho (q)
 +  \int_q \RL (q) \lb - q_\nu \frac{\partial^2}{\partial
    q_\alpha \partial q_\nu} - \frac{D+2}{2} \frac{\partial}{\partial
    q_\alpha} + \frac{1}{2} q_\alpha \frac{\partial^2}{\partial
    q_\beta \partial q_\beta} \rb \G_{-q,p}\Big|_{p=q} = 0.
\end{equation}

Finally, using the identity (\ref{conformal-identity})
\begin{equation}
  \int_q \RL (q) 
\lb q_\nu \frac{\partial^2}{\partial
    q_\alpha \partial q_\nu}  - \frac{1}{2} q_\alpha \frac{\partial^2}{\partial
    q_\beta \partial q_\beta} \rb \G_{-q,p}\Big|_{p=q}
  = - \frac{1}{2} \int_q \left(D + q \cdot \partial_q \right) \RL (q)
  \cdot \frac{\partial}{\partial q_\alpha} \G_{-q,p} \Big|_{p=q},
\end{equation}
(valid if $\G_{-q,p} = \G_{p,-q}$ and $\RL (q)$ is a function of
$q^2$) we obtain the WT identity as
\begin{align}
  \int_q \frac{\delta \Gamma_{I\Lambda}}{\delta \rho (q)}
  D_\alpha^K \rho (q) - \frac{1}{2} \int_q \left(
  2 - q \cdot \partial_q \right) \RL (q) \cdot
  \frac{\partial}{\partial q_\alpha} \G_{-q,p} \Big|_{p=q} = 0.
\end{align}
The implication of this WT identity for special conformal invariance
has been studied in full details in a recent paper by Cabrera et
al. \cite{Cabrera:2026ggg}.



\begin{acknowledgments}
H.S. thanks the Department of Physics and Astronomy, the University of
Iowa, U.S.A. where he had an honorary appointment during the course of
this work.
\end{acknowledgments}

\bibliography{paper}

\end{document}